\documentclass[11pt]{article}

\usepackage{bm}
\usepackage{amsmath}
\usepackage{mathtools}
\usepackage{graphicx}
\usepackage{cite}
\usepackage[margin=1.25in]{geometry}
\usepackage{booktabs}
\usepackage{caption}
\usepackage{subcaption}
\usepackage{nicefrac}
\usepackage{placeins}
\usepackage{afterpage}
\usepackage{float}
\usepackage{relsize}
\usepackage{amsfonts}
\usepackage{mathrsfs}
\usepackage{color}
\usepackage{gensymb}

\graphicspath{{images/}}

\definecolor{olivedrab}{rgb}{0.42,0.56,0.14}
\definecolor{oxfordblue}{rgb}{0.0, 0.13, 0.28}

\newcommand{\sss}[1]{\scriptscriptstyle{#1}}
\newcommand{\sbs}[1]{_{\sss{#1}}}
\newcommand{\sps}[1]{^{\sss{#1}}}

\newcommand{\css}[0]{c\sbs{s}\sps{2}}

\newcommand{\chss}[0]{\hat{c}\sbs{s}\sps{2}}

\newcommand{\tensr}[1]{\bm{\mathsf{#1}}}
\newcommand{\F}{\tensr{F}}
\newcommand{\PP}{\tensr{P}}

\newcommand{\Fi}{\tensr{F}\sps{-1}}

\newcommand{\m}{\mathbf{m}}
\newcommand{\mc}{\mathbf{m\sps{c}}}

\newcommand{\n}{\mathbf{n}}
\newcommand{\nc}{\mathbf{n\sps{c}}}

\newcommand{\f}[1]{f\sbs{#1}}

\newcommand{\g}[1]{g\sbs{#1}}
\newcommand{\gt}[1]{\tilde{g}\sbs{#1}}

\newcommand{\K}{\kappa}
\newcommand{\KMv}[1]{\kappa\sbs{#1}\sps{Mv}}

\newcommand{\Ks}[1]{\kappa\sbs{#1}}
\newcommand{\Kts}[1]{\tilde{\kappa}\sbs{#1}}
\newcommand{\Kps}[1]{\kappa\sps{\prime}\sbs{#1}}
\newcommand{\Ktps}[1]{\tilde{\kappa}\sps{\prime}\sbs{#1}}

\newcommand{\Sig}{\sigma}

\newcommand{\Ss}[1]{\sigma\sbs{#1}}

\newcommand{\N}{\eta}
\newcommand{\NMv}[1]{\eta\sbs{#1}\sps{Mv}}
\newcommand{\Np}{\eta\sps{\prime}}
\newcommand{\Ns}[1]{\eta\sbs{#1}}
\newcommand{\Nts}[1]{\tilde{\eta}\sbs{#1}}
\newcommand{\Nps}[1]{\eta\sps{\prime}\sbs{#1}}
\newcommand{\Ntps}[1]{\tilde{\eta}\sps{\prime}\sbs{#1}}

\newcommand{\ux}{u\sbs{x}}
\newcommand{\uy}{u\sbs{y}}
\newcommand{\uz}{u\sbs{z}}
\newcommand{\uxx}{u\sbs{x}\sps{2}}
\newcommand{\uyy}{u\sbs{y}\sps{2}}
\newcommand{\uzz}{u\sbs{z}\sps{2}}

\DeclareMathSymbol{\shortminus}{\mathbin}{AMSa}{"39}

\newcommand{\sm}{\shortminus}

\title{\vspace{-2.0cm}Fokker-Planck Model Based Central Moment Lattice Boltzmann Method for Effective Simulations of Thermal Convective Flows}
\author{{William Schupbach, Kannan Premnath\footnotemark}\\Department of Mechanical Engineering\\ University of Colorado Denver}

\begin{document}

\maketitle

\begin{abstract}
The Fokker-Planck (FP) equation represents the drift and diffusive processes in kinetic models. It can also be regarded as a model for the collision integral of the Boltzmann-type equation to represent thermo-hydrodynamic processes in fluids. The lattice Boltzmann method (LBM) is a drastically simplified discretization of the Boltzmann equation for simulating complex fluid motions and beyond. We construct new two FP-based LBMs, one for recovering the Navier-Stokes equations for fluid dynamics and the other for simulating the energy equation, where, in each case, the effect of collisions is represented as relaxations of different central moments to their respective attractors. Such attractors are obtained by matching the changes in various discrete central moments due to collision with the continuous central moments prescribed by the FP model. As such, the resulting central moment attractors depend on the lower order moments and the diffusion tensor parameters, and significantly differ from those based on the Maxwell distribution. The diffusion tensor parameters for evolving higher moments in simulating fluid motions at relatively low viscosities are chosen based on a renormalization principle. Moreover, since the number of collision invariants of the FP-based LBMs for fluid motions and energy transport are different, the forms of the respective attractors are quite distinct. The use of such central moment formulations in modeling the collision step offers significant improvements in numerical stability, especially for simulations of thermal convective flows under a wide range of variations in the transport coefficients of the fluid. We develop new FP central moment LBMs for thermo-hydrodynamics in both two- and three-dimensions and demonstrate the ability of our approach to simulate various cases involving thermal convective buoyancy-driven flows especially at high Rayleigh numbers with good quantitative accuracy. Moreover, we show significant improvements in numerical stability of our FP central moment LBMs when compared to other existing central moment LBMs using the Maxwell distribution in achieving high Peclet numbers for mixed convection flows involving shear effects.
\end{abstract}

\let\thefootnote\relax\footnote{*Corresponding author (Email: Kannan.Premnath@ucdenver.edu)}

\newpage

\section{Introduction}
Fluid motions coupled with thermal energy transport arise in a wide range of situations and scales, from atmospheric convection and heat transfer in process equipment and power generation devices to bioengineering systems and microfluidic applications. Macroscopically, the dynamics of fluids is represented by the Navier-Stokes equations (NSE) while the thermal transport is generally modeled using a convective-diffusion equation (CDE) for the temperature field, sometimes classified as having a combination of hyperbolic and parabolic behavior~\cite{minkowycz2009handbook}. As such, the solution of the CDEs is needed not only in understanding or making predictions in heat transfer applications but also in the transport of passive scalar fields involving chemical species, in the interface capturing in multiphase flows using phase field models, and in modeling of surfactants (for recent examples, see~\cite{ tripathi2018motion,elbousefi2023thermocapillary,farsoiya2024coupled,elbousefi2024investigation} that involve a combination of such cases and other related applications~\cite{li2025multi,zheng2025study}). Of specific interest to this work is the computation of buoyancy-driven thermal convective flows~\cite{siggia1994high}, which involve a two-way coupling between the solution of NSE and CDE for the temperature field.

Obtaining solutions to such a set of coupled partial differential equations for realistic situations using analytical methods is infeasible. For that reason, the development and use of numerical techniques plays a fundamental role in simulating and studying the complex behavior of thermal transport phenomena in fluids. Traditionally, methods such as the finite difference, finite element, and especially finite volume methods (see e.g.,~\cite{pletcher2012computational,ferziger2019computational}) have been employed in discretizing the macroscopic governing equations; they are often used in conjunction with other techniques such as multigrid methods (see e.g.,~\cite{mulder1989new,cheikh2007benchmark,moraga2017geometric}) for convergence acceleration and have been investigated previously in detail and are quite commonplace. On the other hand, the focus of this paper is on the development of a novel formulation of lattice Boltzmann (LB) method originating from kinetic theory and in particular as a dramatic simplification of the continuous Boltzmann equation. Much of the recent interest in such methods arise from their computational advantages including their effective implementation on modern parallel computing architectures due the local nature of the algorithms, ease of implementation of boundary conditions using Cartesian grids, and their natural ability to use more fundamental mesocopic models based on kinetic theory to represent complex flow physics~\cite{benzi1992lattice, chen1998lattice, lallemand2021lattice}. Moreover, the LB schemes are flexible in that it can be combined with concepts used in conventional approaches such as unstructured grids via finite volume methods to handle flows in complex geometries (see e.g.,~\cite{xi1999finite,patil2009finite,zarghami2014finite,li2016finite}).

As such, the LB method is an efficient numerical approach for solving the mildly compressible NSE, as well as for convection-diffusion type equations, and many other models for fluid dynamics with multiphysics effects. The underlying idea in the LB method is that a simplification of the continuous Boltzmann equation~\cite{he1997theory} is used to track the density distribution function of fluid particles, whose motion has been restricted to discrete velocity directions, called lattice links, as they move around and collide. More specifically, the collision process is modeled as relaxations of those distribution functions towards their `equilibrium' states, typically represented as a discretized form of the Maxwell distribution, while accounting for any relevant conservation laws. Following this, the distributions are then streamed or advected along the discrete lattice links, and the fluid dynamical variables are then computed via taking the moments of the distribution functions with respect to the powers of the particle velocity components. The LB methods have been applied to simulate a wide range of problems involving multiphase and multicomponent flows, particulate suspension flows, thermal flows, turbulence, combustion, magnetohydrodynamics and aeroacoustics (see e.g.,~\cite{aidun2010lattice,jahanshaloo2013review,li2016lattice,sharma2020current} for some recent reviews).

Modeling the collision process in the LB method is a decisive feature used to extend the physics of flow characteristics in the simulation and, more generally, the numerical stability; to that end, many collision models have been developed.  The first collision model constructed is known as the single relaxation time model (SRT)~\cite{qian1992lattice}, setting the stage for modeling collision as a relaxation process using a single characteristic time scale despite the fact the underlying kinetic phenomena can be multiscale in nature; however, it was found to have serious limitations in terms of numerical stability. To alleviate this issue, the multiple relaxation time (MRT) collision model was developed which relaxes different bare (or raw) moments of the distribution functions at different rates~\cite{d1992generalized,d2002multiple}, which was found to have significant improvements in numerical stability compared to the SRT model. Moreover, as a further improvement, the central moment collision model, which is also referred to as the cascaded collision model, was then constructed~\cite{geier2006cascaded}, which involves performing collisions using central moments. Here, the central moments are formed by taking moments of the distribution functions with respect to the powers of the components of the particle velocity being shifted by the bulk fluid velocity. Such an approach entails the use of additional transformations between raw moments and central moments when performing the collision step; as such, the latter involves the relaxations of central moments to the equilibria defined by the central moments of the Maxwell distribution and this approach resulted in significantly enhancing the numerical stability when compared with the MRT model while naturally maintaining Galilean invariance (see Ref.~\cite{ning2016numerical} for a comparative numerical study of these collision models). However, it was found to have numerical hyperviscosity artifacts when such relaxations are done at different rates for different central moments in simulating high Reynolds number flows. In order to address this issue, a factorized central moment collision model was introduced~\cite{geier2009factorized}, which exploits the factorization property of the Maxwell distribution to construct the attractors in an \emph{ad hoc} manner. Another solution to this issue was provided by developing a collision model that relaxes the cumulants of the distribution function toward their attractors~\cite{geier2015cumulant}, which involves an additional transformations for mappings between central moments to cumulants resulting in added complexity.

More recently, we proposed a new approach to the collision modeling in LB methods that involves using the central moments of the Fokker-Planck (FP) equation to simulate fluid dynamics at high Reynolds numbers, which was shown to naturally resolve the numerical hyperviscosity issues while also showing superior numerical stability properties resembling those observed in the cumulant collision model, but using fewer transformations and simpler formulations~\cite{schupbach2024fokker}. As such, the FP equation is an important formulation in statistical mechanics for evolving the density distribution function of particles under drift-diffusive processes in a variety of applications (see e.g.,~\cite{van1992stochastic,wax2014selected,reichl2016modern,luscombe2021statistical}); it is equivalent to Langevin equation, a stochastic differential equation originally derived to model Brownian dynamics (see e.g.,~\cite{medved2020understanding} for a pedagogical discussion on these topics) and is sometimes also referred to as the Kolmogorov forward equation or as the Smoluchowski equation. It has been used to model the collision term of the Boltzmann equation by Lebowitz \emph{et al}~\cite{lebowitz1960nonequilibrium} to represent fluid dynamics and it has been shown to be a rigorous approximation to the Boltzmann's collision integral under grazing collisions and a diffusion limit~\cite{cercignani1988Boltzmann}. As shown in our recent work, a central moment formulation of the FP collision model results in a local and algebraic operator that is naturally suitable for algorithmic implementation in LB methods for effective simulations of fluid dynamics~\cite{schupbach2024fokker}. One key feature of this FP-based central moment LBM is that the resulting attractors contain products of lower order central moments and a diffusion tensor parameter and latter is specified as fixed points of the renormalization group equation representing the variations of the second order moments under collision when there is scale separation in the relaxation rates of kinetic processes. These considerations were shown to be naturally circumventing the hyperviscosity artifacts while also improving the numerical stability significantly when simulating flows at relatively very low physical viscosities or high Reynolds numbers~\cite{schupbach2024fokker} unlike the Maxwellian based central moment LB formulations which are susceptible to numerical issues. One of the major objectives of this paper is to extend such FP-based central moment LB schemes to simulate thermo-hydrodynamics, i.e., specifically to simulate the energy transport equation in addition to the NSE as done in the previous work. Based on the above discussion, the various key features and numerical attributes associated with different collision models in the LBM are summarized and presented in Table~\ref{comparisons_collision_models}.

\begin{table}[H]
\centering
\begin{tabular}{l l l l l l l l l l l l l}
\hline
\addlinespace
Collision model& &Feature& &Stability      & &Hyperviscosity&  \\
   in LBM      & &       & &characteristics& &effects&  \\
\hline
\hline
\addlinespace
SRT-LBM      & & Relaxes distribution functions  & & Low  & & No  &\\
             & & to their equilibria at same rate      & &      &     &\\[ 0.05in]
MRT-LBM      & & Relaxes raw moments to their    & & Moderate  & & Yes  &\\
             & & equilibria at different rates      & &              & &     &\\[ 0.05in]
MCM-LBM      & & Relaxes central moments to  & & High  & & Yes  &\\
             & & Maxwellian equilibria      & &              & &     &\\
             & & at different rates      & &              & &     &\\[ 0.05in]
FPC-LBM      & & Relaxes central moments to & & Very high  & & No  &\\
             & & FP-based Markovian attractors      & &              & &     &\\
             & & at different rates and using       & &              & &     &\\
             & & a renormalization principle      & &              & &     &\\[ 0.05in]
Cumulant LBM & & Relaxes cumulants to their & & Very high  & & No  &\\
             & & equilibria at different rates     & &              & &     &\\[ 0.05in]
\hline
\hline
\end{tabular}
\caption{Comparative summary of the features, numerical stability characteristics for simulations of flows at high Reynolds numbers, and the emergence of numerical hyperviscosity effects for flows at very low physical viscosities with different collision models in LBM: single relaxation time (SRT)-LBM, multiple relaxation times (MRT)-LBM, Maxwellian-central moments (MCM)-LBM, Fokker-Planck central moment (FPC)-LBM, and cumulant LBM.}
\label{comparisons_collision_models}
\end{table}

Before discussing it in more detail, let us briefly review the prior work on developing LB schemes for flows with heat transfer. As such, they can be classified into the following three major types -- 1) multispeed LB schemes using a single distribution function to compute both the hydrodynamic and temperature fields, 2) the hybrid approaches using a LB scheme for fluid motions and a finite difference method the energy transport, and 3) the double distribution functions (DDF)-based formulations that use one LB scheme for the NSE and another one for the energy equation (see e.g.,~\cite{kruger2017lattice,sharma2020current} for further details). Out of these, the DDF approach has become the most popular due to stability and efficiency considerations. Earlier investigations adopted the SRT collision model in the DDF-LB schemes (see~\cite{he1998novel,guo2002coupled}), which was followed by the use of MRT models to simulate thermal convective flows in this framework~\cite{wang2013lattice,contrino2014lattice}. More recently, the central moments-based collision models were utilized to construct DDF-LB schemes in two-dimensions (2D)~\cite{sharma2018natural,elseid2018cascaded,fei2018modeling}, in three-dimensions (3D)~\cite{hajabdollahi2018central}, and in axisymmetric geometries~\cite{hajabdollahi2019cascaded} with improved stability characteristics when compared to the methods based on SRT and MRT collision models. It should be noted that all such central moments-based formulations used the Maxwell distribution function, which depend only on the conserved hydrodynamic fields, in constructing the collision step.

As further improvements over these developments, in this paper, we will develop new DDF-based LB schemes using the FP central moment (FPC) collision operators in both 2D and 3D. In 2D, we will use the two-dimensional, nine velocity (D2Q9) lattice to construct the FPC-LBMs for both fluid motions and energy transport, and in 3D we will consider the three-dimensional, twenty-seven velocity (D3Q27) lattice for FPC-LBM for hydrodynamics and three-dimensional, fifteen velocity (D3Q15) lattice for FPC-LBM for temperature field. In each case, by matching the variations in the discrete central moments under collision to the corresponding continuous central moments of the FP collision operator, we will construct the equivalent central moment attractors. Since the number of collision invariants of the collision operators for fluid motions and energy transport in the DDF framework are different, it will be shown that the resulting attractors will be quite distinct. We will then validate these new LB schemes for buoyancy-driven thermal convective flows in square and cubic cavities over a wide range of Rayleigh numbers by comparing them against the respective benchmark numerical solutions. Finally, we will then demonstrate the advantages of using these FPC-LBMs for simulating mixed thermal convection at higher Peclet numbers than what can be achieved using the standard central moment LBMs using attractors based on the Maxwell distribution.

This paper is organized as follows. In the next section (Sec.~\ref{sec:macroeqns}), we will summarize the target macroscopic governing equations for thermal convective flows, which is followed by a highlight of the two continuous Boltzmann-type kinetic equations for representing hydrodynamics and energy transport in Sec.~\ref{sec:DDFKineticEqns}. Then, Sec.~\ref{sec:FPC-LBMformulations2D} discusses the 2D continuous formulations and the discrete LB schemes based on the Fokker-Planck central moments-based collision models, while their extensions to 3D are presented in Sec.~\ref{sec:FPC-LBMformulations3D}. Subsequently, the numerical case studies for the validation of the 2D and 3D FPC-LBMs are elaborated in Sec.~\ref{sec:resultsanddiscussion}. Finally, a comparative study of numerical stability of the FPC-LBMs and the prior central moment LBMs that use the Maxwell distribution are discussed in Sec.~\ref{sec:stabilitytest}; this is followed by a summary of the new developments and findings resulting from this paper in Sec.~\ref{sec:summary}. Additional attendant details of this work, including the implementation details of the FP-LBMs in 2D and 3D as well as the guidance on the selection of the model parameters, are collected together in various appendices.

\section{Macroscopic Governing Equations of Thermal Convective Flows\label{sec:macroeqns}}
In this paper, our main goal is to develop and investigate novel LB schemes based on Fokker-Planck central moment-based collision modeling for effective simulations of thermal convective flows; based on stability considerations, the approach will involve the use of dual distribution functions -- one for obtaining the solutions of the hydrodynamic fields and another for solving the temperature field. In this regard, first, we briefly recall the attendant macroscopic governing equations. The mass and momentum equations of the fluid motions read as
\begin{equation}
\frac{\partial \rho}{\partial t} + \bm{\nabla}\cdot(\rho\bm{u})=0,
\label{mass}
\end{equation}
and
\begin{equation}
\frac{\partial (\rho \bm{u}) }{\partial t} + \bm{\nabla}\cdot(\rho\bm{u}\bm{u})= - \bm{\nabla}P + \bm{\nabla}\cdot \tensr{\sigma}\sbs{\nu} + \bm{F},
\label{momentum}
\end{equation}
where $\rho$ is the fluid density and $\bm{u}$ is the bulk fluid velocity. Moreover, $P$ is the fluid pressure, $\bm{\tensr{\sigma}}\sbs{\nu}$ is the viscous stress tensor, and $\bm{F}$ is an external body force (e.g., the buoyancy force). On the other hand, the temperature field satisfies the following energy transport equation that formally has the form of a convection-diffusion equation (CDE):
\begin{equation}
\frac{\partial T}{\partial t} + \bm{\nabla}\cdot(\bm{u}T)=  \bm{\nabla}\cdot\left(\alpha \bm{\nabla}T \right),
\label{energy}
\end{equation}
where  $T$ is the local temperature, and $\alpha = k/(\rho C\sbs{p})$ represents the thermal diffusivity, where $k$ is the thermal conductivity and $C\sbs{p}$ is the specific heat at constant pressure. Note that since any other scalar $\phi$ (such as the concentration field) also obeys the same CDE as above for the temperature field, the novel collision models and LB schemes discussed in this work are directly applicable in such other cases.

\section{Double Distribution Functions-Based Continuous Boltzmann-type Kinetic Equations using for Thermal Convective Flows\label{sec:DDFKineticEqns}}
In general, the continuous Boltzmann equation, which evolves a single particle density distribution function, is a more fundamental representation of the thermo-hydrodynamic processes at a kinetic level than the macroscopic governing equations given in the previous section. It can be written as
\begin{eqnarray}\label{boltzmann_equation}
\frac{\partial f}{\partial t}+ \bm{\xi}\cdot\bm{\nabla}f = \left(\frac{\delta f}{\delta t}\right)\sbs{\!\!\!coll}\sps{\!\!\!Boltz}+\left( \frac{\delta f}{\delta t} \right)\sbs{\!\!\!forcing},
\end{eqnarray}
where $f=f(\bm{x},\bm{\xi},t)$ is the density distribution function at time $t$ with the position and the particle velocity represented by $\bm{x}$ and $\bm{\xi}$, respectively. Here, the left hand side of Eq.~(\ref{boltzmann_equation}) accounts for changes in the distribution function $f$ due to particle advection, sometimes referred to as streaming. On the other hand, the particles interact through collision that redistributes their velocities, while conserving the mass, momentum and energy of the interacting particle pair, and they are taken into consideration in the time variation in $f$ through the Boltzmann's binary collision term $\left(\frac{\delta f}{\delta t}\right)_{\!\!coll}\sps{\!\!\!Boltz}$ given by (see e.g.,~\cite{kremer2010introduction})
\begin{eqnarray}\label{collision_integral}
\left(\frac{\delta f}{\delta t}\right)\sbs{\!\!\!coll}\sps{\!\!\!Boltz} =  \int w b(f\sps{\prime}f\sbs{1}\sps{\prime} - f f\sbs{1} ) \,      \mathrm{d}b \, \mathrm{d}\varepsilon \, \mathrm{d}\xi\sbs{i}.
\end{eqnarray}
Here, the distribution functions of the particle pairs pre- and post-collision are represented by $f \equiv f(\bm{x}, \bm{\xi}, t)$ , $f\sbs{1} \equiv f(\bm{x}, \bm{\xi}\sbs{1}, t)$, and $f\sps{\prime} \equiv f(\bm{x}, \bm{\xi}\sps{\prime}, t)$ , $f\sbs{1}\sps{\prime} \equiv f(\bm{x},\bm{\xi}\sbs{1}\sps{\prime}, t)$, respectively, with $w = |\bm{\xi} - \bm{\xi}\sbs{1} |$ being the relative speed, $\varepsilon$ the azimuthal collision angle, and b the impact parameter. These are accounted for in the first term in the right hand side of Eq.~(\ref{boltzmann_equation}). Moreover, if the fluid is subjected to an external body force $\bm{F}$, the resulting acceleration alters the distribution function $f$ in a manner represented through the second term $\left(\frac{\delta f}{\delta t}\right)_{\!\!forcing}$ on the right hand side of Eq.~(\ref{boltzmann_equation}) given by
\begin{equation}\label{eq:boltzmannforcingterm}
\left(\frac{\delta f}{\delta t}\right)\sbs{\!\!forcing} = - \frac{F}{\rho}\cdot\bm{\nabla}\sbs{\xi} f,
\end{equation}
where $\bm{\nabla}_\xi$ is the gradient in the velocity space. By taking various statistical moments of the distribution function with respect to the certain monomials of the particle velocity of zeroth, first and second order, the fluid mass density, momentum and temperature, respectively, can be recovered. However, on the one hand, the continuous Boltzmann equation is formidable to use since the particle velocity space is infinite and its collision term is a multidimensional integral; on the other hand, if the particle velocity space is discretized using a smaller number of velocities to construct a lattice Boltzmann method (LBM) while using a simpler tractable model for collision, the resulting computation of mass density, momentum and temperature altogether using a single distribution function is known to lead to numerical instabilities. Hence, practical considerations suggest the use of double distribution functions in LBM, where one of them is used to recover the hydrodynamic fields and the other the temperature (see Ref.~\cite{he1998novel}).

Thus, a modified continuous kinetic equation to compute the hydrodynamic fields, which recovers Eqs.~(\ref{mass}) and (\ref{momentum}), is given by
\begin{eqnarray}\label{boltzmann_equation1}
\frac{\partial f}{\partial t}+ \bm{\xi}\cdot\bm{\nabla}f = \left(\frac{\delta f}{\delta t}\right)\sbs{\!\!\!coll}+\left( \frac{\delta f}{\delta t} \right)\sbs{\!\!\!forcing},
\end{eqnarray}
where the collision term satisfies the following constraints reflecting the mass and momentum collision invariants (unlike the Boltzmann's collision term which also respects an additional invariant related to energy):
\begin{equation}
\int \left(\frac{\delta f}{\delta t}\right)\sbs{\!\!\!coll} d\bm{\xi} = 0, \qquad \int \left(\frac{\delta f}{\delta t}\right)\sbs{\!\!\!coll}\bm{\xi}d\bm{\xi} = 0. \label{collisioninvariantf}
\end{equation}
Then, the fluid density $\rho$ and bulk velocity $\bm{u}$ can be obtained by taking the zeroth and first moments, respectively, of the distribution function $f$ obtained from solving Eq.~(\ref{boltzmann_equation1}), which read as
\begin{equation}
\rho = \int fd\bm{\xi}, \qquad \rho\bm{u}= \int f\bm{\xi}d\bm{\xi}.
\end{equation}
The second distribution function $g$ used to recover the energy transport equation (Eq.~(\ref{energy})) can be written as
\begin{eqnarray}\label{boltzmann_equation_two}
\frac{\partial g}{\partial t}+ \bm{\xi}\cdot\bm{\nabla}g = \left(\frac{\delta g}{\delta t}\right)\sbs{\!\!\!coll},
\end{eqnarray}
where the local temperature is obtained by taking the zeroth order moment of $g$. That is,
\begin{equation}
T = \int g d\bm{\xi}.
\end{equation}
Its collision term thus satisfies the following separate invariant:
\begin{equation}
\int \left(\frac{\delta g}{\delta t}\right)\sbs{\!\!\!coll} d\bm{\xi} = 0. \label{collisioninvariantg}
\end{equation}

\section{Fokker-Planck (FP) Central Moments-Based Collision Modeling: 2D Formulations and LB Schemes\label{sec:FPC-LBMformulations2D}}
One of our objectives in this manuscript is to construct robust models for the essential features of the collision operators in Eqs.~(\ref{boltzmann_equation1}) and (\ref{boltzmann_equation_two}), viz., $\left(\frac{\delta f}{\delta t}\right)\sbs{\!\!\!coll}$ and $\left(\frac{\delta g}{\delta t}\right)\sbs{\!\!\!coll}$ based on the Fokker-Planck central moment formulations and subsequently develop effective LB algorithms for thermal convective flows in a double distribution functions (DDF)-based framework, which will be discussed in what follows.

\subsection{Construction of FP Central Moment Collision Operator: Fluid Motions}
Since, as mentioned earlier, the collision term of the Boltzmann equation is computationally intractable, it is generally replaced with simpler models that retain the essential characteristics of the collision process. Many collision models have been constructed over the years; one popular model is the Bhatnagar-Gross-Krook (BGK) model~\cite{bhatnagar1954model} given by
\begin{eqnarray}\label{BGK_collision}
\left(\frac{\delta f}{\delta t} \right)\sbs{\!\!\!coll}\sps{\!\!\!BGK} = \omega\sbs{BGK} (f\sps{M}-f),
\end{eqnarray}
where $f\sps{M}$ is the Maxwell distribution function written as
\begin{eqnarray}\label{maxwell_distribution}
f\sps{M} = \frac{\rho}{(2\pi \css)\sps{\frac{N}{2}}}\exp\left[\frac{-(\bm{\xi}-\bm{u})\sps{2}}{2\css}\right],
\end{eqnarray}
and $\omega\sbs{BGK}$ is the characteristic frequency of collisions. In Eq.~(\ref{maxwell_distribution}), which is essentially the Gaussian or normal distribution, $\rho$ and $\bm{u}$ are the fluid density and bulk velocity, respectively, and $c_s$ is the speed of sound, and N is the number of spatial dimensions. As such, Eq.~(\ref{BGK_collision}) models the effect of collisions as a relaxation process toward an equilibrium state defined in Eq.~(\ref{maxwell_distribution}), and is a dramatically simplified approximation for the collision integral (Eq.~(\ref{collision_integral})). When used in constructing LB methods via suitable discretization in a DDF-based framework, the BGK model leads to numerical instability issues especially in simulating flows with relatively low viscosities or thermal diffusivities (or, equivalently, high Reynolds numbers or Peclet numbers). Thus, we look for alternatives that can address these limitations.

Another simplified model for the collision term is the Fokker-Planck (FP) model~\cite{lebowitz1960nonequilibrium}. While the FP equation was originally derived to model the Brownian dynamics involving drift and diffusion of particles, it is also considered as a model for the Boltzmann's collision integral term when the particle pairs undergo grazing collisions with small velocity changes that commonly occur in scattering processes with inverse power interaction potentials and associated with large impact parameters (see e.g., Refs.~\cite{cercignani1988Boltzmann,gombosi1994gaskinetic,liboff2003kinetic}). The FP collision model can be represented as~\cite{lebowitz1960nonequilibrium}
\begin{eqnarray}\label{FP_collision}
\left( \frac{\delta f}{\delta t}\right)\sbs{\!\!\!coll}\sps{\!\!\!FP} = \omega\sbs{FP} \left[\frac{\partial}{\partial \xi\sbs{i}}((\xi\sbs{i} - u\sbs{i})f) + D\sbs{ij}' \frac{\partial\sps{2} f}{\partial\xi\sbs{i} \partial \xi\sbs{j}} \right].
\end{eqnarray}
Here, $D\sps{\prime}\sbs{ij}$ is the diffusion tensor parameter that is a function of second order central moments or the variance of the distribution function (which will be further elaborated in later sections), whereas the effective diffusion rate coefficient $D_{ij}$ is dependent on the diffusion tensor parameter $D'_{ij}$ via the FP relaxation frequency  $\omega\sbs{FP}$ as $D\sbs{ij}=D\sps{\prime}\sbs{ij}\omega_{FP}$. The FP collision model given in Eq.~(\ref{FP_collision}) represents the effect of collisions in terms of drift and diffusion of the particle distribution in the velocity space. The drift term (or the first term in the right hand of Eq.~(\ref{FP_collision})) represents a process similar to that of dynamic friction and brings to light the idea that collisions act to eliminate gradients in the particle velocities of the fluid; on the other hand, the diffusion term (or the second term in the right hand of Eq.~(\ref{FP_collision})) models the spread of the distribution function under collision via molecular chaos~\cite{gombosi1994gaskinetic}. These mesoscopic processes of particle populations as modeled by the FP formulation, under coarse-graining, i.e., at larger length and time scales, correspond to the fluid dynamics of the bulk continuum as represented by the Navier-Stokes equations which can be established via the Chapman-Enskog multiscale expansions (see e.g.,~\cite{schupbach2024fokker}). Interestingly, the FP collision model (Eq.~(\ref{FP_collision})) maintains the quadratic non-linearity seen in Boltzmann's collision integral (Eq.~(\ref{collision_integral})) since $u\sbs{i}$ and $D\sps{\prime}\sbs{ij}$ appearing in Eq.~(\ref{FP_collision}) could themselves be functions of the distributions, $f$. We also note here that Eq.~(\ref{FP_collision}) is more general than that of the FP model introduced in~\cite{lebowitz1960nonequilibrium} since we consider the diffusion parameter $D\sps{\prime}\sbs{ij}$ to be a tensor rather than a scalar $D\sps{\prime}$.

As discussed in our recent work~\cite{schupbach2024fokker}, using Eq.~(\ref{FP_collision}) directly in constructing LB algorithms is cumbersome since it involves evaluating the derivatives of the distribution function in the velocity space, while they can be effectively and exactly represented by reformulating in terms of their central moments. To that end, we begin with a formal definition of the inner product of any two objects $a$ and $b$ as being the product of $a$ and $b$, integrated over velocity space in 2D as
\begin{eqnarray}
\left<a,b\right>= \int\displaylimits\sbs{-\infty}\sps{\infty} \int\displaylimits\sbs{-\infty}\sps{\infty} a \; b\; \mathrm{d}\xi\sbs{x}\; \mathrm{d}\xi\sbs{y}.\nonumber
\end{eqnarray}
Then, we introduce the central moment weighting factor $W\sbs{mn}$ of order ($m+n$) based on the products of the monomials of the components of the particle velocity shifted by the bulk fluid velocity $\xi\sbs{x} - u\sbs{x}$ and $\xi\sbs{y} - u\sbs{y}$ of order $m$ and $n$, respectively, as
\begin{eqnarray}\label{cm_weights}
W\sbs{mn}=(\xi\sbs{x}-u\sbs{x})\sps{m}(\xi\sbs{y}-u\sbs{y})\sps{n}.
\end{eqnarray}
We can then define the continuous central moment of the distribution function $f$ of order ($m+n$) as the inner product of the distribution function $f$ and $W\sbs{mn}$ as
\begin{eqnarray}\label{central_moments}
\Pi\sbs{mn}=\left<f,W\sbs{mn}\right>.\nonumber
\end{eqnarray}
Next, for convenience, we rewrite the two terms in the FP collision model given in Eq.~(\ref{FP_collision}) in a shorthand notation as
\begin{eqnarray}\label{FP_short}
\left( \frac{\delta f}{\delta t}\right)\sbs{\!\!\!coll}\sps{\!\!\!FP} = \omega\sbs{FP}\left(\frac{\delta f}{\delta t}\sps{FP1}+\frac{\delta f}{\delta t}\sps{FP2}\right),
\end{eqnarray}
where
\begin{eqnarray}\label{dFP}
\frac{\delta f}{\delta t}\sps{FP1} = \frac{\partial}{\partial \xi\sbs{i}}((\xi\sbs{i}-u\sbs{i})f), \quad
\frac{\delta f}{\delta t}\sps{FP2} = D\sps{\prime}\sbs{ij}\frac{\partial\sps{2} f}{\partial\xi\sbs{i}\partial\xi\sbs{j}} =  D\sps{\prime}\sbs{xx}\frac{\partial\sps{2} f}{\partial \xi\sbs{x}\sps{2}}+D\sps{\prime}\sbs{yy}\frac{\partial\sps{2} f}{\partial \xi\sbs{y}\sps{2}} + 2D\sps{\prime}\sbs{xy}\frac{\partial\sps{2} f}{\partial \xi\sbs{x}\partial \xi\sbs{y}}.
\end{eqnarray}
We can then define the changes of central moments of order ($m+n$) under collision, as given by the FP model, by considering the inner product of each term in the right hand side of Eq.~(\ref{FP_short}) along with the attendant terms in Eq.~(\ref{dFP}), with the central moment weighting factors of order $(m+n)$ given in Eq.~(\ref{cm_weights}), as in
\begin{eqnarray}\label{cmchanges}
\left(\frac{\delta \Pi\sbs{mn}}{\delta t}\right)\sbs{\!\!\!coll}\sps{\!\!\!FP} = \omega\sbs{FP}\left[\left<\frac{\delta f}{\delta t}\sps{FP1},W\sbs{mn}\right> + \left<\frac{\delta f}{\delta t}\sps{FP2},W\sbs{mn}\right>\right].
\end{eqnarray}
Then, after integrating them exactly and upon rearrangement, see~\cite{schupbach2024fokker} for details, the refined central moments-based FP collision operator in 2D reads as
\begin{eqnarray}\label{FPcollision}
\left( \frac{\delta}{\delta t} \Pi\sbs{mn}\right)\sbs{\!\!\!coll}\sps{\!\!\!FP} = \omega\sbs{mn}\left[ \Pi\sbs{mn}\sps{Mv}-\Pi\sbs{mn}\right],
\end{eqnarray}
where
\begin{eqnarray}\label{eq:Mvcmattractor2D}
\Pi_{mn}\sps{Mv} = \frac{m(m-1)}{m+n}D\sps{\prime}\sbs{xx}\Pi\sbs{m\sm2,n} + \frac{n(n-1)}{m+n}D\sps{\prime}\sbs{yy}\Pi\sbs{m,n\sm2}+ \frac{2mn}{m+n}D\sps{\prime}\sbs{xy}\Pi\sbs{m\sm1,n\sm1}.
\end{eqnarray}
Some important remarks are in order here. (i) According to Eq.~(\ref{FPcollision}), collision relaxes the central moment of order ($m+n$) to an attractor $\Pi_{mn}\sps{Mv}$ at a rate $\omega\sbs{mn}$. Due to the molecular chaos and repeated randomness or Markovian nature of the collision process, we refer to $\Pi_{mn}\sps{Mv}$ as the Markovian central moment attractor. (ii) Interestingly, Eq.~(\ref{eq:Mvcmattractor2D}) for $\Pi_{mn}\sps{Mv}$ is related to the products of lower central moments of order $(m+n-2)$ and the diffusion tensor parameter $D\sps{\prime}\sbs{ij}$. Since the components of the diffusion tensor parameter $D\sps{\prime}\sbs{ij}$ will be shown in what follows to be dependent on the variance of the distribution function or the second order central moments, when evolving higher moments under collision, their attractors themselves depend nonlinearly on the lower order moments and the second order central moments; by contrast, if a central moment collision operator based on using the continuous Maxwell distribution (see Eq.~(\ref{maxwell_distribution})) is constructed, the resulting attractor can be shown to depend at most only on the zeroth moment (or the density) and the parameter $c_s^2$. In other words, the FP central moments-based collision model utilizing the Markovian attractor can account for more general and flexible representation of the relaxation process of higher moments than that based on the Maxwellian. (iii) Moreover, while original formulation of the FP model given in Eq.~(\ref{FP_collision}) is differential in nature, the FP central moment collision operator presented in Eq.~(\ref{FPcollision}) is algebraic and local, which greatly facilitates its practical numerical implementation within LB framework as shown later. (iv) Finally, the relaxation rate $\omega\sbs{mn}$ for the central moment of order ($m+n$) is a free parameter and can be, in general, different for evolving different moments. This is similar to generalizing the BGK model with a single rate constant~\cite{qian1992lattice} towards using multiple relaxation times for evolving different raw moments by their relaxation to equilibria dependent only on the conserved hydrodynamic fields such as fluid density and velocity, a strategy adopted in some previous LB approaches~\cite{d1992generalized,d2002multiple}. However, our approach uses a different model for collision based on drift-collision processes formulated in terms of central moments and their relaxation to ``equilibria'' that depend on both conserved and non-conserved moments in a recurrent fashion.

\subsubsection{Choice of Diffusion Tensor Parameter}\label{subsec:choiceDij2D}
An important feature of the Markovian attractor given in Eq.~(\ref{eq:Mvcmattractor2D}) appearing in the collision operator in Eq.~(\ref{FPcollision}) is the diffusion tensor parameter $D\sps{\prime}\sbs{ij}$ whose choice plays a primary role in its implementation for achieving accurate and effective numerical solutions. First, we introduce the following convenient notation:
\begin{eqnarray}
D\sps{\prime}\sbs{20}=D\sps{\prime}\sbs{xx}, \quad D\sps{\prime}\sbs{02} = D\sps{\prime}\sbs{yy}, \quad D\sps{\prime}\sbs{11}=D\sps{\prime}\sbs{xy}. \label{difftens}
\end{eqnarray}
Now, if we evaluate Eq.~(\ref{eq:Mvcmattractor2D}) for the second order moments, viz., for ($m,n$) = ($2,0$), ($0,2$), and ($1,1$) and use the above notation for the diffusion tensor parameter, it follows that
\begin{eqnarray*}
\Pi\sbs{20}\sps{Mv}=\Pi\sbs{00} D\sps{\prime}\sbs{20}, \qquad \Pi\sbs{02}\sps{Mv}=\Pi\sbs{00} D\sps{\prime}\sbs{02}, \qquad \Pi\sbs{11}\sps{Mv}=\Pi\sbs{00} D\sps{\prime}\sbs{11},
\end{eqnarray*}
which can be rewritten as $D\sps{\prime}\sbs{20}=\Pi\sbs{20}\sps{Mv}/\Pi\sbs{00}$, $D\sps{\prime}\sbs{02}=\Pi\sbs{02}\sps{Mv}/\Pi\sbs{00}$, and $D\sps{\prime}\sbs{11}=\Pi\sbs{11}\sps{Mv}/\Pi\sbs{00}$. As discussed in our previous work~\cite{schupbach2024fokker}, in order to correctly recover the 2D Navier-Stokes equations (NSE) via a Chapman-Enskog (C-E) analysis, at equilibrium, the diagonal components of the second order moments should be isotropic and equal to the pressure field $P=\rho \css$, while the off-diagonal component should be zero. That is, $\Pi\sbs{20}\sps{Mv}=\Pi\sbs{02}\sps{Mv}=\rho \css$ and $\Pi\sbs{11}\sps{Mv}=0$. Since the zeroth moment is a collision invariant related to the mass density, i.e., $\Pi\sbs{00}=\rho$, then obtain the following expressions for the diffusion tensor parameter: $D\sps{\prime}\sbs{20}=\Pi\sbs{20}\sps{Mv}/\Pi\sbs{00}=\css$, $D\sps{\prime}\sbs{02}=\Pi\sbs{02}\sps{Mv}/\Pi\sbs{00}=\css$, and $D\sps{\prime}\sbs{11}=\Pi\sbs{11}\sps{Mv}/\Pi\sbs{00}=0$. It should be noted that these parameterizations are strictly only needed in relaxing the second order non-conserved moments to recover the desired hydrodynamics represented by the NSE.

On the other hand, for the cases involving flow simulations with high Reynolds numbers, or those with relatively low fluid viscosities that depend on the relaxation frequencies of the second order moments (see the C-E analysis in~\cite{schupbach2024fokker}), we generally have $1/\omega\sbs{II}\ll 1/\omega\sbs{h}$. Here, we consider $\omega\sbs{II}$ as the relaxation rate of the second order moments (i.e., $(m+n)=2$) that is related to the fluid viscosity, whereas $\omega\sbs{h}$ refers to the relaxation rate associated with higher order moments (i.e., $(m+n)>2$) that controls the non-hydrodynamic kinetic processes. In more detail, the second order moments undergo relaxations under collision at a much faster rate than the higher order ones when simulating flows at relatively low fluid viscosities (and we may characterize the former as the fast modes and the latter as the slow modes), and this temporal separation of the scales of the processes should be respected and leveraged in a consistent manner so as to avoid numerical artifacts such as the hyperviscosities that have been observed when the Maxwellian-based central moments are used in LB formulations (see e.g.,~\cite{geier2009factorized}). When evolving those higher moments, the faster second order moments $\Pi\sbs{ab}$ for any $a$ and $b$ satisfying $a+b=2$ would have already approached their post-collision states, and do not change further during the relaxation of higher moments of order $(m+n)>2$. This statement can be written mathematically as
\begin{equation}\label{eq:RGequation}
\left( \frac{\delta}{\delta t} \Pi\sbs{ab}\right)\sbs{\!\!\!coll}\sps{\!\!\!FP} = 0, \quad \text{when}\quad \Pi\sbs{ab} = \tilde{\Pi}\sbs{ab}\quad\text{for}\quad (a,b) = (2,0), (0,2), (1,1),
\end{equation}
where $\tilde{\Pi}\sbs{ab}$ is the post-collision central moment, and the above condition, in turn, becomes
\begin{equation}
\Pi\sbs{00} D\sps{\prime}\sbs{ab}-\tilde{\Pi}\sbs{ab}=0\quad\text{for}\quad (a,b) = (2,0), (0,2), (1,1),
\end{equation}
where the diffusion tensor parameter $D\sps{\prime}\sbs{ab}$ is treated as an unknown, and can be solved to yield $D\sps{\prime}\sbs{ab}=\tilde{\Pi}_{ab}/\Pi\sbs{00}$, or, in terms of their components,
$D\sps{\prime}\sbs{20}=\tilde{\Pi}\sbs{20}/\rho$, $D\sps{\prime}\sbs{02}=\tilde{\Pi}\sbs{02}/\rho$, and $D\sps{\prime}\sbs{11}=\tilde{\Pi}\sbs{11}/\rho$.
It turns out that such an approach for model parameterizations in problems with multiple scales has a rich history in physics and is based on the so-called renormalization group (RG) method, which has been utilized to eliminate singular behavior in various fundamental physics theories and perturbative solutions of differential equations~\cite{chen1996renormalization,goldenfeld2018lectures}. In fact, Eq.~(\ref{eq:RGequation}) can be interpreted as the RG equation whose fixed points provide a principled approach in choosing the model parameters. Moreover, the use of variable diffusion parameters depending on the physical regime has an analogy in the turbulent dispersion of a single particle based on Taylor's theory and a particle pair based on the Richardson's theory, where the eddy diffusivity is parameterized depending on the time relative to certain Lagrangian correlation timescale (see e.g.,~\cite{csanady2012turbulent})

In summary, then, the diffusion tensor parameter in the Markovian central moment attractor in Eq.~(\ref{eq:Mvcmattractor2D}) for evolving moments $\Pi\sbs{mn}$ of order $(m+n)$ under collision is chosen according to
\begin{eqnarray}
D\sps{\prime}\sbs{20}  = \css, \qquad D\sps{\prime}\sbs{02} =\css, \qquad D\sps{\prime}\sbs{11} = 0 \qquad &\mbox{for}&\quad (m+n) \le 2,\nonumber\\[2mm]
D\sps{\prime}\sbs{20} = \frac{\tilde{\Pi}\sbs{20}}{\rho}, \quad D\sps{\prime}\sbs{02} = \frac{\tilde{\Pi}\sbs{02}}{\rho}, \quad D\sps{\prime}\sbs{11} = \frac{\tilde{\Pi}\sbs{11}}{\rho} \quad &\mbox{for}& \quad (m+n) > 2.\label{eq:Dij2Dselection}
\end{eqnarray}

\subsubsection{Selected Continuous Central Moment Markovian Attractors}
For developing a 2D discrete numerical implementation as a LBM, which uses the standard two-dimensional, nine velocity (D2Q9) lattice, based on the above considerations, we need to evolve the following 9 independent moments under collision:  $\Pi\sbs{mn}$ for ($m,n$) = (0,0), (1,0), (0,1), (2,0), (0,2), (1,1), (2,1), (1,2), and (2,2). Of these, the first three are collision invariants corresponding to mass density and momentum, or $\Pi\sbs{00}=\rho$ and $\Pi\sbs{10}=\Pi\sbs{01}=0$, which implies $\Pi\sbs{00}\sps{Mv}=\rho$ and $\Pi\sbs{10}\sps{Mv}=\Pi\sbs{01}\sps{Mv}=0$. The rest of the Markovian attractors corresponding to the non-conserved moments can be evaluated from Eq.~(\ref{eq:Mvcmattractor2D}) via using Eq.~(\ref{eq:Dij2Dselection}). Thus, we can summarize all the 9 Markovian attractors for the central moments corresponding to the D2Q9 lattice as follows:
\begin{eqnarray}\label{eq:MvCM2D}
&&\Pi\sbs{00}\sps{Mv} = \rho, \quad \Pi\sbs{10}\sps{Mv} =0, \quad \Pi\sbs{01}\sps{Mv} = 0, \quad \Pi\sbs{20}\sps{Mv} = \css \rho, \quad \Pi\sbs{02}\sps{Mv} = \css \rho, \nonumber\\[2mm]
&&\Pi\sbs{21}\sps{Mv} = 0, \quad \Pi\sbs{12}\sps{Mv} = 0,\quad \Pi\sbs{22}\sps{Mv}=\frac{1}{\rho}\left[\tilde{\Pi}\sbs{20}\tilde{\Pi}\sbs{02}+2\tilde{\Pi}\sbs{11}\sps{2}\right].
\end{eqnarray}
Note that the fourth order central moment attractor is now a nonlinear function of the post-collision second order central moments.

\subsubsection{Central Moment of Boltzmann's Acceleration Term due to Body Force in 2D}
When an external body force is present, such as the buoyancy force in the case of natural convective flows, the resulting acceleration causes a change in the distribution function at a rate $\left(\frac{\delta f}{\delta t}\right)\sbs{\!\!\!forcing}$ in the continuous Boltzmann equation (Eq.~(\ref{boltzmann_equation1})). Taking its inner product with $W\sbs{mn}$, we can represent the corresponding rate of change of central moments as
\begin{eqnarray}\label{cm_forces}
\Gamma\sbs{mn}= \left<  \left(\frac{\delta f}{\delta t}\right)\sbs{\!\!\!forcing} , W\sbs{mn} \right>,
\end{eqnarray}
Using Eq.~(\ref{eq:boltzmannforcingterm}) in Eq.~(\ref{cm_forces}) we get
\begin{eqnarray}
\Gamma\sbs{mn} = - \frac{F\sbs{x}}{\rho} \left< \frac{\partial f}{\partial \xi\sbs{x}}, W\sbs{mn} \right>  - \frac{F\sbs{y}}{\rho} \left< \frac{\partial f}{\partial \xi\sbs{y}}, W\sbs{mn} \right>.
\end{eqnarray}
Evaluating this exactly, the rate of change of the central moment of order ($m+n$) due to the body force is given by
\begin{eqnarray}\label{forcing:2D}
\Gamma\sbs{mn} = m \frac{F\sbs{x}}{\rho}\Pi\sbs{m\sm1,n} + n \frac{F\sbs{y}}{\rho}\Pi\sbs{m,n\sm1},
\end{eqnarray}
which relates it to products of the body force components and lower order central moments (see~\cite{premnath2012inertial}).
In general, it can be shown that only $\Gamma\sbs{mn}$ for ($m+n)\le2$ are necessary to recover the Navier-Stokes equations via a C-E analysis of the 2D central moment-based continuous Boltzmann kinetic equation (see~\cite{schupbach2024fokker}).

\subsection{Formulating LBM using FP Central Moment Collision Operator for Hydrodynamics}
The continuous Boltzmann equation shown in Eq.~(\ref{boltzmann_equation1}) will now be discretized first in the particle velocity space and then in space and time using the D2Q9 lattice, wherein the FP central moment collision operator given in Eq.~(\ref{FPcollision}) with the attendant attractors (see Eq.~(\ref{eq:MvCM2D})) together with the source term due to the body force (see Eq.~(\ref{forcing:2D})) will be incorporated via suitable transformations. The continuous particle velocity $\bm{\xi}$ is discretized with a finite particle velocity set $\bm{e}_{\alpha}$ for the D2Q9 lattice which arises from a tensor product of the set $\{-1,0,1\}$ with itself whose Cartesian components are then given by
\begin{subequations}
\begin{eqnarray}
\bigr| \mathbf{e}\sbs{x}\bigr> = (0,1,0,\sm1,0,1,\sm1,\sm1,1)^{\dag}, \label{ex}\\
\bigr|\mathbf{e}\sbs{y}\bigr> = (0,0,1,0,\sm1,1,1,\sm1,\sm1)^{\dag}. \label{ey}
\end{eqnarray}
\end{subequations}
Here, and in the rest of this manuscript, $\dag$ denotes the transpose operator and we will use the Dirac's notation of $\bigr|\cdot\bigr>$ and $\bigr<\cdot\bigr|$ for denoting the column and row vectors, respectively. Then the corresponding distribution function along each lattice link direction $\alpha$ can be expressed as $f_\alpha(\bm{x},t)=f(\bm{x},\bm{e}_{\alpha},t)$ and the use of the lattice effectively also discretizes the position space along the link direction with a grid spacing equal to the particle velocity $\bm{e}_{\alpha}$ times the time step $\delta_t$. When Eq.~(\ref{boltzmann_equation1}) is rewritten using $f_\alpha(\bm{x},t)$, it is referred to as the discrete velocity kinetic equation for hydrodynamics, which is then integrated along the particle characteristics over a time step $\delta_t$ (see~He \emph{et al}~\cite{he1998novel,he1999}). In effect, this results in the LBM that can be generically represented as a two-step scheme involving a \emph{collision step} which is then followed by a \emph{streaming step} that advects the post-collision distribution function $\tilde{f}\sbs{\alpha}$ in a lock-step fashion along link direction $\alpha$:
\begin{subequations}
\begin{eqnarray}
  \tilde{f}\sbs{\alpha}(\bm{x},t) &=& f\sbs{\alpha}(\bm{x},t)+\Omega\sbs{\alpha}(\bm{x},t), \quad \alpha = 0,1,2,\ldots, q \label{eq:LBMsteps-collide}\\
  f\sbs{\alpha}(\bm{x},t+\delta\sbs{t}) &=& \tilde{f}\sbs{\alpha}(\bm{x}-\bm{e}\sbs{\alpha}\delta\sbs{t},t),\label{eq:LBMsteps-stream}
\end{eqnarray}
\end{subequations}
where $q=9$. In Eq.~(\ref{eq:LBMsteps-collide}), $\Omega_\alpha(\bm{x},t)$ denotes the overall change in the distribution function $f_\alpha$ due to collision and the effect of body force during a time step $\delta_t$, which needs to reflect the relaxations under collision using the FP central moments-based model given in Eq.~(\ref{FPcollision}) and Eq.~(\ref{eq:MvCM2D}) together with the contribution due to the body force according to Eq.~(\ref{forcing:2D}) through mappings between the distributions and central moments along with the intermediate raw moments.

To construct such a collision step, we first define the \emph{discrete central moments} and \emph{discrete raw moments} of the distribution function $f_{\alpha}$ of order ($m+n$), its corresponding Markovian attractor $f_{\alpha}^{Mv}$, and the source term $S_{\alpha}$ for the body force, respectively, which read as
\begin{subequations}
\begin{eqnarray}
\left( \begin{array}{c}\Ks{mn} \\[2mm] \Ks{mn}\sps{Mv} \\[2mm] \Sig\sbs{mn} \end{array} \right)  &=& \sum\sbs{\alpha = 0}\sps{8} \left( \begin{array}{c}f\sbs{\alpha} \\[2mm] f\sbs{\alpha}\sps{Mv} \\[2mm] S\sbs{\alpha} \end{array} \right)  ( e\sbs{\alpha x} - u\sbs{x})\sps{m}  ( e\sbs{\alpha y} - u\sbs{y})\sps{n}, \;\;\mbox{and}\\[4mm]
\left( \begin{array}{c}\Kps{mn} \\[2mm] \kappa\sps{Mv\prime}\sbs{mn} \\[2mm]  \Sig\sbs{mn}\sps{\prime}\end{array} \right)  &=& \sum\sbs{\alpha = 0}\sps{8} \left( \begin{array}{c}f\sbs{\alpha} \\[2mm] f\sbs{\alpha}\sps{Mv}\\[2mm]  S\sbs{\alpha} \end{array} \right)  e\sbs{\alpha x}\sps{m}  e\sbs{\alpha y}\sps{n} .
\end{eqnarray}
\end{subequations}
Then, the finite sets of independent central moments and raw moments, respectively, for the D2Q9 lattice can be expressed as
\begin{eqnarray}
\mc &=& ( \Ks{00},\Ks{10}, \Ks{01}, \Ks{20}, \Ks{02}, \Ks{11}, \Ks{21}, \Ks{12}, \Ks{22} )\sps{\dag},\nonumber\\
\m  &=& ( \Kps{00}, \Kps{10},\Kps{01}, \Kps{20}, \Kps{02}, \Kps{11},\Kps{21}, \Kps{12},\Kps{22} )\sps{\dag},\nonumber
\end{eqnarray}
and similar sets for the attractors and the sources follow. Moreover, the following 9-dimensional vector $\mathbf{f}$ defines all the discrete distribution functions for the D2Q9 lattice:
\begin{eqnarray}
\mathbf{f} = (\f{0} , \f{1},\f{2}, \dots , \f{8})\sps{\dag},\nonumber
\end{eqnarray}
and analogously one can introduce vectors that collect the components of the discrete attractors  $f_{\alpha}^{Mv}$ and the sources $S_{\alpha}$.

Based on these notations, we can now express the essential steps to incorporate the collision model by means of certain pre- and post-collision transformations. That is, to relax different central moments to their respective attractors together with source term updates, we first map the discrete distribution functions $\mathbf{f}$ to raw moments $\m$ and then to central moments $\mc$; once the post-collision central moments are calculated, they are transformed back to raw moments and then to the discrete distribution functions, which completes the collision step. Here, the mappings between the raw moments and the distribution functions can be represented as
\begin{equation}\label{eq:map-df-rm}
\m = \PP\mathbf{f},  \; \; \; \; \; \; \; \; \;  \; \; \; \; \; \; \; \; \; \mathbf{f} = \PP\sps{\sm1}\m,
\end{equation}
where $\PP$ is a matrix obtained from grouping all the basis vectors for the independent moments for the D2Q9 lattice, i.e., using the components of the monomials $\bigr|\bm{e}\sbs{x}\sps{m}  \bm{e}\sbs{y}\sps{n} \bigr>$ given by
\begin{equation}\label{eq:dfrm}
\tensr{P} =\bigr[ \; \bigr|\bm{1}\bigr>,    \;\;     \bigr|\bm{e}\sbs{x} \bigr>, \;\;\bigr|\bm{e}\sbs{y}\bigr>, \;\; \bigr|\bm{e}\sbs{x}\sps{2}\bigr>, \;\; \bigr|\bm{e}\sbs{y}\sps{2}\bigr>,  \;\; \bigr|\bm{e}\sbs{x} \bm{e}\sbs{y}\bigr>, \;\; \bigr|\bm{e}\sbs{x}\sps{2} \bm{e}\sbs{y}\bigr>, \;\; \bigr|\bm{e}\sbs{x} \bm{e}\sbs{y}\sps{2}\bigr>, \;\;  \bigr|\bm{e}\sbs{x}\sps{2} \bm{e}\sbs{y}\sps{2}\bigr> \;     \bigr]^\dag.
\end{equation}
Here, $\left|\mathbf{1}\right> $ is a 9-dimensional unit vector given by~$\bigr|\bm{1}\bigr> =\left| \; \bm{e}\sbs{x}\sps{0} \bm{e}\sbs{y}\sps{0}\; \right> = (1,1,1,1,1,1,1,1,1)^{\dag}$. Moreover, we write the transformations between central moments and raw moments as
\begin{equation}\label{eq:rmcm}
\mc = \F \m,  \; \; \; \; \; \; \; \; \; \; \; \; \; \; \; \; \; \; \m = \Fi \mc,
\end{equation}
where both the frame transformation matrix $\F$ and its inverse $\F\sps{\sm1}$ are lower triangular matrices that depend on the fluid velocity components $\bm{u} = (u\sbs{x},u\sbs{y})$ and their elements can be readily obtained by collecting together the coefficients of the binomial expansions which relate different central moments to the raw moments (see e.g.,~\cite{yahia2021central}). Before writing the complete implementation details of our 2D FPC-LBM, we will now formally express the discrete central moments of the Markovian attractor $\Ks{mn}\sps{Mv}$ and the source term $\Sig\sbs{mn}$ due to the body force by matching the corresponding continuous versions as given in Eq.~(\ref{eq:MvCM2D}) and Eq.~(\ref{forcing:2D}), respectively. In other words, $\Ks{mn}^{Mv}=\Pi_{mn}^{Mv}$ and
$\Sig\sbs{mn}=\Gamma\sbs{mn}$. Thus, the discrete Markovian central moment attractors for the D2Q9 lattice are given by
\begin{gather}\label{D2Q9attractors}
\KMv{00} = \rho, \; \; \;  \; \; \; \KMv{10}=0, \; \; \;  \; \; \; \KMv{01} = 0, \\[3mm]
\KMv{20} = \css \rho, \; \; \; \; \; \;  \KMv{02}\ =\css \rho, \; \; \;  \; \; \;  \KMv{11} = 0, \nonumber\\[3mm]
\KMv{21} = 0, \; \; \;  \; \; \;   \KMv{12} = 0, \nonumber \\[3mm]
\KMv{22} = \frac{1}{\rho}\left[\Kts{20}\Kts{02} +2 \Kts{11}\Kts{11}\right].\nonumber
\end{gather}
Notice that the fourth order attractor depends on the sum of the products of the post-collision states of the second order central moments, i.e., $\tilde{\K}\sbs{20}$, $\tilde{\K}\sbs{02}$, and  $\tilde{\K}\sbs{11}$. Moreover, using Eq.~(\ref{forcing:2D}), the changes to various central moments due to the source term or the body force can be written as
\begin{gather}\label{eq:sourceCM2D}
\Ss{00} = 0, \; \; \;  \; \; \; \Ss{10} = F\sbs{x},\; \; \; \; \; \;\Ss{01} = F\sbs{y},\\[3mm]
\Ss{20} = 2\frac{F\sbs{x}}{\rho} \Ks{10} = 0, \; \; \; \; \; \; \Ss{02} =2\frac{F\sbs{y}}{\rho} \Ks{01} = 0, \; \; \; \; \; \;\Ss{11} = \frac{F\sbs{x}}{\rho} \Ks{10}+\frac{F\sbs{y}}{\rho} \Ks{01} = 0, \nonumber\\[3mm]
\Ss{21} = 2\frac{F\sbs{x}}{\rho} \Ks{11}+\frac{F\sbs{y}}{\rho} \Ks{20}, \; \; \; \; \; \; \Ss{12} = \frac{F\sbs{x}}{\rho} \Ks{02}+\frac{F\sbs{y}}{\rho} \Ks{11},\nonumber \\[3mm]
\Ss{22} = 2 \left( \frac{F\sbs{x}}{\rho} \Ks{12} + \frac{F\sbs{y}}{\rho} \Ks{21} \right). \nonumber
\end{gather}
For convenience, a step-by-step implementation of the 2D FPC-LBM for computing fluid motions is given in Appendix~\ref{sec:appendixA}.

\subsection{Construction of FP Central Moment Collision Operator: Energy Transport}
Analogous to the treatment of the collision term in the continuous kinetic equation for hydrodynamics (see Eq.~(\ref{boltzmann_equation1})), we represent the corresponding collision term in Eq.~(\ref{boltzmann_equation_two})) for the $g$ distribution function for the energy transport using a FP model like in Eq.~(\ref{FP_collision}) as
\begin{eqnarray}\label{FP_collision_g}
\left( \frac{\delta g}{\delta t}\right)\sbs{\!\!\!coll}\sps{\!\!\!FP} = \omega\sbs{FP} \left[\frac{\partial}{\partial \xi\sbs{i}}((\xi\sbs{i} - u\sbs{i})g) + \hat{D}\sbs{ij}' \frac{\partial\sps{2} g}{\partial\xi\sbs{i} \partial \xi\sbs{j}} \right],
\end{eqnarray}
where $\hat{D}\sbs{ij}'$ is the corresponding diffusion tensor parameter. As before, for practical implementation of such a model in the LB framework, we recast it in a central moment formulation. Thus, defining the central moment of $g$ of order ($m+n$) as
\begin{eqnarray}\label{central_moments_g}
\Lambda\sbs{mn}=\left<g,W\sbs{mn}\right>,\nonumber
\end{eqnarray}
and taking the inner product of Eq.~(\ref{FP_collision_g}) with the weight $W\sbs{mn}$ and performing the various manipulations and rearrangements outlined earlier (see also~\cite{schupbach2024fokker}), we finally obtain a FP central moment collision operator energy transport that reads as
\begin{eqnarray}\label{FPcollision_g}
\left( \frac{\delta}{\delta t} \Lambda\sbs{mn}\right)\sbs{\!\!\!coll}\sps{\!\!\!FP} = \omega\sbs{mn}\left[ \Lambda\sbs{mn}\sps{Mv}-\Lambda\sbs{mn}\right],
\end{eqnarray}
where the corresponding Markovian central moment attractor of order ($m+n$) is given by
\begin{eqnarray}\label{eq:Mvcmattractor2D_g}
\Lambda_{mn}\sps{Mv} = \frac{m(m-1)}{m+n}\hat{D}\sps{\prime}\sbs{xx}\Lambda\sbs{m\sm2,n} + \frac{n(n-1)}{m+n}\hat{D}\sps{\prime}\sbs{yy}\Lambda\sbs{m,n\sm2}+ \frac{2mn}{m+n}\hat{D}\sps{\prime}\sbs{xy}\Lambda\sbs{m\sm1,n\sm1}
\end{eqnarray}
While Eq.~(\ref{eq:Mvcmattractor2D_g}) looks similar to Eq.~(\ref{eq:Mvcmattractor2D}) on the face of it, the resulting higher moments, which are generated via the recurrence relations based on lower moments, resulting from the latter are expected to be different from the former since the number of collision invariants in these two cases differ (see Eqs.~(\ref{collisioninvariantf}) and (\ref{collisioninvariantg})).

\subsubsection{Choice of Diffusion Tensor Parameter}
In order to select the diffusion tensor parameter $\hat{D}\sps{\prime}\sbs{ij}$ present in the Markovian attractor  Eq.~(\ref{eq:Mvcmattractor2D_g}) which is needed in the collision operator in Eq.~(\ref{FPcollision_g}), as before, we first introduce the following notation
\begin{eqnarray*}
\hat{D}\sps{\prime}\sbs{20}=\hat{D}\sps{\prime}\sbs{xx}, \quad \hat{D}\sps{\prime}\sbs{02} = \hat{D}\sps{\prime}\sbs{yy}, \quad \hat{D}\sps{\prime}\sbs{11}=\hat{D}\sps{\prime}\sbs{xy}. \label{difftens_g}
\end{eqnarray*}
and evaluate Eq.~(\ref{eq:Mvcmattractor2D_g}) for the second order moments, viz., for ($m,n$) = ($2,0$), ($0,2$), and ($1,1$) to get
$\Lambda\sbs{20}\sps{Mv}=\Lambda\sbs{00} \hat{D}\sps{\prime}\sbs{20}$, $\Lambda\sbs{02}\sps{Mv}=\Lambda\sbs{00} \hat{D}\sps{\prime}\sbs{02}$, $\Lambda\sbs{11}\sps{Mv}=\Lambda\sbs{00} \hat{D}\sps{\prime}\sbs{11}$; these then provide the following expressions by means of which $\hat{D}\sps{\prime}\sbs{mn}$ can be specified:
$\hat{D}\sps{\prime}\sbs{20}=\Lambda\sbs{20}\sps{Mv}/\Lambda\sbs{00}$, $\hat{D}\sps{\prime}\sbs{02}=\Lambda\sbs{02}\sps{Mv}/\Lambda\sbs{00}$, and $\hat{D}\sps{\prime}\sbs{11}=\Lambda\sbs{11}\sps{Mv}/\Lambda\sbs{00}$. Now, to recover the correct diffusive flux of the energy transport equation, based on a C-E analysis, we require $\Lambda\sbs{20}\sps{Mv}=\Lambda\sbs{02}\sps{Mv}=\chss T$ and $\Lambda\sbs{11}\sps{Mv}=0$, where $\chss$ is a fixed parameter, which is usually equal to $1/3$ in lattice units in LB schemes like the $\css$ used in the kinetic model for hydrodynamics presented earlier, and the zeroth moment is the collision invariant and equal to the temperature field: $\Lambda\sbs{00}=T$. Based on these considerations, it then follows that $\hat{D}\sps{\prime}\sbs{20}=\hat{D}\sps{\prime}\sbs{02}=\chss$ and $\hat{D}\sps{\prime}\sbs{11}=0$. As such, these parameterizations for $\hat{D}\sps{\prime}\sbs{mn}$ based on a fixed constant $\chss$ should suffice in establishing the Markovian attractors for all orders using Eq.~(\ref{eq:Mvcmattractor2D_g}); this is because usually in the kinetic equation for energy transport there is a scale separation between the relaxation parameters for the \emph{first order moments} (which controls the thermal diffusivity and follows from a Chapman-Enskog analysis similar to the one performed in~\cite{schupbach2024fokker}) and \emph{all} the rest of the higher order ones; by contrast, as discussed earlier, for the kinetic equation for hydrodynamics there are disparities in the \emph{second order moments} (which determines the viscosity) and the higher order moments and consequently involved using a renormalization approach for setting the diffusion tensor parameters. Thus, the simplest option, which refer to as the \emph{option A}, reads as
\begin{eqnarray}
\hat{D}\sps{\prime}\sbs{20}  = \chss, \qquad \hat{D}\sps{\prime}\sbs{02} =\chss, \qquad \hat{D}\sps{\prime}\sbs{11} = 0 \qquad \mbox{for all}\quad (m+n)\label{eq:Dij2D_goptionA}
\end{eqnarray}
for evolving moment $\Lambda\sbs{mn}$ under collision.

However, as in the case of simulating hydrodynamics, if we also choose the relaxation parameters for second order moments of the kinetic equation for energy transport to be much smaller than those for the higher order moments, then for evolving the latter under collision, we can invoke renormalization in selecting $\hat{D}\sps{\prime}\sbs{ab}$ for $a$ and $b$ satisfying $a+b=2$: $\hat{D}\sps{\prime}\sbs{20} = \tilde{\Lambda}\sbs{20}/T$, $\hat{D}\sps{\prime}\sbs{02} = \tilde{\Lambda}\sbs{20}/T$, and $\hat{D}\sps{\prime}\sbs{11} = \tilde{\Lambda}\sbs{11}/T$. We refer to this alternative as \emph{option B}, which can be summarized as follows:
\begin{eqnarray}
\hat{D}\sps{\prime}\sbs{20}  = \chss, \qquad \hat{D}\sps{\prime}\sbs{02} =\chss, \qquad \hat{D}\sps{\prime}\sbs{11} = 0 \qquad &\mbox{for}&\quad (m+n) \le 2,\nonumber\\[2mm]
\hat{D}\sps{\prime}\sbs{20} = \frac{\tilde{\Lambda}\sbs{20}}{T}, \quad \hat{D}\sps{\prime}\sbs{02} = \frac{\tilde{\Lambda}\sbs{02}}{T}, \quad \hat{D}\sps{\prime}\sbs{11} = \frac{\tilde{\Lambda}\sbs{11}}{T} \quad &\mbox{for}& \quad (m+n) > 2,\label{eq:Dij2D_g}
\end{eqnarray}
for evolving moment $\Lambda\sbs{mn}$. In the above, we emphasize that $\tilde{\Lambda}\sbs{mn}$ refers to the post-collision state of $\Lambda\sbs{mn}$.

\subsubsection{Selected Continuous Central Moment Markovian Attractors}
Based on the above considerations, we can now derive all the 9 countable Markovian central moments attractors necessary for constructing a LBM using the D2Q9 lattice in the next section. Since the zeroth moment is invariant under collision, we have $\Lambda\sbs{00}=\Lambda\sbs{00}\sps{Mv}=T$. The rest of $\Lambda\sbs{mn}\sps{Mv}$ for ($m,n$) = (1,0), (0,1), (2,0), (0,2), (1,1), (2,1), (1,2), and (2,2) can be obtained from using either the \emph{option A} given in Eq.~(\ref{eq:Dij2D_goptionA}) or the \emph{option B} presented in Eq.~(\ref{eq:Dij2D_g}) in Eq.~(\ref{eq:Mvcmattractor2D_g}).
\
Then, the complete set of central moment attractors for solving the energy transport equation using the D2Q9 lattice based on the \emph{option A} can be written as follows:
\begin{gather}
\Lambda\sbs{00}\sps{Mv} = T, \quad
\Lambda\sbs{10}\sps{Mv} =0, \quad
\Lambda\sbs{01}\sps{Mv} = 0, \quad
\Lambda\sbs{20}\sps{Mv} = \chss T, \quad
\Lambda\sbs{02}\sps{Mv} = \chss T, \nonumber\\[2mm]
\Lambda\sbs{21}\sps{Mv} = \frac{2}{3}\chss \tilde{\Lambda}\sbs{01}, \quad
\Lambda\sbs{12}\sps{Mv} = \frac{2}{3}\chss \tilde{\Lambda}\sbs{10}, \quad
\Lambda\sbs{22}\sps{Mv}=\frac{1}{2}\chss\left(\tilde{\Lambda}\sbs{20}+\tilde{\Lambda}\sbs{02}\right).\label{eq:MvT2D}
\end{gather}
By contrast, if the \emph{option B} is used, the higher order attractors $\Lambda\sbs{21}\sps{Mv}$, $\Lambda\sbs{12}\sps{Mv}$, and $\Lambda\sbs{22}\sps{Mv}$ in Eq.~(\ref{eq:MvT2D}) should be replaced with the following:
\begin{gather}
\Lambda\sbs{21}\sps{Mv} = \frac{2}{3T} \left( \tilde{\Lambda}\sbs{20}\tilde{\Lambda}\sbs{01} + 2\tilde{\Lambda}\sbs{11}\tilde{\Lambda}\sbs{10} \right), \quad
\Lambda\sbs{12}\sps{Mv} = \frac{2}{3T} \left(\tilde{\Lambda}\sbs{02}\tilde{\Lambda}\sbs{10} + 2\tilde{\Lambda}\sbs{11}\tilde{\Lambda}\sbs{01} \right), \nonumber\\[2mm]
\Lambda\sbs{22}\sps{Mv}=\frac{1}{T}\left(\tilde{\Lambda}\sbs{20}\tilde{\Lambda}\sbs{02}+2\tilde{\Lambda}\sbs{11}\sps{2}\right). \label{eq:MvT2DoptionB}
\end{gather}
Interestingly, since the kinetic equation for the energy transport equation has only a single collision invariant (i.e., the temperature field -- see Eq.~(\ref{collisioninvariantg})), unlike that for the hydrodynamics which has three (i.e., the mass density, and momentum components -- see Eq.~(\ref{collisioninvariantf})), the Markovian central moment attractors given in Eq.~(\ref{eq:MvT2D}) (or Eq.~(\ref{eq:MvT2DoptionB})) differ from those in Eq.~(\ref{eq:MvCM2D}); in particular, notice that the former involves non-zero contributions for the third moments $\Lambda\sbs{21}\sps{Mv}$ and $\Lambda\sbs{12}\sps{Mv}$.

Now, an important note on the choice between the \emph{option A} and \emph{option B} is in order. In general, for recovering the energy equation, only the relaxation parameters for the first order moments (related to the thermal diffusivity) are physically relevant and \emph{all} the rest of the higher order ones can be chosen freely. In practice, it would be sufficient to set the relaxation parameters for all such higher moments, including the second order moments to have the same values, and hence the separation of time scales for relaxations under collision as considered for \emph{option B} is not necessary in solving the energy equation (unlike that for solving the NSE); moreover, the \emph{option A} is simpler and in numerical experiments it was found have very similar numerical stability characteristics as the \emph{option B}. For these reasons, in this work, we adopt the \emph{option A} and the resulting attractors given in Eq.~(\ref{eq:MvT2D}) in constructing the LBM discussed next.

\subsection{Formulating LBM using FP Central Moment Collision Operator for Energy Transport}
Discretizing the kinetic equation for the energy transport given in Eq.~(\ref{boltzmann_equation_two}) on the D2Q9 lattice following the approach outlined earlier, we get a LB equation for the discrete distribution $\g{\alpha}$, which can be written in terms of the following collision and streaming steps:
\begin{subequations}
\begin{eqnarray}
\gt{\alpha}(\bm{x},t) &=&\g{\alpha}(\bm{x},t)+\hat{\Omega}\sbs{\alpha}(\bm{x},t), \quad \alpha = 0,1,2,\ldots, q \label{eq:LBMsteps-collideg}\\
 \g{\alpha}(\bm{x},t+\delta_t) &=& \gt{\alpha}(\bm{x}-\bm{e}\sbs{\alpha}\delta\sbs{t},t),\label{eq:LBMsteps-streamg}
\end{eqnarray}
\end{subequations}
where $q=9$. Then, in order to express the effect of collisions $\hat{\Omega}\sbs{\alpha}$ in terms of relaxation of different central moments to their Markovian attractors, we first define the following central moments and raw moments, respectively:
\begin{subequations}
\begin{eqnarray}
\left( \begin{array}{c}\Ns{mn} \\[3mm] \Ns{mn}\sps{Mv} \end{array} \right)  &=& \sum\sbs{\alpha = 0}\sps{8} \left( \begin{array}{c}g\sbs{\alpha} \\[3mm] g\sbs{\alpha}\sps{Mv}  \end{array} \right)  ( e\sbs{\alpha x} - u\sbs{x})\sps{m}  ( e\sbs{\alpha y} - u\sbs{y})\sps{n}, \;\;\mbox{and}\\[0.2in]
\left( \begin{array}{c}\Np\sbs{mn} \\[3mm] \N\sps{Mv\prime}\sbs{mn}  \end{array} \right)  &=& \sum\sbs{\alpha = 0}\sps{8} \left( \begin{array}{c}g\sbs{\alpha} \\[3mm] g\sbs{\alpha}\sps{Mv} \end{array} \right)  e\sbs{\alpha x}\sps{m}  e\sbs{\alpha y}\sps{n} .
\end{eqnarray}
\end{subequations}
Then, the 9 countable central moments $\Ns{mn}$ and raw moments $\Np\sbs{mn}$ for the D2Q9 lattice can be listed together in vectors $\nc$ and $\n$, respectively, as
\begin{eqnarray}
\nc&=& ( \Ns{00},\Ns{10}, \Ns{01}, \Ns{20}, \Ns{02}, \Ns{11}, \Ns{21}, \Ns{12}, \Ns{22} )^\dag,\nonumber\\
\n &=& ( \Nps{00}, \Nps{10},\Nps{01}, \Nps{20}, \Nps{02}, \Nps{11},\Nps{21}, \Nps{12},\Nps{22} )^\dag,\nonumber
\end{eqnarray}
while the 9 discrete distribution functions in the vector $\mathbf{g}$ as
\begin{eqnarray}
\mathbf{g} = (\g{0} , \g{1},\g{2}, \dots , \g{8})\sps{\dag}.\nonumber
\end{eqnarray}
Next, to accomplish the relaxations of different central moments to their attractors at their own rates under collision, we first need to transform the discrete distribution functions to raw moments and then to central moments, which can be written as
\begin{equation}
\n = \PP\mathbf{g},  \; \; \; \; \; \; \; \; \;  \; \; \; \; \; \; \; \; \; \nc = \F\n.
\end{equation}
where $\PP$ and $\F$ correspond to the respective transformation matrices (see the discussion on the formulation of the 2D FPC-LBM for fluid motions). Then, following the relaxations under collision,
\begin{equation}
\tilde{\eta}_{mn}=\eta_{mn} + \omega_{mn}(\eta_{mn}^{Mv}-\eta_{mn}),
\end{equation}
for ($m,n$) = (0,0), (1,0), (0,1), (2,0), (0,2), (1,1), (2,1), (1,2), and (2,2), with $\omega_{mn}$ representing the relaxation rate of $\eta_{mn}$, the post-collision central moments are first mapped to the respective raw moments and then the distribution functions, which can be expressed as follows:
\begin{equation}
\tilde{\n} = \Fi\tilde{\nc},  \; \; \; \; \; \; \; \; \;  \; \; \; \; \; \; \; \; \; \tilde{\mathbf{g}} = \PP^{-1}\tilde{\n},
\end{equation}
where tilde over the symbols represent the post-collision states, and $\Fi$ and $\PP^{-1}$ refer to inverse transformations. Here, the discrete Markovian central moment attractor $\eta_{mn}^{Mv}$ used in the collision step is based on the respective continuous counterpart $\Lambda_{mn}^{Mv}$ via the matching principle $\eta_{mn}^{Mv}=\Lambda_{mn}^{Mv}$, from which we have
\begin{gather}\label{D2Q9attractors}
\Ns{00}\sps{Mv} = T, \; \; \;  \; \; \;
\Ns{10}\sps{Mv}=0, \; \; \;  \; \; \;
\Ns{01}\sps{Mv} = 0, \\[3mm]
\Ns{20}\sps{Mv} =\chss T, \; \; \; \; \; \;
\Ns{02}\sps{Mv} =\chss T, \; \; \;  \; \; \;
\Ns{11}\sps{Mv} = 0, \nonumber\\[3mm]
\Ns{21}\sps{Mv} = \frac{2}{3}\chss\tilde{\N}\sbs{01}, \; \; \;  \; \; \;
\Ns{12}\sps{Mv} = \frac{2}{3}\chss\tilde{\N}\sbs{10}, \; \; \;  \; \; \;
\Ns{22}\sps{ Mv} = \frac{1}{2}\chss\left(\tilde{\N}\sbs{20}+\tilde{\N}\sbs{02}\right).\nonumber
\end{gather}
Once the post-collision distribution functions are obtained, they are then streamed along the lattice links to the neighboring nodes according to Eq.~(\ref{eq:LBMsteps-streamg}). Then, the temperature is calculated from the updated distribution functions via their zeroth moment, or $T = \sum\sbs{\alpha=0}\sps{8} g\sbs{\alpha}$. Complete implementation details of the 2D FPC-LBM for solving the energy transport equation is presented in Appendix~\ref{sec:appendixB}.

\section{Fokker-Planck (FP) Central Moments-Based Collision Modeling: 3D Formulations and LB Schemes\label{sec:FPC-LBMformulations3D}}
In this section, we will extend the previous developments to three-dimensions (3D) suitable for simulating thermal convective flows. In particular, we will develop FP central moment collision operators and the LB schemes using the D3Q27 lattice for the fluid motions and D3Q15 lattice for energy transport.

\subsection{Construction of FP Central Moment Collision Operator: Fluid Motions}
To begin with, the FP collision model given in Eq.~(\ref{FP_collision}) can be conveniently rewritten in 3D in a compact notation as
\begin{eqnarray}\label{FP_short3D}
\left( \frac{\delta f}{\delta t}\right)\sbs{\!\!\!coll}\sps{\!\!\!FP} = \omega\sbs{FP}\left(\frac{\delta f}{\delta t}\sps{FP1}+\frac{\delta f}{\delta t}\sps{FP2}\right),
\end{eqnarray}
where the component-wise contributions of its two separate terms can be expressed as
\begin{eqnarray}\label{dFP3D}
\frac{\delta f}{\delta t}\sps{FP1} = \frac{\partial}{\partial \xi\sbs{i}}((\xi\sbs{x}-u\sbs{i})f) &=& \frac{\partial}{\partial \xi\sbs{x}}((\xi\sbs{y}-u\sbs{x})f)+\frac{\partial}{\partial \xi\sbs{y}}((\xi\sbs{y}-u\sbs{y})f)+\frac{\partial}{\partial \xi\sbs{z}}((\xi\sbs{z}-u\sbs{z})f), \nonumber\\
\frac{\delta f}{\delta t}\sps{FP2} = D\sps{\prime}\sbs{ij}\frac{\partial\sps{2} f}{\partial\xi\sbs{i}\partial\xi\sbs{j}} &=& D\sps{\prime}\sbs{xx}\frac{\partial\sps{2} f}{\partial \xi\sbs{x}\sps{2}}+D\sps{\prime}\sbs{yy}\frac{\partial\sps{2} f}{\partial \xi\sbs{y}\sps{2}}+D\sps{\prime}\sbs{zz}\frac{\partial\sps{2} f}{\partial \xi\sbs{z}\sps{2}} \nonumber\\
\qquad\qquad\qquad\qquad&&+ 2D\sps{\prime}\sbs{xy}\frac{\partial\sps{2} f}{\partial \xi\sbs{x}\partial \xi\sbs{y}} + 2D\sps{\prime}\sbs{xz}\frac{\partial\sps{2} f}{\partial \xi\sbs{x}\partial \xi\sbs{z}} + 2D\sps{\prime}\sbs{yz}\frac{\partial\sps{2} f}{\partial \xi\sbs{y}\partial \xi\sbs{z}},
\end{eqnarray}
As a first step towards developing a central moment formulation of this model, defining the inner product over the 3D velocity space of any two objects $a$ and $b$ as
\begin{eqnarray}
\left<a,b\right>= \int\displaylimits\sbs{-\infty}\sps{\infty} \int\displaylimits\sbs{-\infty}\sps{\infty} \int\displaylimits\sbs{-\infty}\sps{\infty} a \; b\; \mathrm{d}\xi\sbs{x}\; \mathrm{d}\xi\sbs{y}\; \mathrm{d}\xi\sbs{z}
\end{eqnarray}
and the central moment weighting factor as
\begin{eqnarray}\label{cm_weights3D}
W\sbs{mnp}=(\xi\sbs{x}-u\sbs{x})\sps{m}(\xi\sbs{y}-u\sbs{y})\sps{n}(\xi\sbs{z}-u\sbs{z})\sps{p},
\end{eqnarray}
we can then introduce the central moment of the distribution function $f$ of order ($m+n+p$) as
\begin{eqnarray}
\Pi\sbs{mnp}=\left<f,W\sbs{mnp}\right>.
\end{eqnarray}
As in the 2D formulations before, it is useful to express the diffusion tensor parameter in Eq.~(\ref{dFP3D}) in the following notation in order to clarify its natural relationship with central moments as defined above:
\begin{eqnarray}
D\sps{\prime}\sbs{200}=D\sps{\prime}\sbs{xx}, \quad D\sps{\prime}\sbs{020}=D\sps{\prime}\sbs{yy}, \quad D\sps{\prime}\sbs{002}=D\sps{\prime}\sbs{zz}, \quad D\sps{\prime}\sbs{110}=D\sps{\prime}\sbs{xy}, \quad D\sps{\prime}\sbs{011}=D\sps{\prime}\sbs{yz}, \quad D\sps{\prime}\sbs{101}=D\sps{\prime}\sbs{xz}.
\end{eqnarray}
Then taking the inner product of Eq.~(\ref{FP_short3D}) with the weighting factor $W\sbs{mnp}$, that is,
\begin{eqnarray}\label{cmchanges3D}
\left(\frac{\delta \Pi\sps{mnp}}{\delta t}\right)\sbs{\!\!\!coll}\sps{\!\!\!FP} = \omega\sbs{FP}\left[\left<\frac{\delta f}{\delta t}\sps{FP1},W\sbs{mnp}\right> + \left<\frac{\delta f}{\delta t}\sps{FP2},W\sbs{mnp}\right>\right]=\omega\sbs{FP}\left[\frac{\delta \Pi\sbs{mnp}\sps{FP1}}{\delta t}+\frac{\delta \Pi\sbs{mnp}\sps{FP2}}{\delta t}\right].
\end{eqnarray}
and upon simplification of each term in the right side of this last equation and further rearrangement as detailed in our recent work~\cite{schupbach2024fokker}, we can finally express the 3D FP central moment collision model's rate equation in terms of prescribing the relaxation of the central moment $\Pi\sbs{mnp}$ towards its Markovian attractor $\Pi\sbs{mnp}\sps{Mv}$ at a rate $\omega\sbs{mnp}$ as
\begin{equation}\label{FPcoll3d}
\left(\frac{\delta}{\delta t} \Pi\sbs{mnp}\right)\sps{\!\!\! FP}\sbs{\!\!\! coll} = \omega\sbs{mnp} \left[ \Pi\sbs{mnp}\sps{Mv}-\Pi\sbs{mnp} \right],
\end{equation}
where $\Pi\sbs{mnp}\sps{Mv}$ is given by the following recurrence equation:
\begin{eqnarray}\label{eq:Mvcmattractor3D}
\Pi\sbs{mnp}\sps{ Mv} &=&  D\sps{\prime}\sbs{200}\frac{m(m\sm1)}{(m+n+p)}\Pi\sbs{m\sm2,n,p} + D\sps{\prime}\sbs{020}\frac{n(n-1)}{(m+n+p)}\Pi\sbs{m,n\sm2,p} +   \\[2mm]
                    && D\sps{\prime}\sbs{002}\frac{p(p\sm1)}{(m+n+p)}\Pi\sbs{m,n,p\sm2} + 2D\sps{\prime}\sbs{110}\frac{mn}{(m+n+p)}\Pi\sbs{m\sm1,n\sm1,p} +\nonumber \\[2mm]
                    && 2D\sps{\prime}\sbs{011}\frac{np}{(m+n+p)}\Pi\sbs{m,n\sm1,p\sm1} + 2D\sps{\prime}\sbs{101}\frac{mp}{(m+n+p)}\Pi\sbs{m\sm1,n,p\sm1}. \nonumber
\end{eqnarray}

\subsubsection{Choice of Diffusion Tensor Parameter\label{subsec:choice_of_diffusion_paramter}}
Then, evaluating Eq.~(\ref{eq:Mvcmattractor3D}) for the second order moments $\Pi\sbs{abc}\sps{Mv}$ with $(a+b+c) = 2$, i.e., for ($a,b,c$) = (2,0,0), (0,2,0), (0,0,2), (1,1,0), (0,1,1), and (1,0,1), we can relate it to the diffusion tensor parameter as
\begin{equation*}
\Pi\sbs{abc}\sps{Mv}=\Pi\sbs{000} D\sps{\prime}\sbs{abc} \qquad \text{for}\;\; (a+b+c) = 2.
\end{equation*}
Since the zeroth moment, which is equal to mass density, is conserved under collision, i.e., $\Pi\sbs{000}=\Pi\sbs{000}\sps{Mv}=\rho$, and the recovery of the 3D NSE via C-E analysis~\cite{schupbach2024fokker} requires that the diagonal components of the second order central moments be isotropic and equal to the pressure field, i.e., $\Pi\sbs{abc}\sps{Mv}=\css \rho$ for ($a,b,c$) = (2,0,0), (0,2,0), and (0,0,2), and the off-diagonal components be zero, i.e., $\Pi\sbs{abc}\sps{Mv}=0$ for ($a,b,c$) = (1,1,0), (0,1,1), and (1,0,1), it follows that $D\sps{\prime}\sbs{abc}= \css$ in the former case and $D\sps{\prime}\sbs{abc}=0$ in the latter case. However, as argued earlier for the 2D formulation, when there are large disparities in the relaxation rates of the second order moments that are related to the physical fluid viscosity and the higher order moments, with the former being relatively much smaller for high Reynolds number flows, in order to simulate such cases accurately and without any numerical artifacts such as hyperviscosities, it is essential to renormalize the diffusion tensor parameter in evolving such higher moments $\Pi\sbs{mnp}$ under collision with $(m+n+p)>2$. The condition for renormalization is obtained as a fixed point of the following equation:
$\left( \frac{\delta}{\delta t} \Pi\sbs{abc}\right)\sbs{\!coll}\sps{\!FP} = 0$ when $\Pi\sbs{abc} = \tilde{\Pi}\sbs{abc}$ for $(a+b+c) = 2$ or equivalently, $\Pi\sbs{000} D\sps{\prime}\sbs{abc}-\tilde{\Pi}\sbs{abc}=0$ for ($a,b,c$) = (2,0,0), (0,2,0), (0,0,2), (1,1,0), (0,1,1), and (1,0,1), where $\tilde{\Pi}\sbs{abc}$ refers to the post-collision state of the second central moments. Simply put, this condition implies quasi-equilibrium states for the second order moments when evolving higher moments. These then parameterize $D\sps{\prime}\sbs{abc}$ in such cases as $D\sps{\prime}\sbs{abc}=\tilde{\Pi}\sbs{abc}/\rho$ for any $a$, $b$, and $c$ satisfying $(a+b+c) = 2$. Thus, in summary, we have the following selection rule for the diffusion tensor parameter for the 3D model for evolving the central moment $\Pi\sbs{mnp}$ under collision:
\begin{eqnarray}\label{eq:diffusion-tensor-parameter-3D}
&& D\sps{\prime}\sbs{200} = D\sps{\prime}\sbs{020} = D\sps{\prime}\sbs{002} = \css, \quad \quad\quad\quad\quad\quad\quad\;\;\mbox{for}\quad (m+n+p) \le 2, \nonumber\\[2mm]
&& D\sps{\prime}\sbs{110} =  D\sps{\prime}\sbs{101} =  D\sps{\prime}\sbs{011} = 0,  \nonumber \\[2mm]
&& D\sps{\prime}\sbs{200} = \frac{\tilde{\Pi}\sbs{200}}{\rho}, D\sps{\prime}\sbs{020} = \frac{\tilde{\Pi}\sbs{020}}{\rho}, D\sps{\prime}\sbs{002} = \frac{\tilde{\Pi}\sbs{002}}{\rho}, \quad \mbox{for}\quad (m+n+p) > 2, \\[2mm]
&& D\sps{\prime}\sbs{110} = \frac{\tilde{\Pi}\sbs{110}}{\rho}, D\sps{\prime}\sbs{101} = \frac{\tilde{\Pi}\sbs{101}}{\rho}, D\sps{\prime}\sbs{011} = \frac{\tilde{\Pi}\sbs{011}}{\rho}. \nonumber
\end{eqnarray}
These choices, especially the renormalized parameterization for evolving higher moments, and the resulting attractors given below were numerically demonstrated to be crucial in eliminating hyperviscosity artifacts arise when there is a scale separation in the relaxation rates in simulating high $\mbox{Re}$ flows~\cite{schupbach2024fokker}. However, the underlying physical basis and its relation to the renormalization group formulation was not explicitly clarified in~\cite{schupbach2024fokker}, which has been done here.

\subsubsection{Selected Continuous Central Moment Markovian Attractors}
Our next goal is to determine the continuous central moment Markovian attractors that inform an LB scheme discussed later that uses the D3Q27 lattice, or equivalently, evolves 27 independent moments under collision. Since the zeroth and first central moments are conserved under collision, it follows that $\Pi\sbs{000}\sps{Mv} = \Pi\sbs{000}=\rho$, $\Pi\sbs{100}\sps{Mv}=\Pi\sbs{100}=0$, $\Pi\sbs{010}\sps{Mv}=\Pi\sbs{010}=0$, and $\Pi\sbs{001}\sps{Mv} = \Pi\sbs{001}= 0$. The rest of 23 central moment attractors follow evaluating Eq.~(\ref{eq:Mvcmattractor3D}) together with Eq.~(\ref{eq:diffusion-tensor-parameter-3D}). Then, in summary, we list the complete set of Markovian central moment attractors as follows:
\begin{eqnarray}
\Pi\sbs{000}\sps{Mv} = \rho, \quad \Pi\sbs{100}\sps{Mv} = \Pi\sbs{010}\sps{Mv} = \Pi\sbs{001}\sps{Mv} = 0,
\end{eqnarray}
\begin{eqnarray}
\Pi\sbs{110}\sps{Mv} = \Pi\sbs{101}\sps{Mv} = \Pi\sbs{011}\sps{Mv} = 0, \quad \Pi\sbs{200}\sps{Mv} = \Pi\sbs{020}\sps{Mv} = \Pi\sbs{002}\sps{Mv} = \css \rho,\nonumber
\end{eqnarray}
\begin{eqnarray}
\Pi\sbs{120}\sps{Mv} = \Pi\sbs{102}\sps{Mv} = \Pi\sbs{210}\sps{Mv} = \Pi\sbs{012}\sps{Mv} = \Pi\sbs{201}\sps{Mv} = \Pi\sbs{021}\sps{Mv} = \Pi\sbs{111}\sps{Mv} = 0.\nonumber
\end{eqnarray}

\begin{eqnarray}
\Pi\sbs{220}\sps{Mv}=\frac{1}{\rho}\left[\tilde{\Pi}\sbs{200}\tilde{\Pi}\sbs{020}+2\tilde{\Pi}\sbs{110}\sps{2}\right], \qquad
\Pi\sbs{202}\sps{Mv}=\frac{1}{\rho}\left[\tilde{\Pi}\sbs{200}\tilde{\Pi}\sbs{002}+2\tilde{\Pi}\sbs{101}\sps{2}\right], \nonumber
\end{eqnarray}
\vspace{-3mm}
\begin{eqnarray}
\Pi\sbs{022}\sps{Mv}=\frac{1}{\rho}\left[\tilde{\Pi}\sbs{020}\tilde{\Pi}\sbs{002}+2\tilde{\Pi}\sbs{011}\sps{2}\right], \qquad
\Pi\sbs{211}\sps{Mv}=\frac{1}{\rho}\left[\tilde{\Pi}\sbs{200}\tilde{\Pi}\sbs{011}+2\tilde{\Pi}\sbs{110}\tilde{\Pi}\sbs{101}\right], \nonumber
\end{eqnarray}
\vspace{-3mm}
\begin{eqnarray}
\Pi\sbs{121}\sps{Mv}=\frac{1}{\rho}\left[\tilde{\Pi}\sbs{020}\tilde{\Pi}\sbs{101}+2\tilde{\Pi}\sbs{110}\tilde{\Pi}\sbs{011}\right], \qquad
\Pi\sbs{112}\sps{Mv}=\frac{1}{\rho}\left[\tilde{\Pi}\sbs{002}\tilde{\Pi}\sbs{110}+2\tilde{\Pi}\sbs{101}\tilde{\Pi}\sbs{011}\right], \nonumber
\end{eqnarray}
\vspace{-3mm}
\begin{eqnarray}
\Pi\sbs{122}\sps{ Mv}=\frac{2}{5\rho}\left[\tilde{\Pi}\sbs{020}\tilde{\Pi}\sbs{102}+\tilde{\Pi}\sbs{002}\tilde{\Pi}\sbs{120}+4\tilde{\Pi}\sbs{011}\tilde{\Pi}\sbs{111}+2(\tilde{\Pi}\sbs{101}\tilde{\Pi}\sbs{021}+\tilde{\Pi}\sbs{110}\tilde{\Pi}\sbs{012})\right], \nonumber
\end{eqnarray}
\vspace{-3mm}
\begin{eqnarray}
\Pi\sbs{212}\sps{ Mv} = \frac{2}{5\rho}[\tilde{\Pi}\sbs{200}\tilde{\Pi}\sbs{012}+\tilde{\Pi}\sbs{002}\tilde{\Pi}\sbs{210} + 4\tilde{\Pi}\sbs{101}\tilde{\Pi}\sbs{111}+2(\tilde{\Pi}\sbs{110}\tilde{\Pi}\sbs{102}+\tilde{\Pi}\sbs{011}\tilde{\Pi}\sbs{201})], \nonumber
\end{eqnarray}
\vspace{-3mm}
\begin{eqnarray}
\Pi\sbs{221}\sps{ Mv} = \frac{2}{5\rho}[\tilde{\Pi}\sbs{200}\tilde{\Pi}\sbs{021}+\tilde{\Pi}\sbs{020}\tilde{\Pi}\sbs{201} + 4\tilde{\Pi}\sbs{110}\tilde{\Pi}\sbs{111}+2(\tilde{\Pi}\sbs{011}\tilde{\Pi}\sbs{210}+\tilde{\Pi}\sbs{101}\tilde{\Pi}\sbs{120})], \nonumber
\end{eqnarray}
\begin{eqnarray}
\Pi\sbs{222}\sps{Mv}\! \! \!&=& \! \! \! \frac{1}{3\rho}[\tilde{\Pi}\sbs{200}\tilde{\Pi}\sbs{022}+\tilde{\Pi}\sbs{020}\tilde{\Pi}\sbs{202}+\tilde{\Pi}\sbs{002}\tilde{\Pi}\sbs{220}\nonumber \\
 & & \qquad \qquad+ 4(\tilde{\Pi}\sbs{110}\tilde{\Pi}\sbs{112}+\tilde{\Pi}\sbs{101}\tilde{\Pi}\sbs{121}+\tilde{\Pi}\sbs{011}\tilde{\Pi}\sbs{211})].\label{eq:continouous-Markovian-CM-3D}
\end{eqnarray}

\subsubsection{Central Moment of Boltzmann's Acceleration Term due to Body Force in 3D}
To account for the effect of any external body force $\bm{F}=(F_x,F_y,F_z)$, we take the inner product of the Boltzmann's acceleration term $\left(\frac{\delta f}{\delta t}\right)\sbs{\!\!\!forcing}$ (see Eq.~(\ref{eq:boltzmannforcingterm})) that appears as an additional source contribution to the continuous kinetic equation for hydrodynamics given in Eq.~(\ref{boltzmann_equation1}) with the central moment weighting factor $W\sbs{mnp}$, which can be represented as\begin{eqnarray}\label{cm_forces3D}
\Gamma\sbs{mnp}= \left<  \left(\frac{\delta f}{\delta t}\right)\sbs{\!\!\!forcing} , W\sbs{mnp} \right>.
\end{eqnarray}
Evaluating this last equation, we get the source contribution to the central moment of order $(m+n+p)$ as
\begin{eqnarray}\label{forcing:3D}
\Gamma\sbs{mnp} = m \frac{F\sbs{x}}{\rho}\Pi\sbs{m\sm1,n,p} + n \frac{F\sbs{y}}{\rho}\Pi\sbs{m,n\sm1,p}+p \frac{F\sbs{z}}{\rho}\Pi\sbs{m,n,p\sm1},
\end{eqnarray}

\subsection{Formulating LBM using FP Central Moment Collision Operator for Hydrodynamics}
For brevity, we will only highlight main ideas behind formulating the FPC-LBM for hydrodynamics in 3D using the D3Q27 lattice as a detailed discussion is found in our recent work~\cite{schupbach2024fokker}. The Cartesian components of the particle velocity of the latter are obtained from forming the tensor products of the set $\{-1,0,1\}$ with itself in 3D, which are given by
\begin{subequations}
\begin{eqnarray}
\bigr|\mathbf{e}\sbs{x}\bigr>  &=& (0,1,\sm1,0,0,0,0,1,\sm1,1,\sm1,1,\sm1,1,\sm1,0,0,0,0,1,\sm1,1,\sm1,1,\sm1,1,\sm1)\sps{\dag},\\[2mm]
\bigr|\mathbf{e}\sbs{y}\bigr>  &=& (0,0,0,1,\sm1,0,0,1,1,\sm1,\sm1,0,0,0,0,1,\sm1,1,\sm1,1,1,\sm1,\sm1,1,1,\sm1,\sm1)\sps{\dag},\\[2mm]
\bigr|\mathbf{e}\sbs{z}\bigr>  &=& (0,0,0,0,0,1,\sm1,0,0,0,0,1,1,\sm1,\sm1,1,1,\sm1,\sm1,1,1,1,1,\sm1,\sm1,\sm1,\sm1)\sps{\dag}.
\end{eqnarray}
\end{subequations}
The LB scheme using this particle velocity set is based on Eqs.~(\ref{eq:LBMsteps-collide}) and (\ref{eq:LBMsteps-stream}) with $q=27$. However, to express the collisions step in terms of relaxation of central moments, we first define the various central moments as
\begin{eqnarray}
\left( \begin{array}{c}\Ks{mnp} \\[2mm] \KMv{mnp} \\[2mm] \Ss{mnp} \end{array} \right)  &=& \sum\sbs{\alpha = 0}\sps{26} \left( \begin{array}{c}\f{\alpha} \\[2mm] \f{\alpha}\sps{Mv} \\[2mm] S\sbs{\alpha} \end{array} \right)  ( e\sbs{\alpha x} - u\sbs{x})\sps{m}  ( e\sbs{\alpha y} - u\sbs{y})\sps{n}( e\sbs{\alpha z} - u\sbs{z})\sps{p}, \;\;
\end{eqnarray}
from which the 27 independent moments for the D3Q27 lattice can be formed which are given by
\begin{align*}
\mc= ( & \Ks{000}, \Ks{100}, \Ks{010}, \Ks{001}, \Ks{110}, \Ks{101}, \Ks{011}, \Ks{200}, \Ks{020}, \Ks{002},\Ks{120}, \Ks{102}, \Ks{210},\Ks{012},  \\[5pt] &  \Ks{201}, \Ks{021}, \Ks{111}, \Ks{220}, \Ks{202}, \Ks{022},  \Ks{211}, \Ks{121}, \Ks{112}, \Ks{122}, \Ks{212}, \Ks{221}, \Ks{222} )^\dag.
\end{align*}
%
%
%
%
The corresponding discrete Markovian central moment attractors $\KMv{mnp}$ are obtained via matching with their continuous counterparts $\Pi\sbs{mnp}\sps{Mv}$. Then, we have
\begin{equation}
\KMv{000} = \rho, \; \; \; \; \; \; \KMv{100} = 0, \; \; \; \; \; \; \Ks{010}\sps{ Mv} = 0, \; \; \; \; \; \; \KMv{001} = 0, \nonumber
\end{equation}
\begin{equation}
\KMv{110} = 0, \; \; \; \; \; \;\KMv{101} = 0, \; \; \; \; \; \;\KMv{ 011} = 0, \; \; \; \; \; \; \KMv{ 200} = \rho \css, \; \; \; \; \; \;\KMv{ 020} = \rho \css, \; \; \; \; \; \;\KMv{ 002} =\rho \css, \nonumber
\end{equation}
\begin{equation}
\KMv{ 120} = 0, \; \; \; \; \; \; \KMv{ 102} =  0, \; \; \; \; \; \;\KMv{ 210} = 0, \; \; \; \; \; \;\KMv{ 012} = 0,  \; \; \; \; \; \;\KMv{ 201} = 0, \; \; \; \; \; \;\KMv{021} = 0, \; \; \; \; \; \;\KMv{111} = 0, \nonumber
\end{equation}
\begin{equation}
\KMv{220} = \frac{1}{\rho} (\Kts{200}\Kts{020}+2\Kts{110}\Kts{110}),\; \; \;  \; \; \; \KMv{202} = \frac{1}{\rho} (\Kts{200}\Kts{002}+2\Kts{101}\Kts{101}), \nonumber
\end{equation}
\begin{equation}
\KMv{022} = \frac{1}{\rho}(\Kts{020}\Kts{002}+2\Kts{011}\Kts{011}), \nonumber
\end{equation}
\begin{equation}
\KMv{211} = \frac{1}{\rho}(\Kts{200}\Kts{011}+2\Kts{110}\Kts{101}),\; \; \;  \; \; \; \KMv{121} = \frac{1}{\rho}(\Kts{020}\Kts{101}+2\Kts{110}\Kts{011}), \nonumber
\end{equation}
\begin{equation}
\KMv{112} = \frac{1}{\rho}(\Kts{002}\Kts{110}+2\Kts{011}\Kts{101}), \nonumber
\end{equation}
\begin{equation}
\KMv{122} = \frac{2}{5\rho}(\Kts{020}\Kts{102}+\Kts{002}\Kts{120} + 4\Kts{011}\Kts{111}+2(\Kts{101}\Kts{021}+\Kts{011}\Kts{012})), \nonumber
\end{equation}
\begin{equation}
\KMv{212} = \frac{2}{5\rho}(\Kts{200}\Kts{012}+\Kts{002}\Kts{210} + 4\Kts{101}\Kts{111}+2(\Kts{110}\Kts{102}+\Kts{011}\Kts{201})), \nonumber
\end{equation}
\begin{equation}
\KMv{221} = \frac{2}{5\rho}(\Kts{200}\Kts{021}+\Kts{020}\Kts{201} + 4\Kts{110}\Kts{111}+2(\Kts{011}\Kts{210}+\Kts{101}\Kts{120})), \nonumber
\end{equation}
\begin{equation}
\KMv{222} = \frac{1}{3\rho}(\Kts{200}\Kts{022}+\Kts{020}\Kts{202}+\Kts{002}\Kts{220}+4(\Kts{110}\Kts{112}+\Kts{101}\Kts{121}+\Kts{011}\Kts{211})), \nonumber
\end{equation}
where tilde in these expressions refer to post-collision states of the moments.
Similarly, matching with the changes in continuous central moments due to body forces given in Eq.~(\ref{forcing:3D}), the corresponding changes in the discrete central moments $\sigma\sbs{mnp}$ can be expressed as
\begin{eqnarray}
\sigma\sbs{mnp} = m \frac{F\sbs{x}}{\rho}\kappa\sbs{m\sm1,n,p} + n \frac{F\sbs{y}}{\rho}\kappa\sbs{m,n\sm1,p}+p \frac{F\sbs{z}}{\rho}\kappa\sbs{m,n,p\sm1}
\end{eqnarray}
for all 27 possible combinations of $m$, $n$, $p$. The overall implementation steps are similar to the 2D FPC-LBM discussed earlier and full details of the algorithm for implementing the 3D FPC-LBM for computing fluid motions using the D3Q27 lattice are presented in~\cite{schupbach2024fokker}.

\subsection{Construction of FP Central Moment Collision Operator: Energy Transport}
The FP collision model of the continuous kinetic equation for energy transport (see Eq.~(\ref{boltzmann_equation_two})) in 3D is obtained by replacing $f$ with $g$ for the distribution function and using $\hat{D}\sps{\prime}\sbs{ij}$ in place of $D\sps{\prime}\sbs{ij}$ in for the diffusion tensor parameter in Eqs.~(\ref{FP_short3D}) and (\ref{dFP3D}). Then, upon defining the central moment of the distribution function $g$ of order ($m+n+p$) as
\begin{eqnarray}
\Lambda\sbs{mnp}=\left<g,W\sbs{mnp}\right>,
\end{eqnarray}
and taking the inner product of the attendant FP collision model as mentioned above and simplifying and rearranging we get the following collision operator
\begin{equation}\label{FPcoll3dg}
\left(\frac{\delta}{\delta t} \Lambda\sbs{mnp}\right)\sps{\!\!\! FP}\sbs{\!\!\! coll} = \omega\sbs{mnp} \left[ \Lambda\sbs{mnp}\sps{Mv}-\Lambda\sbs{mnp} \right],
\end{equation}
where $\Lambda\sbs{mnp}\sps{Mv}$ reads as
\begin{eqnarray}\label{eq:Mvcmattractor3Dg}
\Lambda\sbs{mnp}\sps{ Mv} &=&  \hat{D}\sps{\prime}\sbs{200}\frac{m(m\sm1)}{(m+n+p)}\Lambda\sbs{m\sm2,n,p} + \hat{D}\sps{\prime}\sbs{020}\frac{n(n-1)}{(m+n+p)}\Lambda\sbs{m,n\sm2,p} +   \\[2mm]
                    && \hat{D}\sps{\prime}\sbs{002}\frac{p(p\sm1)}{(m+n+p)}\Lambda\sbs{m,n,p\sm2} + 2\hat{D}\sps{\prime}\sbs{110}\frac{mn}{(m+n+p)}\Lambda\sbs{m\sm1,n\sm1,p} +\nonumber \\[2mm]
                    && 2\hat{D}\sps{\prime}\sbs{011}\frac{np}{(m+n+p)}\Lambda\sbs{m,n\sm1,p\sm1} + 2\hat{D}\sps{\prime}\sbs{101}\frac{mp}{(m+n+p)}\Lambda\sbs{m\sm1,n,p\sm1}. \nonumber
\end{eqnarray}

\subsubsection{Choice of Diffusion Tensor Parameter}
Based on an approach similar to the 2D formulation, the selection rule for the diffusion tensor parameter, based on the \emph{option A} reads as
\begin{eqnarray}\label{eq:diffusion-tensor-parameter-3DgoptionA}
&& \hat{D}\sps{\prime}\sbs{200} = \hat{D}\sps{\prime}\sbs{020} = \hat{D}\sps{\prime}\sbs{002} = \chss,  \nonumber\\[2mm]
&& \hat{D}\sps{\prime}\sbs{110} =  \hat{D}\sps{\prime}\sbs{101} =  \hat{D}\sps{\prime}\sbs{011} = 0, \quad \quad\quad\quad\quad\quad\quad\;\;\mbox{for all}\quad (m+n+p)
\end{eqnarray}
for evolving the central moment $\Lambda\sbs{mnp}$ under collision. Here, $\chss$ is a fixed constant equal to $1/3$. On the other hand, the \emph{option B}, which assumes a separation of time scales for relaxations of second order moments with those of the higher order ones leading to a renormalization of the diffusion tensor parameter for evolving such higher moments, specifies
\begin{eqnarray}\label{eq:diffusion-tensor-parameter-3Dg}
&& \hat{D}\sps{\prime}\sbs{200} = \hat{D}\sps{\prime}\sbs{020} = \hat{D}\sps{\prime}\sbs{002} = \chss, \quad \quad\quad\quad\quad\quad\quad\;\;\mbox{for}\quad (m+n+p) \le 2, \nonumber\\[2mm]
&& \hat{D}\sps{\prime}\sbs{110} =  \hat{D}\sps{\prime}\sbs{101} =  \hat{D}\sps{\prime}\sbs{011} = 0,  \nonumber \\[2mm]
&& \hat{D}\sps{\prime}\sbs{200} = \frac{\tilde{\Lambda}\sbs{200}}{T}, \hat{D}\sps{\prime}\sbs{020} = \frac{\tilde{\Lambda}\sbs{020}}{T}, \hat{D}\sps{\prime}\sbs{002} = \frac{\tilde{\Lambda}\sbs{002}}{T}, \quad \mbox{for}\quad (m+n+p) > 2, \\[2mm]
&& \hat{D}\sps{\prime}\sbs{110} = \frac{\tilde{\Lambda}\sbs{110}}{T}, \hat{D}\sps{\prime}\sbs{101} = \frac{\tilde{\Lambda}\sbs{101}}{T}, \hat{D}\sps{\prime}\sbs{011} = \frac{\tilde{\Lambda}\sbs{011}}{T}, \nonumber
\end{eqnarray}
where tilde refers to the post-collision state, for evolving the central moment $\Lambda\sbs{mnp}$.

\subsubsection{Selected Continuous Central Moment Markovian Attractors}
Next, in anticipation of formulating a LB scheme using the collision operator with the D3Q15 lattice in the next section, we can obtain the attractors for all the 15 independent moments as follows: The zeroth moment is a collision invariant, i.e., $\Lambda\sbs{000}\sps{Mv} = \Lambda\sbs{000} = T$, the first order moment attractors should be related to the temperature fluxes in order to recover the energy transport equation via a C-E analysis, or equivalently, their central moments are null, i.e., $\Lambda\sbs{100}\sps{Mv}=\Lambda\sbs{010}\sps{Mv}=\Lambda\sbs{001}\sps{Mv}=0$, and the higher order attractors are determined by evaluating Eq.~(\ref{eq:Mvcmattractor3Dg}) together with either Eq.~(\ref{eq:diffusion-tensor-parameter-3DgoptionA}) or Eq.~(\ref{eq:diffusion-tensor-parameter-3Dg}). With \emph{option A} (based on Eq.~(\ref{eq:diffusion-tensor-parameter-3DgoptionA})), the results are summarized as follows:
\begin{gather}
\Lambda\sbs{000}\sps{Mv} = T, \quad \Lambda\sbs{100}\sps{Mv} = \Lambda\sbs{010}\sps{Mv} = \Lambda\sbs{001}\sps{Mv} = 0, \nonumber \\[2mm]
\Lambda\sbs{110}\sps{Mv} = \Lambda\sbs{101}\sps{Mv} = \Lambda\sbs{011}\sps{Mv} = 0, \quad \Lambda\sbs{200}\sps{Mv} = \Lambda\sbs{020}\sps{Mv} = \Lambda\sbs{002}\sps{Mv} = \chss T,\nonumber \\[2mm]
\Lambda\sbs{120}\sps{Mv} = \frac{2}{3}\chss\tilde{\Lambda}\sbs{100}, \qquad
\Lambda\sbs{012}\sps{Mv} = \frac{2}{3}\chss\tilde{\Lambda}\sbs{010}, \qquad
\Lambda\sbs{201}\sps{Mv} = \frac{2}{3}\chss\tilde{\Lambda}\sbs{001}, \qquad
\Lambda\sbs{111}\sps{Mv} = 0,\nonumber \\[2mm]
\Lambda\sbs{220}\sps{Mv}=\frac{1}{2}\chss\left(\tilde{\Lambda}\sbs{200}+\tilde{\Lambda}\sbs{020}\right). \label{eq:g-attractorsD3Q15optionA}
\end{gather}
If the \emph{option B} is utilized (see Eq.~(\ref{eq:diffusion-tensor-parameter-3Dg})), the third and the fourth order attractors in the last equation should be replaced with
\begin{gather*}
\Lambda\sbs{120}\sps{Mv} =  \frac{2}{3T} \left( \tilde{\Lambda}\sbs{020}\tilde{\Lambda}\sbs{100} + 2\tilde{\Lambda}\sbs{110}\tilde{\Lambda}\sbs{010}   \right), \qquad
\Lambda\sbs{012}\sps{Mv} = \frac{2}{3T} \left( \tilde{\Lambda}\sbs{002}\tilde{\Lambda}\sbs{010} + 2\tilde{\Lambda}\sbs{011}\tilde{\Lambda}\sbs{001}   \right), \nonumber \\[2mm]
\Lambda\sbs{201}\sps{Mv} = \frac{2}{3T} \left( \tilde{\Lambda}\sbs{200}\tilde{\Lambda}\sbs{001} + 2\tilde{\Lambda}\sbs{101}\tilde{\Lambda}\sbs{100}   \right), \qquad
\Lambda\sbs{111}\sps{Mv} = \frac{2}{3T} \left( \tilde{\Lambda}\sbs{110}\tilde{\Lambda}\sbs{001} + \tilde{\Lambda}\sbs{011}\tilde{\Lambda}\sbs{100} + \tilde{\Lambda}\sbs{101}\tilde{\Lambda}\sbs{010}\right),\nonumber \\[2mm]
\Lambda\sbs{220}\sps{Mv}=\frac{1}{T}\left(\tilde{\Lambda}\sbs{200}\tilde{\Lambda}\sbs{020}+2\tilde{\Lambda}\sbs{110}\sps{2}\right). \nonumber
\end{gather*}
For the same reasons as those discussed as in the 2D case, we will use the \emph{option A} and its associated attractors given in Eq.~(\ref{eq:g-attractorsD3Q15optionA}) in constructing the LBM on the D3Q15 lattice discussed next.

\subsection{Formulating LBM using FP Central Moment Collision Operator for Energy Transport}
We will now discuss the formulation of the 3D FPC-LBM for energy transport using the D3Q15 lattice, which is a subset of the D3Q27 lattice and the Cartesian components of its particle velocity is given by
\begin{subequations}
\begin{eqnarray}
\bigr|\mathbf{e}\sbs{x}\bigr>  &=& (0,1,-1,0,0,0,0,1,\sm1,1,\sm1,1,\sm1,1,\sm1)\sps{\dag},\\[2mm]
\bigr|\mathbf{e}\sbs{y}\bigr>  &=& (0,0,0,1,\sm1,0,0,1,1,\sm1,\sm1,1,1,\sm1,\sm1)\sps{\dag},\\[2mm]
\bigr|\mathbf{e}\sbs{z}\bigr>  &=& (0,0,0,0,0,1,\sm1,1,1,1,1,\sm1,\sm1,\sm1,\sm1)\sps{\dag}.
\end{eqnarray}
\end{subequations}
Based on this, a LB scheme for the discrete distribution function $g_\alpha$ can be expressed in the form of Eq.~(\ref{eq:LBMsteps-collideg}) and (\ref{eq:LBMsteps-streamg}) with $q = 15$. Then, to model the collision term $\hat{\Omega}\sbs{\alpha}$ as a relaxation process based on central moments to their Markovian attractors, we introduce the following discrete central moments and raw moments, respectively:
\begin{subequations}
\begin{eqnarray}
\left( \begin{array}{c}\Ns{mnp} \\[2mm] \NMv{mnp} \end{array} \right)  &=& \sum\sbs{\alpha = 0}\sps{14} \left( \begin{array}{c}\g{\alpha} \\[2mm] \g{\alpha}\sps{Mv} \\[2mm]  \end{array} \right)  ( e\sbs{\alpha x} - u\sbs{x})\sps{m}  ( e\sbs{\alpha y} - u\sbs{y})\sps{n}( e\sbs{\alpha z} - u\sbs{z})\sps{p}, \;\;\mbox{and}\\[0.2in]
\left( \begin{array}{c}\Nps{mnp} \\[2mm] \Ns{mnp}\sps{\prime Mv} \\[2mm] \end{array} \right)  &=& \sum\sbs{\alpha = 0}\sps{14} \left( \begin{array}{c}\g{\alpha} \\[2mm] \g{\alpha}\sps{Mv}\\[2mm]  \end{array} \right)  e\sbs{\alpha x}\sps{m}  e\sbs{\alpha y}\sps{n} e\sbs{\alpha z}\sps{p}.
\end{eqnarray}
\end{subequations}
The list of 15 independent central moments $\Ns{mnp}$ and raw moments $\Np\sbs{mnp}$ for the D3Q15 lattice are then given in terms vectors $\nc$ and $\n$, respectively, as
\begin{align}
\nc= ( \Ns{000}, \Ns{100}, \Ns{010}, \Ns{001}, \Ns{110}, \Ns{101}, \Ns{011}, \Ns{200}, \Ns{020}, \Ns{002},\Ns{120}, \Ns{012},\Ns{201}, \Ns{111}, \Ns{220})^\dag,\nonumber
\end{align}
\begin{align}
\n= ( \Nps{000}, \Nps{100}, \Nps{010}, \Nps{001}, \Nps{110}, \Nps{101}, \Nps{011}, \Nps{200}, \Nps{020}, \Nps{002},\Nps{120}, \Nps{012}, \Nps{201}, \Nps{111}, \Nps{220})^\dag,\nonumber
\end{align}
whereas the 15 discrete distribution functions as vector $\mathbf{g}$:
\begin{eqnarray}
\mathbf{g} = (\g{0} , \g{1},\g{2}, \dots , \g{15})\sps{\dag}.\nonumber
\end{eqnarray}

Then, to formulate the effect of collisions as a set of relaxations of different central moments to their attractors at their own rates, the discrete distribution functions are first mapped to raw moments and then to central moments, as in $\n = \PP\mathbf{g}$ and $\nc = \F\n$, respectively, where the transformation matrix $\PP$ is obtained from grouping together the 15 basis vectors as
\begin{eqnarray}
\tensr{P}=\bigr[\bigr|\bm{1}\bigr>,\;\;
\bigr|\bm{e}_x \bigr>,\;\;
\bigr|\bm{e}_y\bigr>,\;\;
\bigr|\bm{e}_z\bigr>,\;\;
\bigr|\bm{e}_x \bm{e}_y\bigr>,\;\;
\bigr|\bm{e}_x \bm{e}_z\bigr>,\;\;
\bigr|\bm{e}_y \bm{e}_z\bigr>,\;\;
\bigr|\bm{e}_x^2\bigr>,\;\;
\bigr|\bm{e}_y^2\bigr>,\;\;
\bigr|\bm{e}_z^2\bigr>,\;\; \nonumber \\[2mm]
\bigr|\bm{e}_x \bm{e}_y^2\bigr>,\;\;
\bigr|\bm{e}_y \bm{e}_z^2\bigr>,\;\;
\bigr|\bm{e}_x^2 \bm{e}_z\bigr>,\;\;
\bigr|\bm{e}_x \bm{e}_y \bm{e}_z\bigr>,\;\;
\bigr|\bm{e}_x^2 \bm{e}_y^2\bigr>\;\;
\nonumber
\bigr]^\dag,
\end{eqnarray}
where $\bigr|\bm{1}\bigr>$ is a 15-dimensional unit vector $\bigr|\bm{1}\bigr>=(1,1,\ldots,1)^\dag$
and the frame transformation matrix $\F$ follows from identifying the coefficients involved in the binary transforms relating the various
central moments and raw moments and is thus a function of the fluid velocity components. While the explicit expressions of these matrices are not shown here for brevity, they are utilized as part of the implementation details shown in Appendix~\ref{sec:appendixC}. Subsequently, the relaxations in terms of central moments under collision can be written as
\begin{equation}
\tilde{\eta}_{mnp}=\eta_{mnp} + \omega_{mnp}(\eta_{mnp}^{Mv}-\eta_{mnp})
\end{equation}
for ($m,n,p$) = (0,0,0), (1,0,0), (0,1,0), (0,0,1), (1,1,0), (1,0,1), (0,1,1), (2,0,0), (0,2,0), (0,0,2), (1,2,0), (0,1,2), (2,0,1), (1,1,1), and (2,2,0), where $\omega_{mnp}$ denotes the relaxation rate of $\eta_{mnp}$. Here, discrete Markovian central moment attractors $\eta_{mnp}^{Mv}$ are obtained from matching with their continuous counterparts $\Lambda_{mnp}^{Mv}$ and they read as
\begin{equation}
\NMv{000} = T, \; \; \; \; \; \; \NMv{100} = 0, \; \; \; \; \; \; \Ns{010}\sps{ Mv} = 0, \; \; \; \; \; \; \NMv{001} = 0, \nonumber
\end{equation}
\begin{equation}
\NMv{110} = 0, \; \; \; \; \; \;\NMv{101} = 0, \; \; \; \; \; \;\NMv{ 011} = 0, \; \; \; \; \; \; \NMv{ 200} =  \chss T, \; \; \; \; \; \;\NMv{ 020} =\chss T, \; \; \; \; \; \;\NMv{ 002} =\chss T, \nonumber
\end{equation}
\begin{equation}
\NMv{120} = \frac{2}{3}\chss\Nts{100},\; \; \; \; \; \; \NMv{012} = \frac{2}{3}\Nts{010}, \; \; \; \; \; \;
\NMv{201} = \frac{2}{3}\chss\Nts{001},\; \; \; \; \; \; \NMv{111} = 0, \nonumber
\end{equation}
\begin{equation}
\NMv{220} = \frac{1}{2}\chss\left(\Nts{200}+\Nts{020}\right).
\end{equation}
Then, the post-collision distribution functions  needed before performing the streaming step (see Eq.~(\ref{eq:LBMsteps-streamg})) are obtained via the following successive inverse mappings: $\tilde{\n} = \Fi\tilde{\nc}$ and $\tilde{\mathbf{g}} = \PP^{-1}\tilde{\n}$. Finally, the temperature field is updated through the zeroth moment of $g_\alpha$, i.e., $T = \sum\sbs{\alpha=0}\sps{14} g\sbs{\alpha}$. We conclude this section by noting that the algorithmic details of the 3D FPC-LBM for solving the energy transport equation using the D3Q15 lattice are given in Appendix~\ref{sec:appendixC}.

\section{Results and Discussion\label{sec:resultsanddiscussion}}
In this study, through simulations of natural convection in a cavity in both two and three dimensions, we first demonstrate that our Fokker-Planck central moments-based collision model can be utilized for accurate simulation of thermal convective flows. The physics of this problem is modulated primarily by the Rayleigh number $\mbox{Ra}$, which represent the relative effects of buoyancy forces to the opposing effects of viscous and thermal diffusions, as well as the Prandtl number $\mbox{Pr}$, which characterizes the relative strength of viscous to thermal diffusion. We will compare the numerical results of simulations using our FPC-LBMs, across a range of Rayleigh numbers at a fixed $\mbox{Pr}$ for air with a variety of references including both prior numerical solutions~\cite{de1983natural,hortmann1990finite,dixit2006simulation,sharma2018natural, fusegi1991numerical,wang2017numerical} and experimental measurements~\cite{tian2000low,mergui1993experimental, krane1983some,bilski1986experimental}. In two dimensions, we will also perform a grid convergence study in obtaining the Nusselt numbers in the simulations involving higher $\mbox{Ra}$. It is well known that above a Rayleigh number of about $\mbox{Ra}=10\sps{8}$, the flow becomes transitional and turbulent natural convection characteristics emerge, and the use of adequately refined grids and the sufficient duration for the collection of statistics are required. While in the 2D case, we will consider $\mbox{Ra}$ as high as $10\sps{10}$, since the standard LBM is restricted to uniform grids and due to the limited availability of computing resources, we will restrict $\mbox{Ra}$ to be up to $10\sps{8}$ in 3D; in future work, we plan to incorporate grid clustering techniques using coordinate transformations to simulate higher $\mbox{Ra}$ in 3D. In addition, we will perform a comparative numerical stability study to show the benefits of using the FPC-LBM over the state-of-the-art LBM utilizing the central moments-based collision models based on the Maxwell distribution, which has been used in several recent investigations~\cite{sharma2018natural,elseid2018cascaded,fei2018modeling,hajabdollahi2018central,hajabdollahi2019cascaded}, for simulations of mixed thermal convection driven by buoyancy forces and an imposed shear.

\subsection{Natural Convection in a Square Cavity}
\begin{figure}[H]
\centering
\includegraphics[trim = 0 0 0 0, clip, width =70mm]{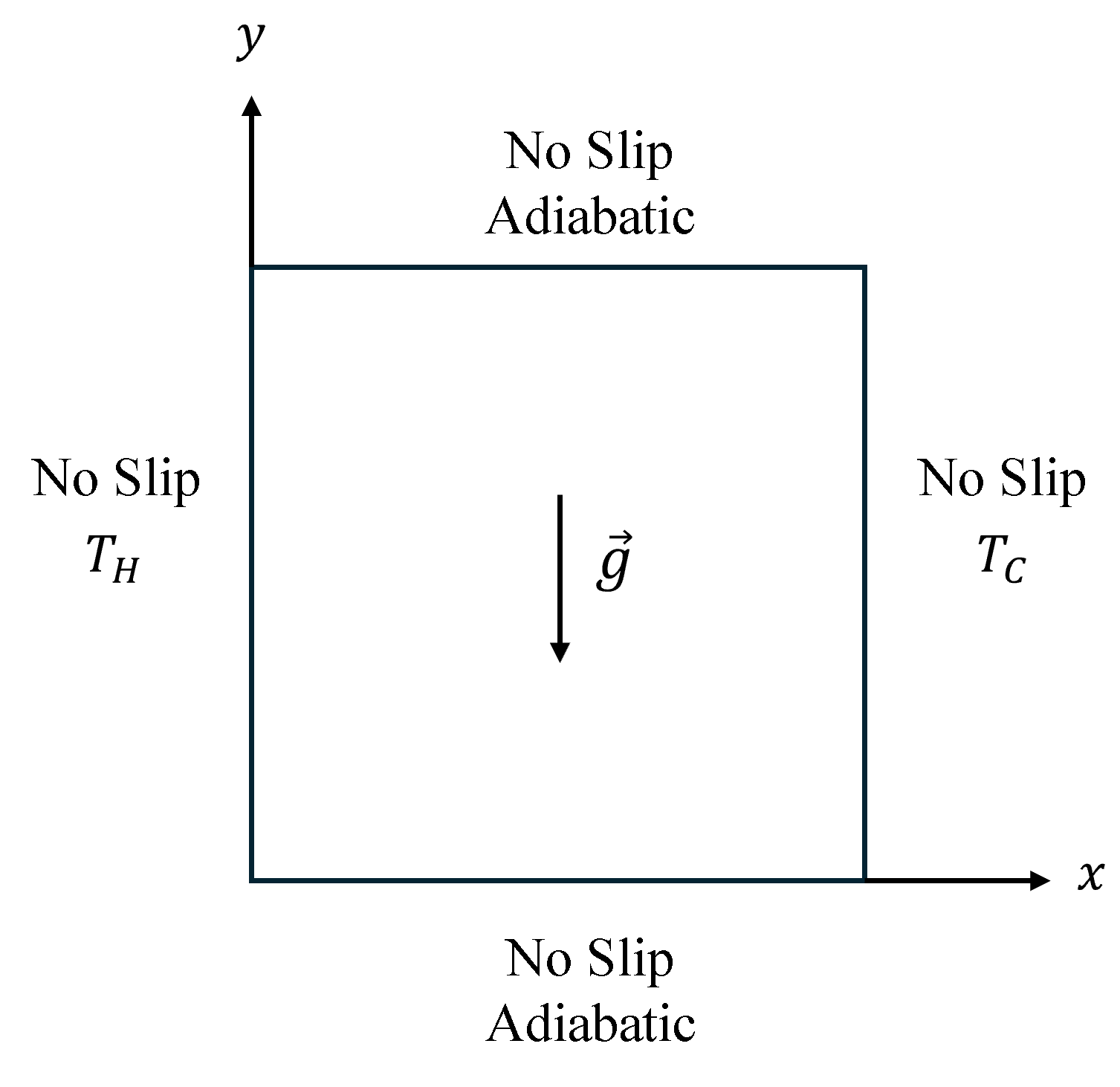}
\caption{Schematic diagram for the buoyancy-driven thermal convection in a square cavity with the attendant boundary conditions.}
\label{fig:schematicNC2D}
\end{figure}

We perform simulations of the canonical natural convection in a square cavity benchmark problem at various $\mbox{Ra}$ to provide a validation of the Fokker-Planck central moments-based collision  model applied to both the distribution functions solving for fluid flow as well as the temperature field. Applying and quantitatively studying this collision model to the second distribution function has not been done before and is a natural extension of our previous FPC-LBM for hydrodynamics~\cite{schupbach2024fokker}. In essence, the 2D FPC-LBM for fluid motions given Appendix~\ref{sec:appendixA} and another 2D FPC-LBM for energy transport provided in Appendix~\ref{sec:appendixB}. A schematic diagram of the problem set up in a square cavity of side length $L_o$ is shown in Fig.~\ref{fig:schematicNC2D}.
First, we now briefly mention the boundary conditions utilized for this problem whose implementation details are given, for example, in Ref.~\cite{kruger2017lattice}. The no-slip condition is applied on all four walls for the fluid velocity using the half-way bounce back scheme for the distribution function $f_\alpha$. On the other hand, the upper and lower walls have zero temperature gradient (adiabatic) which are also specified using a half-way bounce back method for $g_\alpha$, while the left wall has a constant and uniform hot temperature, $T\sbs{H}$, and the right wall has a constant and uniform cold temperature applied, $T\sbs{C}$, and they are implemented using the half-way anti-bounce back approach for the distribution function $g_\alpha$. As is standard in performing simulations using LB schemes, we will utilize the lattice units in setting up the parameters of the problem, i.e., unit lattice spacing and time step are used (see Ref.~\cite{kruger2017lattice}). Then, the characteristic length or the side of the square cavity $L_o$ determines the number of grid nodes used and the characteristic velocity is fixed to $U\sbs{o}=0.01$ so that the Mach number is relatively small and the resulting flow can be regarded as incompressible. Furthermore, the characteristic force per unit mass used to simulate gravity is $F\sbs{o} = U\sbs{o}\sps{2}/L\sbs{o} = g$. The hot temperature imposed on the left wall is fixed at $T\sbs{H}=2$, while the cold temperature on the right wall is set at $T\sbs{C}=1$ in lattice units.

The FPC-LBM used to solve the temperature field is coupled to that for the hydrodynamics via an external body force due to buoyancy. Under the usual Boussinesq approximation, we can write the vertical component of the force as $F\sbs{y} =g\beta(T-T_o)$, while its horizontal component as $F\sbs{x}=0$, so that $\bm{F}=(F_x,F_y)$ is the force locally impressed on the fluid. Here, $\beta$ is the thermal expansion coefficient, $T$ is the local temperature and the mean temperature defined as $T\sbs{o} = (T\sbs{H}+T\sbs{C})/2$ represents the characteristic temperature scale. We perform simulations across a range of Rayleigh numbers $\mbox{Ra}$ at a fixed Prandtl number $\mbox{Pr}$, which are defined as
\begin{eqnarray}
Ra = \frac{g\beta(T\sbs{H} - T\sbs{C})L\sbs{o}\sps{3}}{\alpha \nu},\qquad Pr = \frac{\nu}{\alpha}, \nonumber
\end{eqnarray}
where $\nu$ and $\alpha$ are the kinematic viscosity (which controls the momentum diffusivity) and thermal diffusivity, respectively, and they are related to the relaxation parameters used in the collision steps (see the last paragraphs of Appendices~\ref{sec:appendixA} and~\ref{sec:appendixB} for details). Here, unless otherwise stated, we fix $\mbox{Pr}$ as $0.71$ corresponding to that of air.

The heat transfer rate is determined by the dimensionless Nusselt number, which requires an estimation of the normal temperature gradient on walls. Since we use a half-way bounce back boundary technique, our true boundary location is half way located between the first fluid node and that of the ghost node inside the wall. To that end, we utilize a second order Lagrange interpolating polynomial for unequally spaced data to estimate the first derivative of the temperature field, which reads as
\begin{eqnarray}
f\sps{\prime}(x) = f(x\sbs{0}) \frac{2x-x\sbs{1}-x\sbs{2}}{(x\sbs{0}-x\sbs{1})(x\sbs{0}-x\sbs{2})} +  f(x\sbs{1}) \frac{2x-x\sbs{0}-x\sbs{2}}{(x\sbs{1}-x\sbs{0})(x\sbs{1}-x\sbs{2})} +  f(x\sbs{2}) \frac{2x-x\sbs{0}-x\sbs{1}}{(x\sbs{2}-x\sbs{0})(x\sbs{2}-x\sbs{1})}, \nonumber
\end{eqnarray}
where the wall location is at $x\sbs{0}=0$, and the first two fluid nodes are located at $x\sbs{1} = 1/2$ and $x\sbs{2}=3/2$, respectively. Subsequently, expressing the function values corresponding to the temperature on the wall and the temperature at the first two fluid nodes as $f(x\sbs{0}) = T\sbs{H}$, $f(x\sbs{1}) = T\sbs{1}$, and $f(x\sbs{2}) = T\sbs{2}$, respectively, and then evaluating the derivative at $x=x\sbs{0} = 0$, we get
\begin{eqnarray}
\frac{\partial T}{\partial x}\Bigr|\sbs{x\sbs{0}} = -\frac{8}{3}T\sbs{H} + 3 T\sbs{1} -\frac{1}{3}T\sbs{2}. \nonumber
\end{eqnarray}
Then, the local Nusselt number $\mbox{Nu}\sbs{0}$ at the hot wall, where $T=T\sbs{H}$, is defined as
\begin{eqnarray}
\mbox{Nu}\sbs{0}= \left(\frac{L\sbs{o}}{T\sbs{H}-T\sbs{C}}\right) \frac{\partial T}{\partial x}\Bigr|\sbs{x\sbs{0}}\nonumber
\end{eqnarray}
and the average Nusselt number $\left<\mbox{Nu}\sbs{0}\right>$ at this wall as
\begin{eqnarray}
\left<\mbox{Nu}\sbs{0}\right>= \frac{1}{L\sbs{o}} \int\displaylimits\sbs{y=0}\sps{L\sbs{o}}\mbox{Nu}\sbs{0} \, \mathrm{d} y \nonumber
\end{eqnarray}

\begin{figure}[H]
\centering
\begin{subfigure}{0.48\textwidth}
\includegraphics[trim = 0 0 0 0, clip, width =80mm]{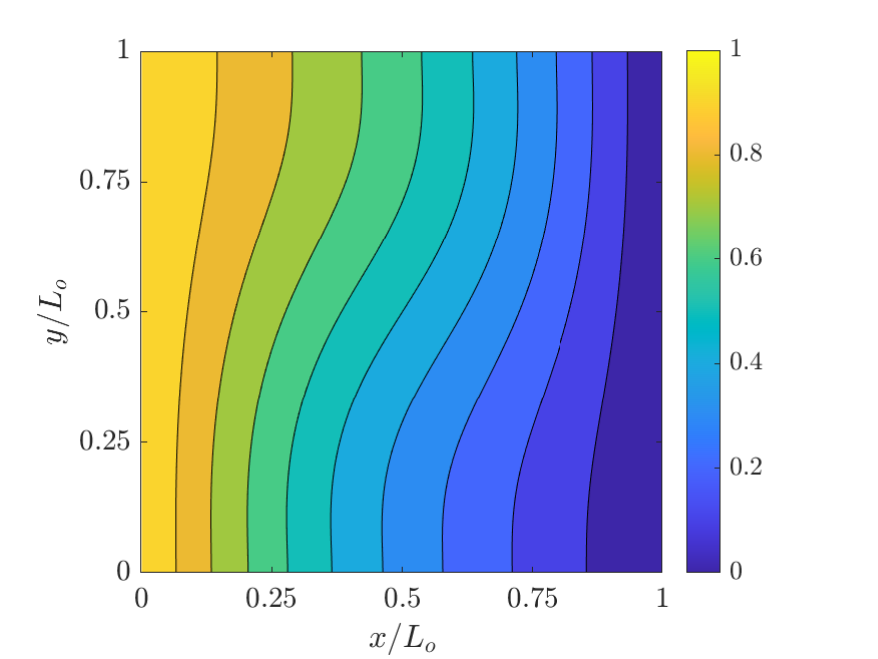}
\end{subfigure}
\begin{subfigure}{0.48\textwidth}
\includegraphics[trim = 0 0 0 0, clip, width =80mm]{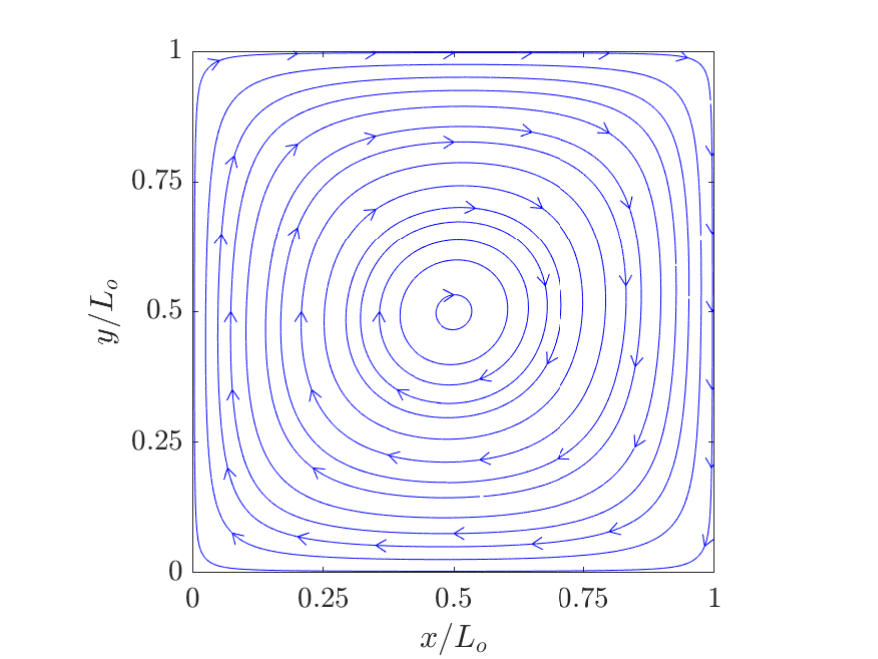}
\end{subfigure}
\caption{Dimensionless temperature contours (\emph{left}) and streamlines (\emph{right}) of natural convection in square cavity computed using the 2D FPC-LBM with $\mbox{Ra}=10\sps{3}$, $\mbox{Pr}=0.71$, and a resolution of $512\times512$ grid nodes.}
\label{ra103contours}
\end{figure}

\begin{figure}[H]
\centering
\begin{subfigure}{0.48\textwidth}
\includegraphics[trim = 0 0 0 0, clip, width =80mm]{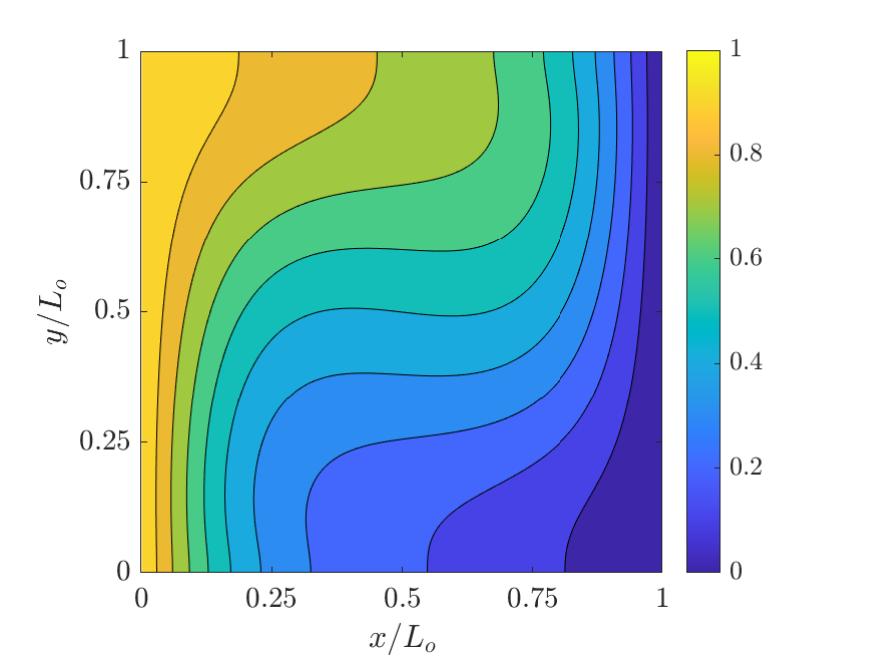}
\end{subfigure}
\begin{subfigure}{0.48\textwidth}
\includegraphics[trim = 0 0 0 0, clip, width =80mm]{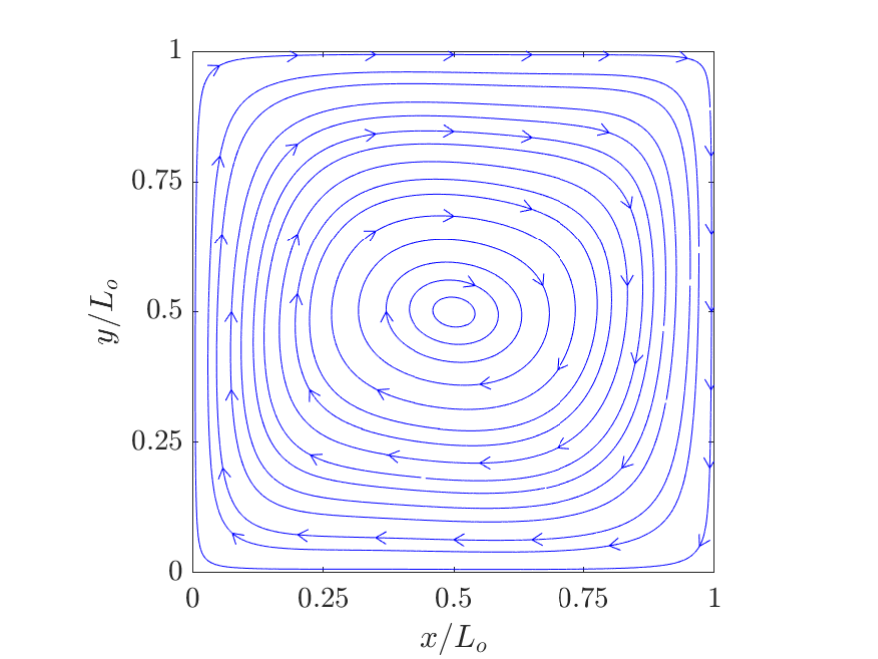}
\end{subfigure}
\caption{Dimensionless temperature contours (\emph{left}) and streamlines (\emph{right}) of natural convection in square cavity computed using the 2D FPC-LBM with $\mbox{Ra}=10\sps{4}$, $\mbox{Pr}=0.71$, and a resolution of $512\times512$ grid nodes.  }
\label{ra104contours}
\end{figure}

\begin{figure}[H]
\centering
\begin{subfigure}{0.48\textwidth}
\includegraphics[trim = 0 0 0 0, clip, width =80mm]{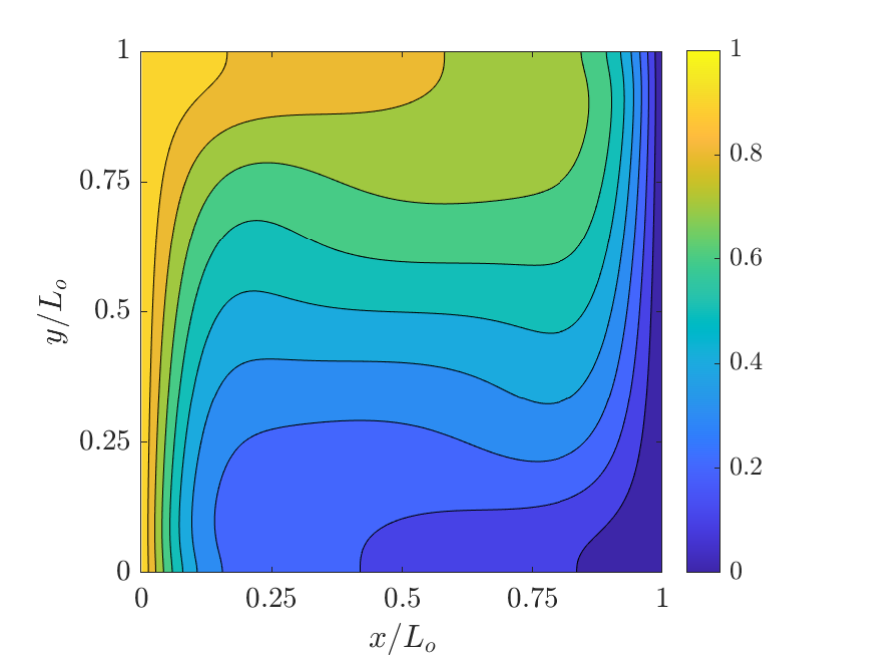}
\end{subfigure}
\begin{subfigure}{0.48\textwidth}
\includegraphics[trim = 0 0 0 0, clip, width =80mm]{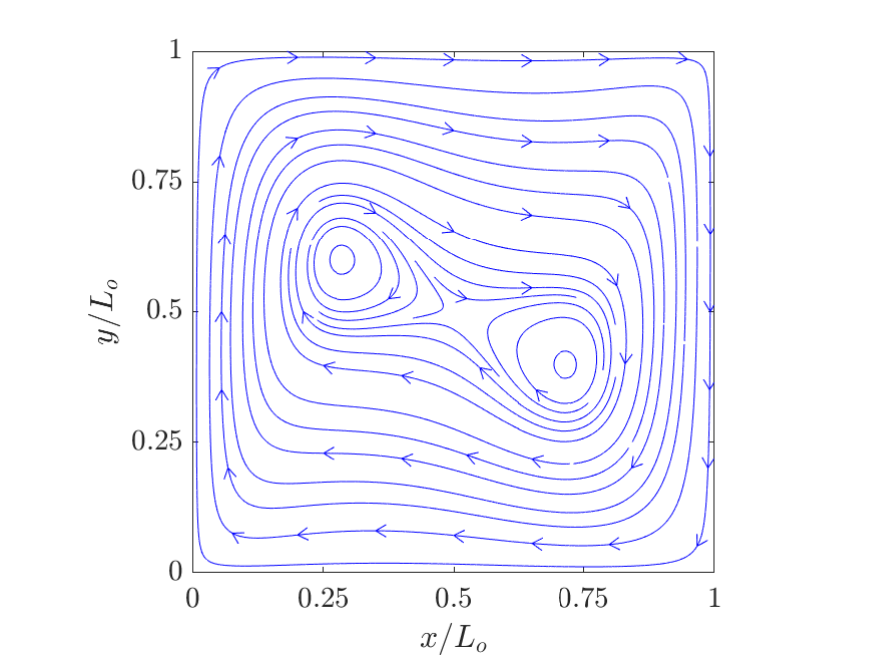}
\end{subfigure}
\caption{Dimensionless temperature contours (\emph{left}) and streamlines (\emph{right}) of natural convection in square cavity computed using the 2D FPC-LBM with $\mbox{Ra}=10\sps{5}$, $\mbox{Pr}=0.71$, and a resolution of $512\times512$ grid nodes. }
\label{ra105contours}
\end{figure}

\begin{figure}[H]
\centering
\begin{subfigure}{0.48\textwidth}
\includegraphics[trim = 0 0 0 0, clip, width =80mm]{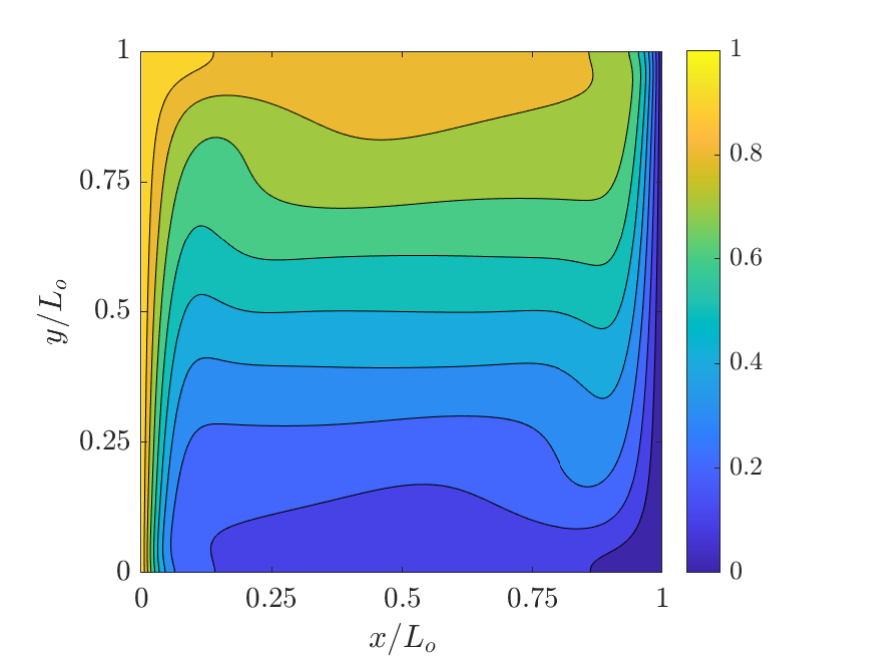}
\end{subfigure}
\begin{subfigure}{0.48\textwidth}
\includegraphics[trim = 0 0 0 0, clip, width =80mm]{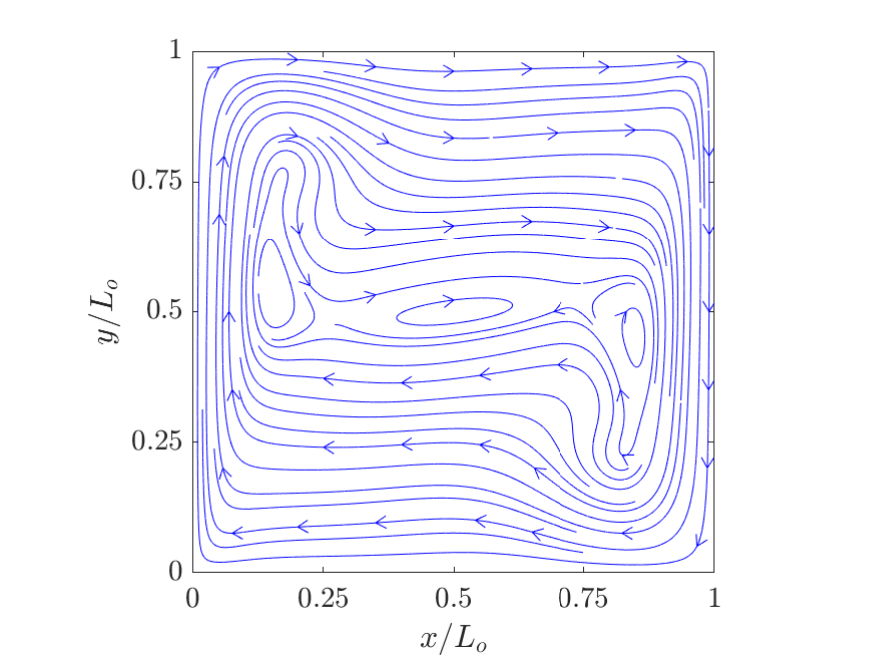}
\end{subfigure}
\caption{Dimensionless temperature contours (\emph{left}) and streamlines (\emph{right}) of natural convection in square cavity computed using the 2D FPC-LBM with $\mbox{Ra}=10\sps{6}$, $\mbox{Pr}=0.71$, and a resolution of $512\times512$ grid nodes.  }
\label{ra106contours}
\end{figure}

The simulation results of the dimensionless temperature contours $\theta = (T-T\sbs{C})/(T\sbs{H}-T\sbs{C}$) and streamlines for the velocity fields at steady state computed using the 2D FPC-LBMs using a resolution of $512\times512$ grid nodes for $\mbox{Ra}=10\sps{3}$, $10\sps{4}$, $10\sps{5}$, and $10\sps{6}$ are shown in Figs.~\ref{ra103contours}-\ref{ra106contours}. Clearly, as $\mbox{Ra}$ increases, the temperature boundary layers at the hot and cold walls becomes thinner with a concomitant increase in the natural convective flows in their vicinities; moreover, as the relative effects of the buoyancy forces over the counteracting viscous forces and thermal diffusion increase, the vortical structures, which are formed as a result of clockwise flow circulation due to thermal exchanges between the side walls, become increasingly more complex. These observations are found to be qualitatively similar to those shown in the literature.

\begin{figure}[H]
\begin{subfigure}{0.98\textwidth}
\centering
\includegraphics[trim = 0 0 0 0, clip, width =100mm]{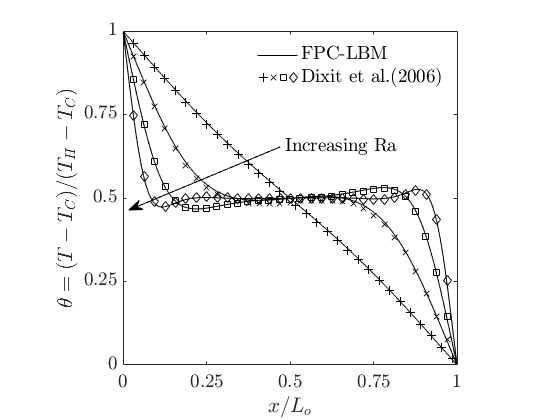}
\end{subfigure}
\caption{Comparison of computed dimensionless temperatures along the horizontal centerline for natural convection in a square cavity computed using the 2D FPC-LBM (lines) with the prior numerical results of Dixit and Babu (2006)~\cite{dixit2006simulation} for Rayleigh numbers $\mbox{Ra}=10\sps{3}$, $10\sps{4}$, $10\sps{5}$ and $10\sps{6}$ at a fixed Prandtl number $\mbox{Pr}=0.71$ and a resolution of $512\times512$ grid nodes.}
\label{Tcomp}
\end{figure}

However, to get a more quantitative comparison, next we look at the temperature profiles along the horizontal centerlines at $y/L\sbs{o}=0.5$ for the above low to moderate Rayleigh numbers. Figure~\ref{Tcomp} shows those profiles computed using the FPC-LBMs overlayed with the prior reference numerical data of Ref.~\cite{dixit2006simulation} for $\mbox{Ra}=10\sps{3}$ through $\mbox{Ra}=10\sps{6}$. Good agreement is seen for these cases.

Moreover, to get further validation quantitatively, some relevant parameters such as the Nusselt numbers (maximum, minimum and average) along the heated wall and maximum natural convection velocity components and their locations are computed using the 2D FPC-LBMs and then compared with reference data available in the literature. The definitions of those parameters are as follows: $u\sbs{max}$ is the maximum horizontal velocity on the vertical centerline of the cavity and $v\sbs{max}$ is the maximum vertical velocity on the horizontal centerline of the cavity with their locations given as $y$ and $x$, respectively; $\langle \mbox{Nu}\sbs{0} \rangle$, $(\mbox{Nu}\sbs{0})\sbs{max}$, and $(\mbox{Nu}\sbs{0})\sbs{min}$  are the average Nusselt number, maximum Nusselt number and minimum Nusselt number, respectively, along the hot wall  at $x = 0$, while the location of the last two quantities are identified to be at $y$. Here, the coordinate locations are normalized by the characteristic length $L\sbs{o}$.
%
The results for these parameters at $\mbox{Ra}=10\sps{3}$, $10\sps{4}$, $10\sps{5}$, and $10\sps{6}$ are tabulated in Tables~\ref{con1}-\ref{con4}, where the comparisons are made against the prior numerical solutions reported in Refs.~\cite{de1983natural,hortmann1990finite,dixit2006simulation,sharma2018natural}. Evidently, the FPC-LBM results are generally in good quantitative agreement with the available data in the literature.
\begin{table}[H]
\centering
\begin{tabular}{c c c c c c c c c c c}
\hline
\addlinespace
       $\mbox{Ra}$   &  Parameter   &  FPC-LBM && Ref. 1 && Ref. 2 && Ref. 3 && Ref. 4\\ [1mm]
\hline
\addlinespace
$10\sps{3}$  & $u\sbs{max}$                   &  $3.6418$    &&  $3.634$     &&  $-$   &&  $3.6529$     &&    $3.64945$     \\
 	   	& $y$                                    & $0.8131$    &&  $0.813$     &&  $-$   &&  $0.8125$     &&    $0.81324$     \\ [2mm]
	   	& $v\sbs{max}$                   & $3.6900$    &&  $3.697$     &&  $-$   &&  $3.6820$     &&    $ 3.69744$      \\
 	   	& $x$                                    &  $0.1787$    &&  $0.179$     &&  $-$   &&  $ 0.1718$     &&    $0.17832$     \\ [2mm]
		& $\langle \mbox{Nu}\sbs{0} \rangle$  &  $1.1153$    &&  $1.105$     &&  $-$   && $1.1272$ &&  $-$       \\ [2mm]

 	  	& $(\mbox{Nu}\sbs{0})\sbs{max}$    &  $1.5089$    &&  $1.501$     &&  $-$   &&  $-$     &&    $-$      \\
 	   	& $y$                                    & $0.0793$    &&  $0.087$     &&  $-$   &&  $-$     &&    $-$     \\ [2mm]
	   	& $(\mbox{Nu}\sbs{0})\sbs{min}$    & $0.6898$    &&  $0.694$     &&  $-$   &&  $-$     &&    $-$     \\
 	   	& $y$                                   & $0.9990$    &&  $1.000$     &&  $-$   &&  $-$     &&    $-$     \\ [2mm]
\hline
\end{tabular}
\caption{Comparison of computed Nusselt numbers (maximum, minimum and average) along the heated wall and maximum natural convection velocity components and their locations computed using the 2D FPC-LBM with the reference numerical data of Ref. 1 (de Vahl Davis (1983))~\cite{de1983natural}, Ref. 2 (Hortmann et al. (1990))~\cite{hortmann1990finite}, Ref. 3 (Dixit and Babu et al. (2006))~\cite{dixit2006simulation}, and Ref. 4 (Sharma et al. (2018))~\cite{sharma2018natural} for natural convection in a square cavity with $\mbox{Ra}=10\sps{3}$, $\mbox{Pr}=0.71$, and a resolution of $512\times512$ of grid nodes.}
\label{con1}
\end{table}

\begin{table}[H]
\centering
\begin{tabular}{c c c c c c c c c c c}
\hline
\addlinespace
       $\mbox{Ra}$   &  Parameter   &  FPC-LBM && Ref. 1 && Ref. 2 && Ref. 3 && Ref. 4\\ [1mm]
\hline
\addlinespace
$10\sps{4}$ & $u\sbs{max}$                  & $16.1537$ &&   $16.182$     &&  $16.1759$    &&  $16.163$    &&  $16.18340$     \\
 	   	& $y$                                  & $0.8229$   &&   $0.823$     &&  $0.82551$    &&   $ 0.828$   &&    $ 0.82323$    \\ [2mm]
	   	& $v\sbs{max}$                 & $19.5845$  &&   $19.509$     &&   $19.6242 $   &&   $19.569$   &&  $19.62830$      \\
 	   	& $x$                                  &  $0.1182$  &&   $0.120$     &&   $0.12009 $   &&   $ 0.125$   &&  $ 0.11886$   \\ [2mm]
		& $\langle \mbox{Nu}\sbs{0} \rangle$  & $2.2431$      &&  $2.242$    &&  $-$    &&  $2.247$   &&  $-$     \\ [2mm]

 	  	& $(\mbox{Nu}\sbs{0})\sbs{max}$  & $3.5343$   &&  $3.545$      && $-$    &&   $-$   &&  $-$      \\
 	   	& $y$                                  & $0.1360$   &&  0.149$$      && $-$    &&   $-$   &&  $-$    \\ [2mm]
	   	& $(\mbox{Nu}\sbs{0})\sbs{min}$   & $0.5835$   &&  $0.592$      &&  $-$    &&  $-$    &&  $-$    \\
 	   	& $y$                                  &  $0.9990$  &&  $1.000$      && $-$    &&   $-$   &&  $-$      \\ [2mm]
\hline
\end{tabular}
\caption{Comparison of computed Nusselt numbers (maximum, minimum and average) along the heated wall and maximum natural convection velocity components and their locations computed using the 2D FPC-LBM with the reference numerical data of Ref. 1 (de Vahl Davis (1983))~\cite{de1983natural}, Ref. 2 (Hortmann et al. (1990))~\cite{hortmann1990finite}, Ref. 3 (Dixit and Babu et al. (2006))~\cite{dixit2006simulation}, and Ref. 4 (Sharma et al. (2018))~\cite{sharma2018natural} for natural convection in a square cavity with $\mbox{Ra}=10\sps{4}$, $\mbox{Pr}=0.71$, and a resolution of $512\times512$ of grid nodes. }
\label{con2}
\end{table}

\begin{table}[H]
\centering
\begin{tabular}{c c c c c c c c c c c}
\hline
\addlinespace
       $\mbox{Ra}$   &  Parameter   &  FPC-LBM && Ref. 1 && Ref. 2 && Ref. 3 && Ref. 4\\ [1mm]
\hline
\addlinespace
$10\sps{5}$ & $u\sbs{max}$                                     &  $34.7823$    &&  $34.18$     &&  $34.7385 $   &&  $35.521$     &&    $34.74050$     \\
 	   	& $y$                                                     & $0.8546$    &&  $0.855$     &&  $0.85535$   &&  $0.8554$     &&    $0.85461 $     \\ [2mm]
	   	& $v\sbs{max}$                                    & $68.6323$    &&  $68.22$     &&  $68.6359 $   &&  $68.655$     &&    $68.63500$      \\
 	   	& $x$                                                     &  $0.0654$    &&  $0.066$     &&  $0.06602 $   &&  $0.0664$     &&    $ 0.06585$     \\ [2mm]
                     & $\langle \mbox{Nu}\sbs{0} \rangle$  &   $4.5188$    &&  $4.523$     &&  $4.52188 $   &&  $ 4.5226$     &&    $-$               \\ [2mm]
 	  	& $(\mbox{Nu}\sbs{0})\sbs{max}$      &  $7.7272$    &&  $7.761$     &&  $7.7214$   &&  $-$     &&    $-$      \\
 	   	& $y$                                                     & $0.0788$    &&  $0.085$     &&  $-$   &&  $-$     &&    $-$     \\ [2mm]
	   	& $(\mbox{Nu}\sbs{0})\sbs{min}$       & $0.7220$    &&  $0.736$     &&  $-$   &&  $-$     &&    $-$     \\
 	   	& $y$                                                    & $0.9954$    &&  $1.000$     &&  $-$   &&  $-$     &&    $-$     \\ [2mm]
\hline
\end{tabular}
\caption{Comparison of computed Nusselt numbers (maximum, minimum and average) along the heated wall and maximum natural convection velocity components and their locations computed using the 2D FPC-LBM with the reference numerical data of Ref. 1 (de Vahl Davis (1983))~\cite{de1983natural}, Ref. 2 (Hortmann et al. (1990))~\cite{hortmann1990finite}, Ref. 3 (Dixit and Babu et al. (2006))~\cite{dixit2006simulation}, and Ref. 4 (Sharma et al. (2018))~\cite{sharma2018natural} for natural convection in a square cavity with $\mbox{Ra}=10\sps{5}$, $\mbox{Pr}=0.71$, and a resolution of $512\times512$ of grid nodes. }
\label{con3}
\end{table}

\begin{table}[H]
\centering
\begin{tabular}{c c c c c c c c c c c}
\hline
\addlinespace
       $\mbox{Ra}$   &  Parameter   &  FPC-LBM && Ref. 1 && Ref. 2 && Ref. 3 && Ref. 4\\ [1mm]
\hline
\addlinespace
$10\sps{6}$ & $u\sbs{max}$                                     &  $64.8670$    &&  $65.33$     &&  $64.8340$   &&  $ 64.630$     &&    $64.83190$     \\
 	   	& $y$                                                     & $0.8506$    &&  $0.851$     &&  $0.85036$   &&  $ 0.8496$     &&    $ 0.84990$     \\ [2mm]
	   	& $v\sbs{max}$                                    & $220.5401$    &&  $216.75$     &&  $220.473$   &&  $ 219.866$     &&    $ 220.56600$      \\
 	   	& $x$                                                     &  $0.0381$    &&  $0.0387$     &&  $0.03887$   &&  $ 0.0371$     &&    $0.03775$     \\ [2mm]
                     & $\langle \mbox{Nu}\sbs{0} \rangle$  &   $8.8116$    &&  8.928$$     &&  $-$   &&  $8.805$     &&    $-$               \\ [2mm]
 	  	& $(\mbox{Nu}\sbs{0})\sbs{max}$      &  $17.5230$    &&  $18.076$     &&  $-$   &&  $-$     &&    $-$      \\
 	   	& $y$                                                     & $0.0400$    &&  $0.0456$     &&  $-$   &&  $-$     &&    $-$     \\ [2mm]
	   	& $(\mbox{Nu}\sbs{0})\sbs{min}$       & $0.9755$    &&  $1.005$     &&  $-$   &&  $-$     &&    $-$     \\
 	   	& $y$                                                    & $0.9990$    &&  $1.000$     &&  $-$   &&  $-$     &&    $-$     \\ [2mm]
\hline
\end{tabular}
\caption{Comparison of computed Nusselt numbers (maximum, minimum and average) along the heated wall and maximum natural convection velocity components and their locations computed using the 2D FPC-LBM with the reference numerical data of Ref. 1 (de Vahl Davis (1983))~\cite{de1983natural}, Ref. 2 (Hortmann et al. (1990))~\cite{hortmann1990finite}, Ref. 3 (Dixit and Babu et al. (2006))~\cite{dixit2006simulation}, and Ref. 4 (Sharma et al. (2018))~\cite{sharma2018natural} for natural convection in a square cavity with $\mbox{Ra}=10\sps{6}$, $\mbox{Pr}=0.71$, and a resolution of $512\times512$ of grid nodes. }
\label{con4}
\end{table}

Moving beyond the moderate Rayleigh numbers, from $\mbox{Ra}$ at $10^6$ to higher values at $10^7$ and $10^8$, it is expected that to resolve the progressively thinner boundary layers, resolutions greater than $512^2$ grid points are necessary to resolve the heat transfer rates adequately across the heated wall. Hence, we performed a systematic grid convergence study in the prediction of the Nusselt numbers at  $\mbox{Ra}=10^6$, $10^7$ and $10^8$ for a wide range of following grid resolutions: $512^2$, $768^2$, $1024^2$, $1280^2$, $1536^2$, $1792^2$, and $2048^2$. All the simulations were run for $20\times 10^6$ time steps. The results are shown in Table~\ref{Convergence_Study_1} for the average, maximum and minimum Nusselt numbers. Looking at this table, we note that up to $\mbox{Ra}=10\sps{7}$ the parameters listed tend to converge generally to the first or second decimal place as the grid resolution is increased. For $\mbox{Ra}=10\sps{7}$, a resolution of $1536^2$ seems to provide sufficiently converged heat transfer rate parameters, while the same grid for $\mbox{Ra}=10\sps{8}$ are not sufficient and higher resolutions are required. Moreover, it should be noted that the latter case is near the threshold or critical Rayleigh number to transition to turbulent natural convection, as, for example, Ref.~\cite{le1998onset} reports the onset to turbulence to be associated with $\mbox{Ra}\approx 1.84\times10\sps{8}$. Hence, it is not guaranteed that steady state results exists at such a high $\mbox{Ra}$.
\begin{table}[H]
\centering
\begin{tabular}{c c c c c c c c c}
\hline
\addlinespace
$Ra$ & $N\sps{2} $ & $\langle \mbox{Nu} \rangle$ & $\langle \mbox{Nu}\sbs{0} \rangle$ & $\langle \mbox{Nu}\sbs{1/2} \rangle$ & $(\mbox{Nu}\sbs{0})\sbs{max}$ & $y$ & $(\mbox{Nu}\sbs{0})\sbs{min}$ & $y$ \\[0.05mm]
\hline
\addlinespace
	   & $512^2$   & $8.8031$ & $8.8102$ & $8.8222$ & $17.5217$ & $0.0400$ & $0.9742$ & $0.9990$  \\ 
 	   & $768^2$   & $8.8110$ & $8.8142$ & $8.8236$ & $17.5368$ & $0.0397$ & $0.9731$ & $0.9993$  \\
 	   & $1024^2$ & $8.8147$ & $8.8156$ & $8.8243$ & $17.5398$ & $0.0396$ & $0.9714$ & $0.9995$  \\
$10\sps{6}$ & $1280^2$ & $8.8159$ & $8.8153$ & $8.8234$ & $17.5351$ & $0.0395$ & $0.9711$ & $0.9996$  \\
 	   & $1536^2$ & $8.8176$ & $8.8122$ & $8.8260$ & $17.5186$ & $0.0394$ & $0.9723$ & $0.9997$  \\
 	   & $1792^2$ & $8.8172$ & $8.8212$ & $8.8224$ & $17.5443$ & $0.0394$ & $0.9737$ & $0.9980$  \\
 	   & $2048^2$ & $8.7914$  & $8.7960$ & $8.7865$ & $17.4449$ & $0.0374$ & $0.9769$ & $0.9978$ \\  
\hline
\addlinespace
	   & $512^2$ & $16.4756$  &$16.4907$ &$16.5109$ &$38.987$6 &$0.0186$ &$1.3838$ & $0.9971$  \\ 
 	   & $768^2$ & $16.4943$ & $16.5022$ & $16.5175$&$39.2576$ &$0.0176$ &$1.3655$ &$ 0.9993$ \\
 	   & $1024^2$ & $16.5025$ &$16.5076$ &$16.5197$ &$39.3428$&$0.0181$ &$1.3616$ &$0.9995$  \\
$10\sps{7}$ & $1280^2$ & $16.5071$  & $ 16.5105$ & $16.5208 $ & $39.3771 $ & $ 0.0176$ & $1.3581 $ & $ 0.9996$  \\
 	   & $1536^2$ & $16.5099$  & $16.5124$ & $16.5213$ & $39.3928$ & $0.0179$ & $1.3604$ & $0.9997$  \\
 	   & $1792^2$ &  $16.5115$  &  $16.5133$  &  $16.5212$  &  $39.3935$  &  $0.0176$  &  $1.3565$  &$0.9997$  \\
 	   & $2048^2$ & $16.5121$  & $16.5123$  & $16.5210$  & $39.3919$  & $0.0178$  & $1.3585$  & $0.9993$   \\
\hline
\addlinespace
	   & $512^2$ &$30.0925$  & $30.1365$ &$30.1631$ &$90.8228$ &$9.7656\times 10^{\sm4}$ & $1.9434$  &$0.9971$ \\
 	   & $768^2$ &$30.1548$  &$30.1733$  & $30.1999$&  $84.9057$&$6.5104\times 10^{\sm4}$ &$1.9448$ &$0.9980$ \\
 	   & $1024^2$ & $30.1772$  & $30.1872$ &  $30.2102$ & $86.0334$ & $0.0083$ & $1.9354$ & $0.9985$  \\
$10\sps{8}$ & $1280^2$ & $30.1851$  & $30.1903$ & $30.2113$ & $86.5318$ & $0.0082$ & $1.9298$ & $0.9988$  \\
 	   & $1536^2$ & $30.1833$  & $30.1831$ & $30.2061$ & $86.7108$ & $0.0081$ & $1.9236$ & $0.9990$  \\
 	   & $1792^2$ & $30.1876$  & $30.2005$ & $30.2073$ & $86.9228$ & $0.0081$ & $1.9222$ & $0.9997$  \\
 	   & $2048^2$ & $29.9287$  & $30.1593$ & $29.8197$ & $86.6836$ & $0.0085$ & $1.9299$ & $0.9998$  \\
\hline
\end{tabular}
\caption{Grid refinement study for simulation of natural convection in a square cavity at moderate Rayleigh numbers computed using the 2D FPC-LBM with resolutions of $512^2$, $768^2$, $1024^2$, $1536^2$, $1792^2$ and $2048^2$ grid nodes showing the results of the Nusselt numbers (maximum, minimum, and average).}
\label{Convergence_Study_1}
\end{table}
For convenience, we summarize the results for most relevant average Nusselt numbers at the hot wall computed using the 2D FPC-LBM at $\mbox{Ra}=10^7$, $10^8$ and $10^9$ and compared with other numerical studies~\cite{le1998onset,dixit2006simulation,sharma2018natural} in Table~\ref{CompareNu2D}. We used a grid resolution of $2048^2$, which is generally higher than those used in such other investigations. At any rate, our results are largely consistent with those reported by such earlier studies.
\begin{table}[H]
\centering
\begin{tabular}{c c c c }
\hline
\addlinespace
 & $\mbox{Ra}=10^7$  & $\mbox{Ra}=10^8$ & $\mbox{Ra}=10^9$  \\[ 0.05mm]
\hline
\addlinespace
FPC-LBM   & 16.5123 & 30.1593 & 53.9812   \\[ 0.05in]
Sharma et al. (2018)~\cite{sharma2018natural} & 16.52328  & 30.2246  &54.7531  \\[ 0.05in]
Dixit and Babu (2006)~\cite{dixit2006simulation}  & 16.7900 & 30.5060 & 57.3500  \\[ 0.05in]
Le Qu{\'e}r{\'e} and Behnia (1998)~\cite{le1998onset}  & 16.5230 & 30.2250 & -  \\[ 0.05in]
\hline
\end{tabular}
\caption{Comparison of the average Nusselt numbers computed using the 2D FPC-LBM with a resolution of $2048\times2048$ grid nodes at $\mbox{Ra}=10^7, 10^8$, and $10^9$ against reference numerical data of Le Qu{\'e}r{\'e} and Behnia (1998)~\cite{le1998onset}, Dixit and Babu (2006)~\cite{dixit2006simulation} and Sharma et al. (2018)~\cite{sharma2018natural}.}
\label{CompareNu2D}
\end{table}

\begin{figure}[H]
\begin{subfigure}{0.49\textwidth}
\centering
\includegraphics[trim = 0 0 0 0, clip, width =75mm]{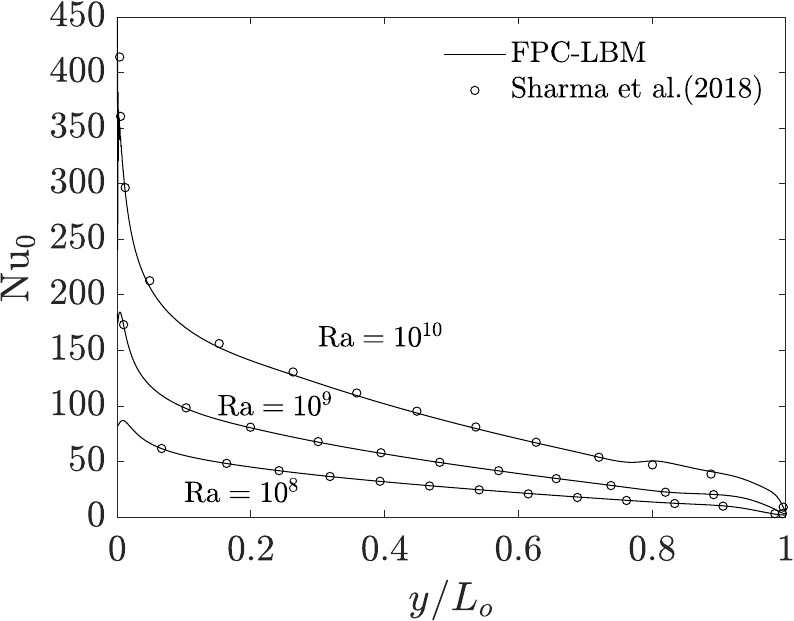}
\caption{ }
\end{subfigure}
\begin{subfigure}{0.48\textwidth}
\centering
\includegraphics[trim = 0 0 0 0, clip, width =75mm]{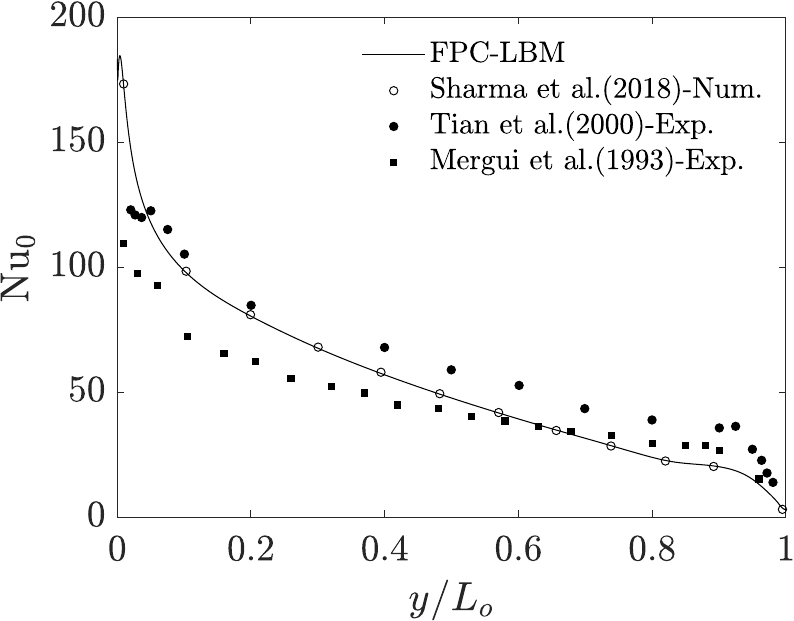}
\caption{ }
\end{subfigure}
\caption{(a) Variation of the Nusselt number on the hot wall (left) for in the natural convection in a square cavity at high Rayleigh numbers of $\mbox{Ra}=10\sps{8}$, $10\sps{9}$ and $10\sps{10}$. A grid resolution of $2048\times 2048$ was used and time averaged statistics for the temperature field computed using the 2D FPC-LBM (lines) were collected from $10\times10\sps{6}$ to $100\times10\sps{6}$ time steps before estimating the Nusselt numbers. Comparison is made with the prior numerical results of Sharma et al. (2018)~\cite{sharma2018natural}. (b) Moreover, the FPC-LBM results for the case of $\mbox{Ra}=10\sps{9}$ is compared with the available experimental data from Mergui et al. (1993)~\cite{mergui1993experimental} Tian et al. (2000)~\cite{tian2000low} in addition to the prior numerical solution of Sharma et al. (2018)~\cite{sharma2018natural}.}
\label{highracomp}
\end{figure}

\noindent
Pushing the Rayleigh numbers into the transitional and fully turbulent regime, we performed additional computations at $\mbox{Ra} = 10^9$ and $10^{10}$ using a grid resolution of $2048^2$. For these cases, numerical results were recently reported in Ref.~\cite{sharma2018natural}, which we used for comparison. Figure~\ref{highracomp}a shows the variation of the average Nusselt number along the hot wall from top to bottom for $\mbox{Ra}=10\sps{8}$, $10\sps{9}$ and $10\sps{10}$ computed using the 2D FPC-LBMs and the results were time averaged over a duration of 90 million time steps. Shown in this figure are the prior numerical results of Ref.~\cite{sharma2018natural} and our results are found to be in good quantitative agreement with it. Moreover, in Fig.~\ref{highracomp}b our results at $\mbox{Ra}=10^9$ are also compared with the available experimental data of Refs.~\cite{tian2000low} and \cite{mergui1993experimental}, which show satisfactory agreement given the measurement uncertainty inherent in the latter. In summary, these results indicate the FPC-LBM is a viable and accurate approach for thermal convective flows at high Rayleigh numbers.
%


\subsection{Natural Convection in a Cubic Cavity}
\begin{figure}[H]
\centering
\includegraphics[trim = 0 0 0 0, clip, width =70mm]{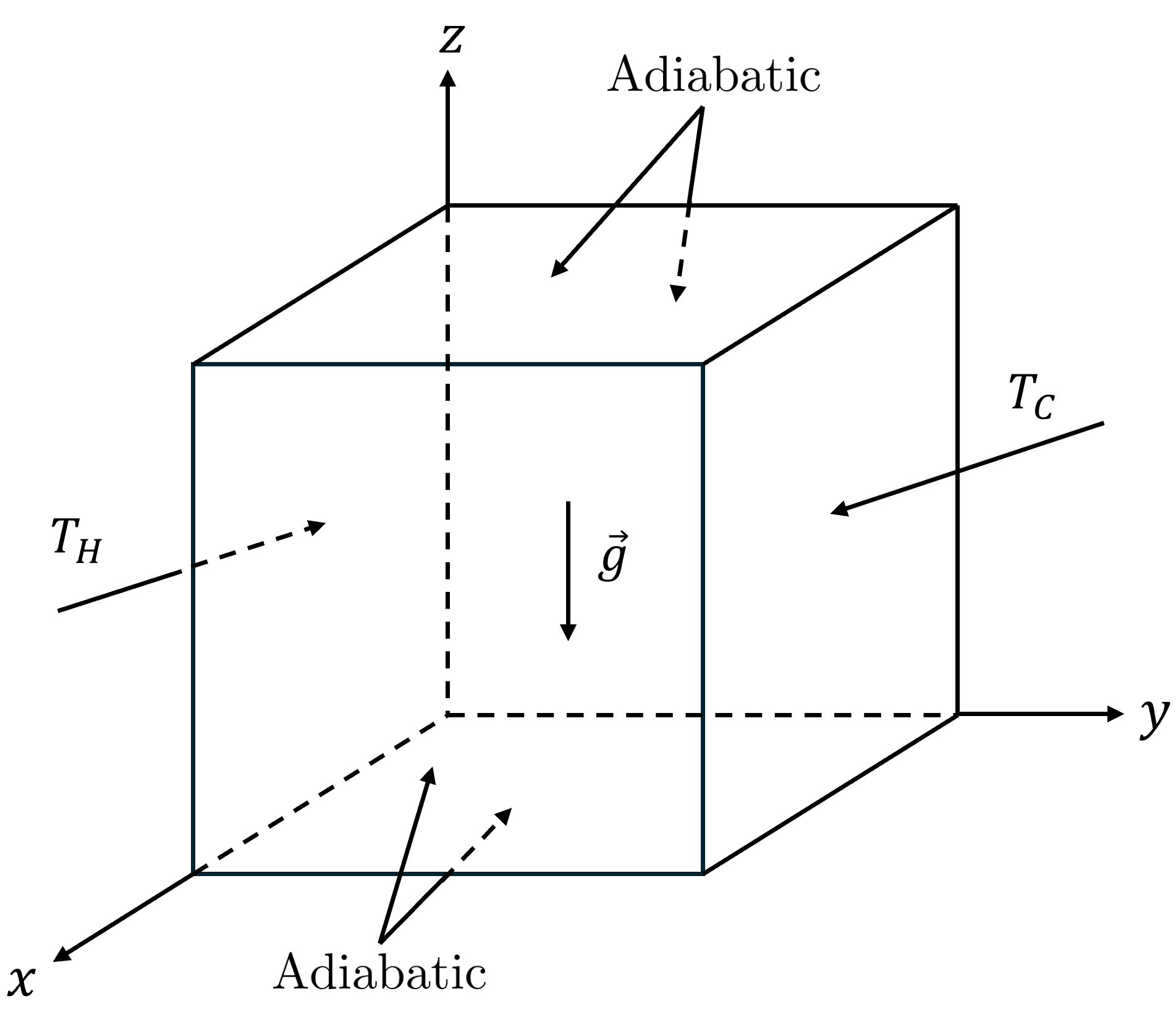}
\caption{Schematic diagram for the buoyancy-driven thermal convection in a cubic cavity with the attendant thermal boundary conditions. No slip velocity conditions are imposed on all wall surfaces.}
\label{natcon3d}
\end{figure}

Next, we discuss the simulation results of the natural convection inside a cubic cavity of volume $L\sbs{o}\sps{3}$ using the FPC-LBM on the D3Q27 lattice for hydrodynamics presented in our recent work~\cite{schupbach2024fokker} and another FPC-LBM on the D3Q15 lattice for the energy equation discussed in Appendix~\ref{sec:appendixC}. The no-slip boundary condition is applied on the velocity field on all wall surfaces; moreover, a constant and uniform hot temperature, $T\sbs{H}=2$, is applied on the left side wall and a constant and uniform cold temperature, $T\sbs{C}=1$, is applied to the right wall. All other walls are considered to be insulated and have zero temperature gradient in the normal direction as indicated in the schematic diagram in Fig.~\ref{natcon3d}. The Boussinesq approximation is used to apply a buoyancy force $\bm{F}=(F_x, F_y, F_z)$ due to local changes in density that would otherwise occur due to local changes in temperature, resulting in natural convection; the local components of this body force are $F_x = 0$, $F_y = 0$, and $F_z = g\beta(T-T\sbs{o})$. In this work, we perform 3D simulations of natural convection in a cubic cavity at moderately high Rayleigh numbers.

\begin{figure}[H]
\centering
\begin{subfigure}{0.48\textwidth}
\includegraphics[trim = 0 0 0 0, clip, width =80mm]{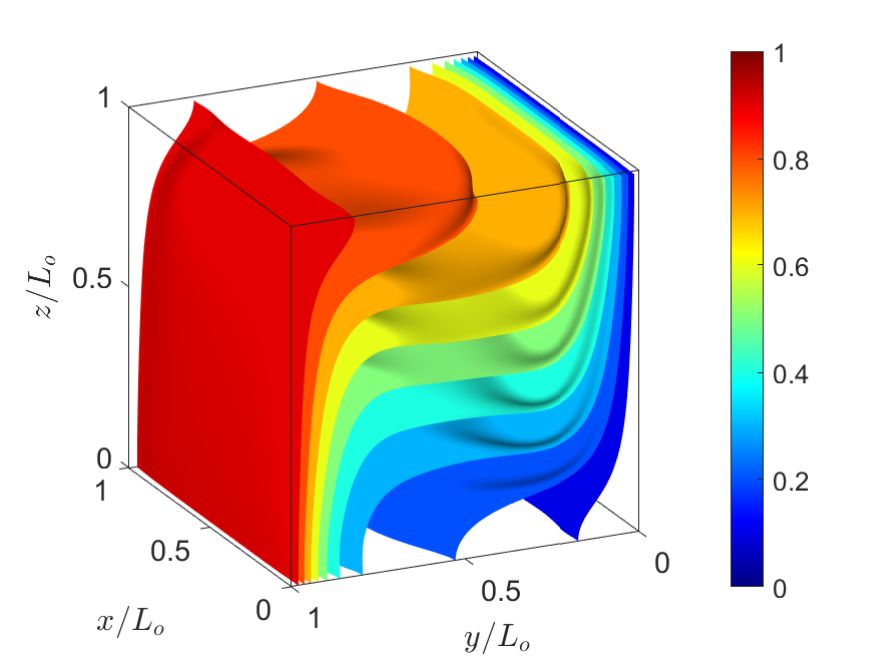}
\caption{$Ra=10^5$}
\end{subfigure}
\begin{subfigure}{0.48\textwidth}
\includegraphics[trim = 0 0 0 0, clip, width =80mm]{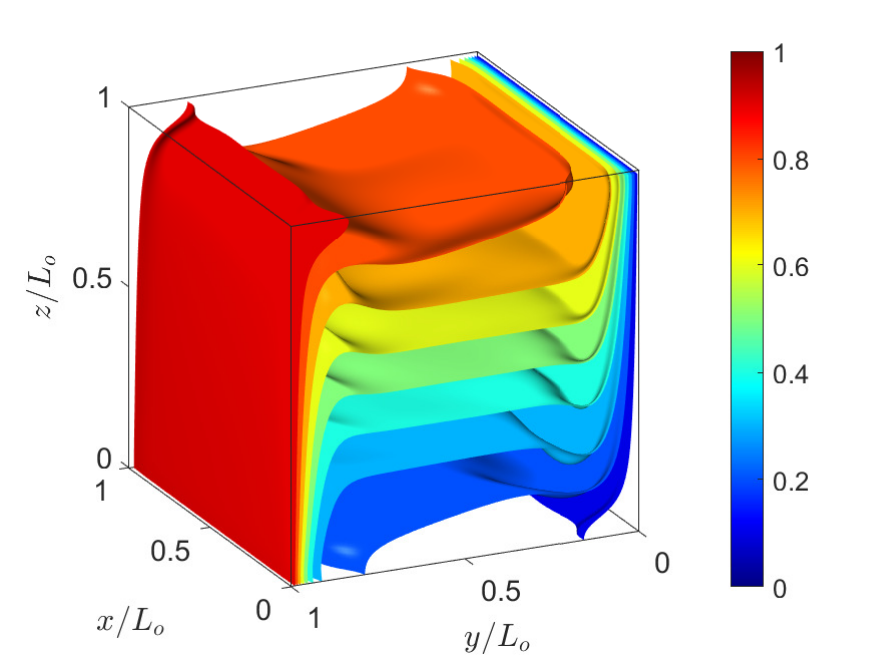}
\caption{$Ra=10^6$}
\end{subfigure}

\begin{subfigure}{0.48\textwidth}
\includegraphics[trim = 0 0 0 0, clip, width =80mm]{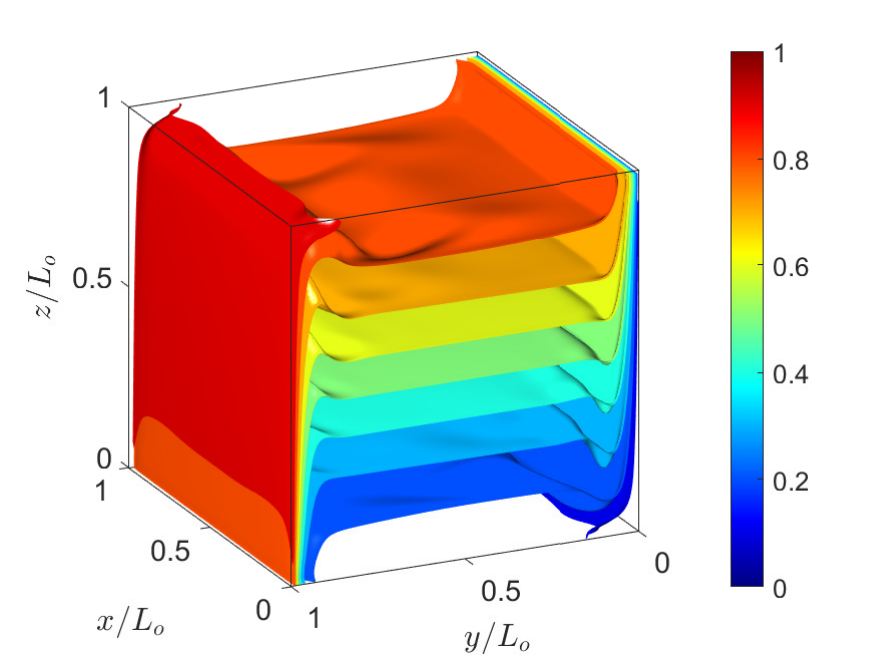}
\caption{$Ra=10^7$}
\end{subfigure}
\begin{subfigure}{0.48\textwidth}
\includegraphics[trim = 0 0 0 0, clip, width =80mm]{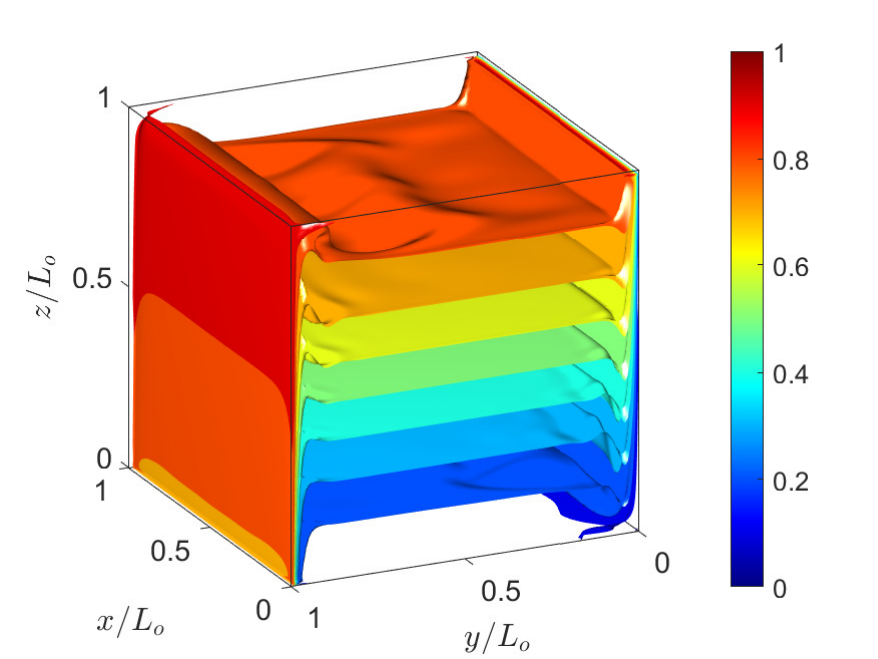}
\caption{$Ra=10^8$}
\end{subfigure}
\caption{Isosurfaces of dimensionless temperatures between hot and cold reference values in natural convection in a cubic cavity computed using the 3D FPC-LBM for Rayleigh numbers from $\mbox{Ra}=10\sps{5}$, $10\sps{6}$, $10\sps{7}$, and $10\sps{8}$ with $\mbox{Pr}=0.71$. All cases used a resolution of $128\times 128 \times 128$ grid nodes.}
\label{natcon3diso}
\end{figure}

\begin{figure}[H]
\centering
\begin{subfigure}{0.48\textwidth}
\includegraphics[trim = 0 0 0 0, clip, width =70mm]{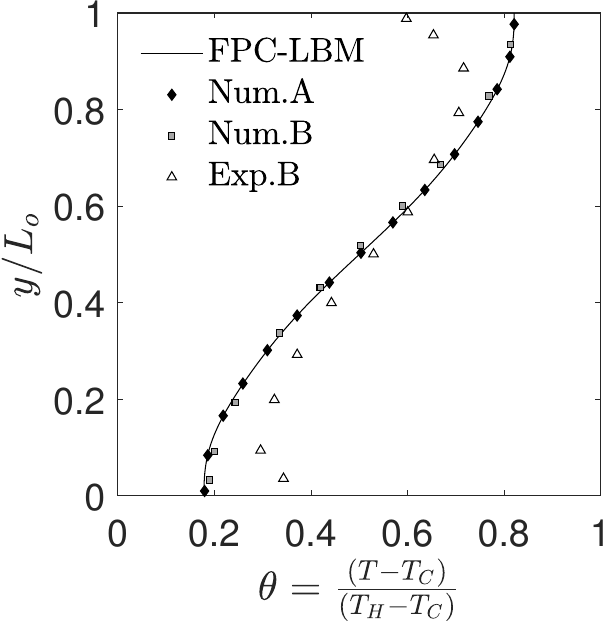}
\end{subfigure}
\begin{subfigure}{0.48\textwidth}
\includegraphics[trim = 0 -14 0 0, clip, width =70mm]{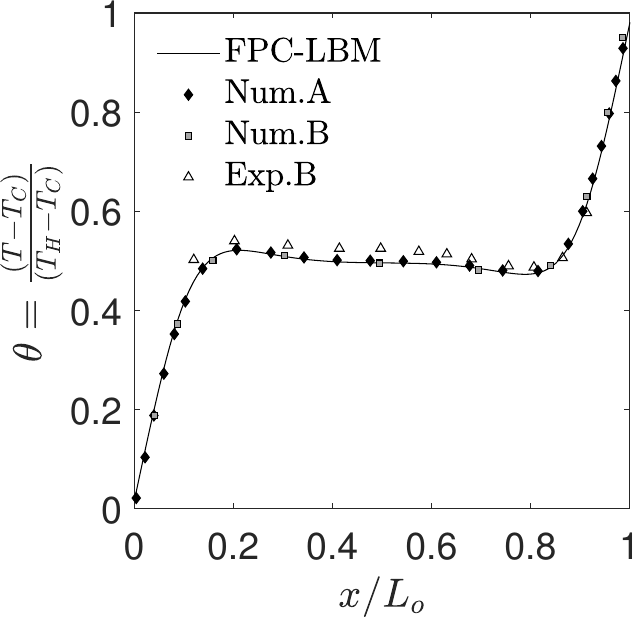}
\end{subfigure}

\begin{subfigure}{0.48\textwidth}
\includegraphics[trim = 0 0 0 0, clip, width =70mm]{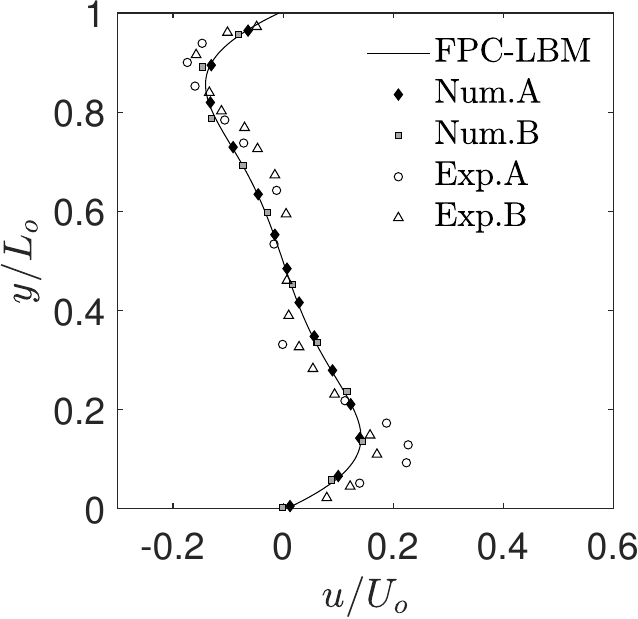}
\end{subfigure}
\begin{subfigure}{0.48\textwidth}
\includegraphics[trim = 0 0 0 0, clip, width =70mm]{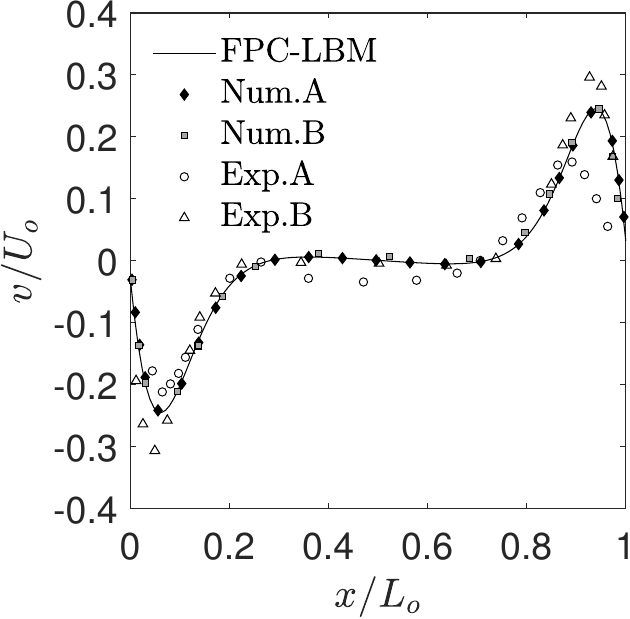}
\end{subfigure}
\caption{Dimensionless temperature and velocity profiles along the horizontal and vertical centerlines in the primary mid-planes in natural convection in a cubic cavity at $\mbox{Ra}=10\sps{5}$ and $\mbox{Pr}=0.71$ computed using the 3D FPC-LBM (lines). Comparisons are made with the prior numerical data of Num.A (Wang et al. (2017))~\cite{wang2017numerical}, Num.B (Fusegi et al. (1991))~\cite{fusegi1991numerical}, and available experimental data of Exp.A  (Bilski et al. (1986) at $\mbox{Ra}=1.89\times10\sps{5}$)~\cite{bilski1986experimental}, and Exp.B (Crane (1983) at $\mbox{Ra}=1.03\times10\sps{5}$)~\cite{krane1983some}.}
\label{natcon3dcomp}
\end{figure}

\begin{figure}[H]
\centering
\begin{subfigure}{0.48\textwidth}
\includegraphics[trim = 0 0 0 0, clip, width =70mm]{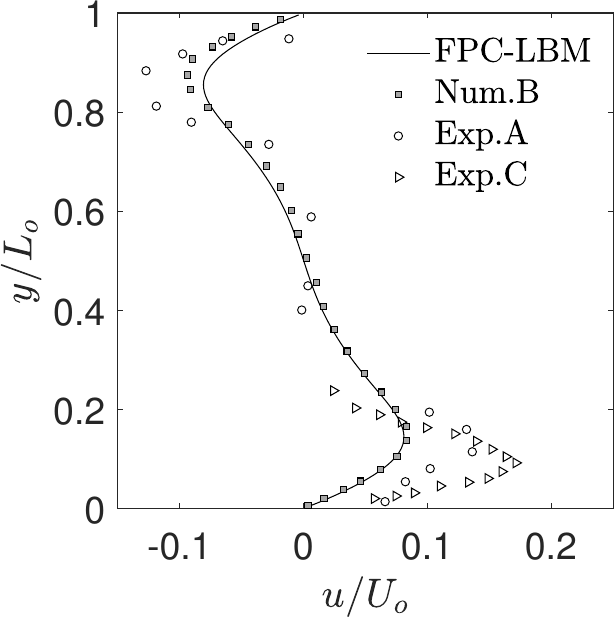}
\end{subfigure}
\begin{subfigure}{0.48\textwidth}
\includegraphics[trim = 0 0 0 0, clip, width =70mm]{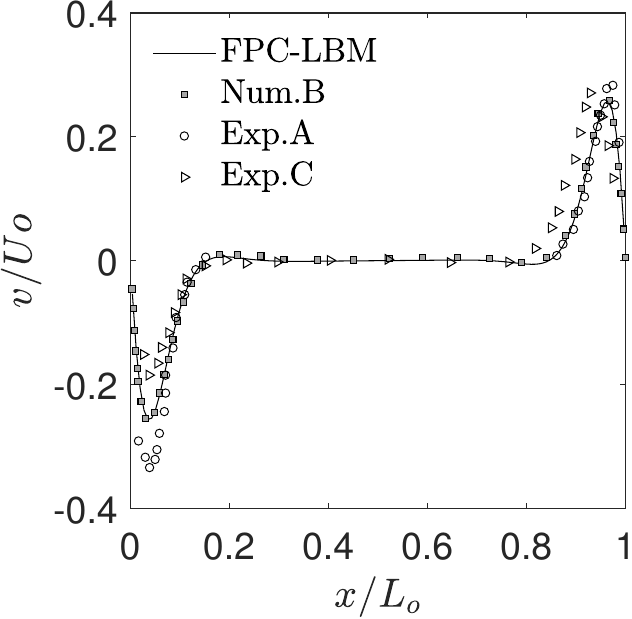}
\end{subfigure}
\caption{Dimensionless velocity profiles along the horizontal and vertical centerlines in the primary mid-plane in natural convection in a cubic cavity at $\mbox{Ra}=10\sps{6}$ and $\mbox{Pr}=0.71$ computed using the 3D FPC-LBM (lines) . Comparisons are made with the prior numerical data of Num.B (Fusegi et al. (1991))~\cite{fusegi1991numerical}, and available experimental data of Exp.A  (Bilski et al. (1986))~\cite{bilski1986experimental}, and Exp.C (Briggs and Jones (1985))~\cite{briggs1985two}. }
\label{natcon3dcomp2}
\end{figure}

Figure~\ref{natcon3diso} shows the simulations results for $\mbox{Ra} = 10^5$, $10^6$, $10^7$, and $10^8$ indicating the isosurfaces of constant dimensionless temperature profiles between the values of $\theta = 0$ and $\theta = 1$ in increments of $\delta \theta = 0.1$, where $\theta = (T-T\sbs{C})/(T\sbs{H}-T\sbs{C})$. They indicate that the formation of progressively thinner boundary layers near the walls with the imposed temperatures as the Rayleigh number is increased, and due to the presence of buoyancy forces they result in the vortex generation within the cavity. The natural convective flows thus formed cause increase in the heat transfer rates as $\mbox{Ra}$ is increased, which will be measured and evaluated in terms of the Nusselt number in what follows.
Before discussing further simulation results, we now provide an indication of the computational run times involved. The simulations were performed on the INCLINE parallel computer cluster hosted at the University of Colorado at Colorado Springs (UCCS). As an example, the turnaround time for performing $5\times 10^6$ time steps of the simulation for the $\mbox{Ra} = 10^8$ case using $128^3$ grid nodes on 512 cores was about $30.7$ hours.
Then, to get a quantitative comparison of the simulation results using the FPC-LBM, we look at the dimensionless temperature and natural convection velocity profiles along the vertical and horizontal center-lines in the primary mid-plane, and numerical~\cite{wang2017numerical,fusegi1991numerical} and experimental~\cite{bilski1986experimental,krane1983some,briggs1985two} data available for the cases of $\mbox{Ra}=10\sps{5}$ and  $\mbox{Ra}=10\sps{6}$ will be utilized as reference solutions in this regard. While such numerical solutions use the same velocity and thermal boundary conditions, it should be noted that the experimental data are subject to variations in the boundary conditions as the perfectly insulated condition is not always completely met; in particular, Krane (1983)~\cite{krane1983some}, Bilski et al. (1986) and Briggs and Jones (1985)~\cite{briggs1985two} consider thermally conducting top and bottom surfaces of the cavity in lieu of insulated boundaries. For example, this is illustrated by the experimental set up shown in Fig.~2 in Briggs and Jones (1985)~\cite{briggs1985two}. Figures~\ref{natcon3dcomp} and \ref{natcon3dcomp2} present the results and comparisons at $\mbox{Ra}=10\sps{5}$ and  $\mbox{Ra}=10\sps{6}$, respectively. While the FPC-LBM results are consistent with the experimental data across the domain associated with the constant wall temperature boundaries, some differences are noticed across the direction associated with the insulated boundaries, where, as noted above, the experimental set up does not have perfectly insulated surfaces, but conducting surfaces. Nevertheless, the FPC-LBM results are in very good agreement with the available numerical data for these cases.
%
Moreover, by quantifying the dimensionless heat transfer rates, we tabulate the average Nusselt number on the hot wall computed using the 3D FPC-LBMs and compare with prior numerical solutions available in the literature for $\mbox{Ra}=10^5$, $10^6$, $10^7$, and $10^8$ in Table~\ref{CompareNu3D}. Our results are seen to be in very good agreement with these different sources of reference data.
%
\begin{table}[H]
\centering
\begin{tabular}{c c c c c c c c c c c c c}
\hline
\addlinespace
$Ra$ & & $10^5$  & & $10^6$ & & $10^7$ & & $10^{8}$ \\
\hline
\addlinespace
FPC- LBM       & & 4.3500 & & 8.6473  & & 16.2419  & &  29.6172    \\[ 0.05in]
Fugesi et al. (1991)~\cite{fusegi1991numerical}  & & 4.3610 & & 8.7700  & &  -             & & - \\[ 0.05in]
Tric et al. (2000)~\cite{tric2000first}      & &  4.3370 & & 8.6407  & &16.3427   & & - \\[ 0.05in]
Wang et al. (2017)~\cite{wang2017numerical}  & & -           & & -            & & 16.4153  & & 29.9169     \\[ 0.05in]
Xu et al. (2019)~\cite{xu2019lattice}        & & -           & & 8.6435  & & 16.4032  & &   - \\[ 0.05in]
\hline
\end{tabular}
\caption{Comparison of the average Nusselt numbers computed using the 3D FPC-LBM against reference numerical data of Fugesi et al. (1991)~\cite{fusegi1991numerical}, Tric et al. (2000)~\cite{tric2000first}, Wang et al. (2017)~\cite{wang2017numerical}, and Xu et al. for various Rayleigh numbers ($\mbox{Ra}=10^5$, $10^6$, $10^7$, and $10^{8}$) with $\mbox{Pr}=0.71$. All cases were simulated with a grid resolution of $128\times 128\times 128$ except for the final case of $\mbox{Ra}=10\sps{8}$ which had a resolution of $160\times 160\times 160$.}
\label{CompareNu3D}
\end{table}


\section{Numerical Stability Tests: Comparison of 3D FPC-LBM with 3D Central Moment LBM using Maxwell Distribution\label{sec:stabilitytest}}

\begin{figure}[H]
\centering
\includegraphics[trim = 0 0 0 0, clip, width =70mm]{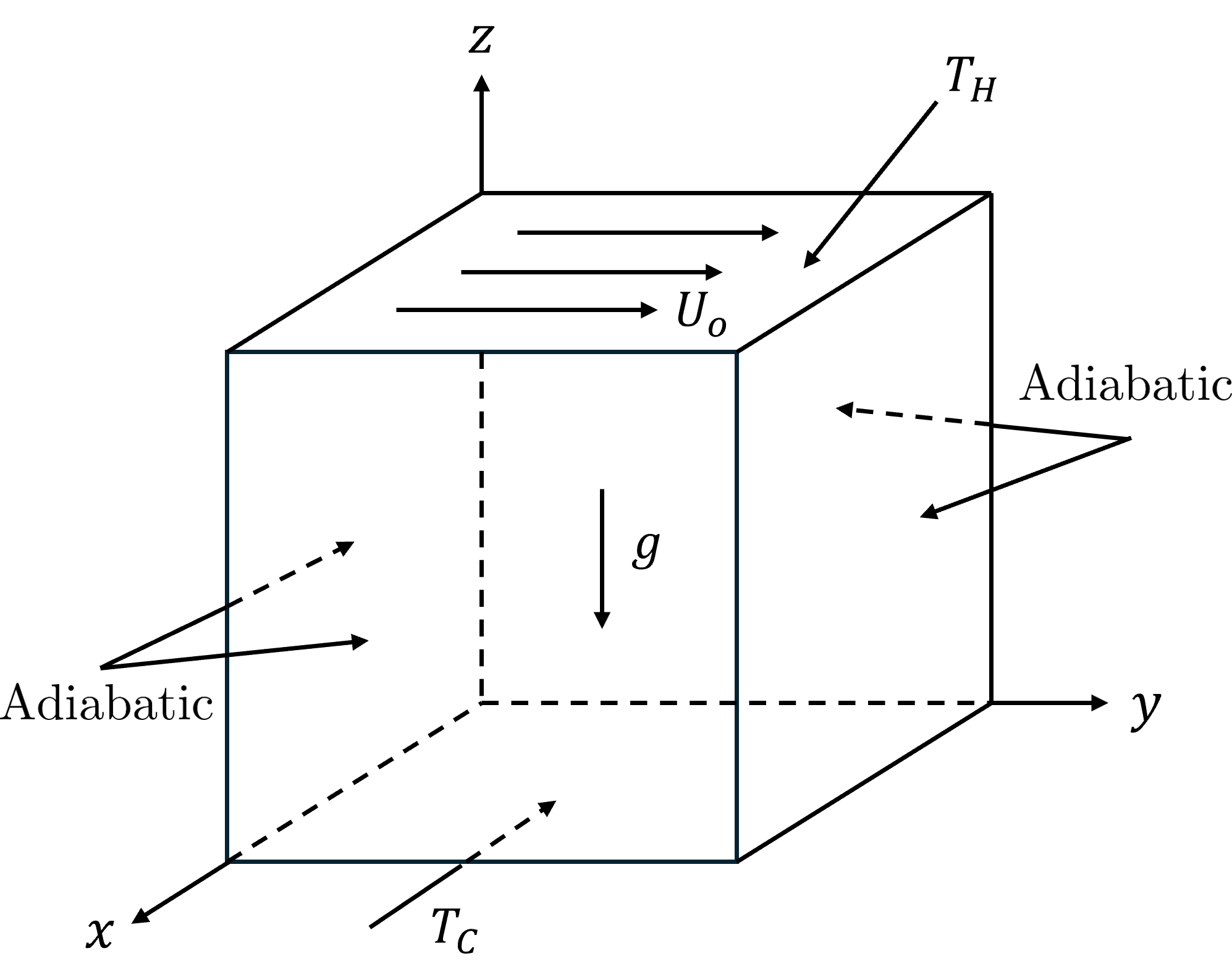}
\caption{Schematic diagram for the numerical stability test involving mixed thermal convection due to shear and buoyancy forces in a cubic cavity with the attendant thermal boundary conditions. No slip velocity conditions are imposed on all wall surfaces, except for the top which is specified to be moving at a velocity $U_o$.}
\label{stabdiag}
\end{figure}

Finally, we now demonstrate the numerical advantages of using the FPC-LBM for simulating thermo-hydrodynamics when compared to other recently developed advanced collision models-based LBM in terms of significant stability improvements. We do this via the simulations of mixed convection involving shear flow combined with the presence of buoyancy forces in a cubic cavity of side $L\sbs{o}$, where the top wall is set into motion with a uniform velocity $U\sbs{o}$, while the other walls are stationary (see Fig.~\ref{stabdiag} for the problem setup). In addition, the top wall is imposed with a uniform hot temperature while the bottom wall is maintained at uniform cold temperature, and the remaining walls are considered adiabatic. Moreover, we apply a gravitational body force under the usual Boussinesq approximation so that both of the 3D FPC-LBMs for hydrodynamics and the energy equation are coupled. This configuration is meant to initiate numerical instability in simulations under shear at higher Mach numbers and with relatively coarse grid resolutions.

We will perform a systematic stability study for three different scenarios as will be discussed below. It involves using the 3D FPC-LBM and comparing it with the more recently developed LBM using the central moments-based collision operators using the Maxwell distribution function as the attractor~\cite{sharma2018natural,elseid2018cascaded,fei2018modeling,hajabdollahi2018central,hajabdollahi2019cascaded} extended to 3D, i.e., which use Eq.~(\ref{maxwell_distribution}) to prescribe the discrete central moment equilibria for the D3Q27 lattice for hydrodynamics via the matching principle and by replacing $\rho$ with $T$ in the same equation to construct the corresponding discrete equilibrium central moments for the D3Q15 lattice for the energy equation, again via matching. We do not include the older SRT and MRT collision models as part of this comparative study as they are already known to be less stable compared to the central moments-based collision models using the Maxwellian attractors (see e.g.,~\cite{elseid2018cascaded}).

The determination of stability/instability in numerical simulations are determined as follows. For a fixed grid resolution $L\sbs{o}^3$, Mach number $\mbox{Ma}=U\sbs{o}/c_s=\sqrt{3}U\sbs{o}$, and Prandtl number $\mbox{Pr}=\nu/\alpha$, when performing simulations using the FPC-LBM or the LBM using Maxwellian-based central moment collision operator (henceforth referred to as the MCM-LBM), we reduce the thermal diffusivity $\alpha$, which is related to the relaxation parameters of the first order moments for the LB solver for the temperature field, until the simulations remain stable for 1 million ($10^6$) time steps; on the other hand, the simulations are deemed unstable if the errors for the hydrodynamic or temperature fields grow large quickly (often exponentially). Based on the minimum achievable thermal diffusivity $\alpha$ for a chosen $L\sbs{o}$ and $U\sbs{o}$, we determine the maximum stable dimensionless characteristic parameter for mixed convection, viz., the Peclet number defined as $\mbox{Pe} = U\sbs{o}L\sbs{o}/\alpha$ in reporting the results of the comparative study.

\begin{figure}[H]
\centering
\begin{subfigure}{0.48\textwidth}
\includegraphics[trim = 0 0 0 0, clip, width =70mm]{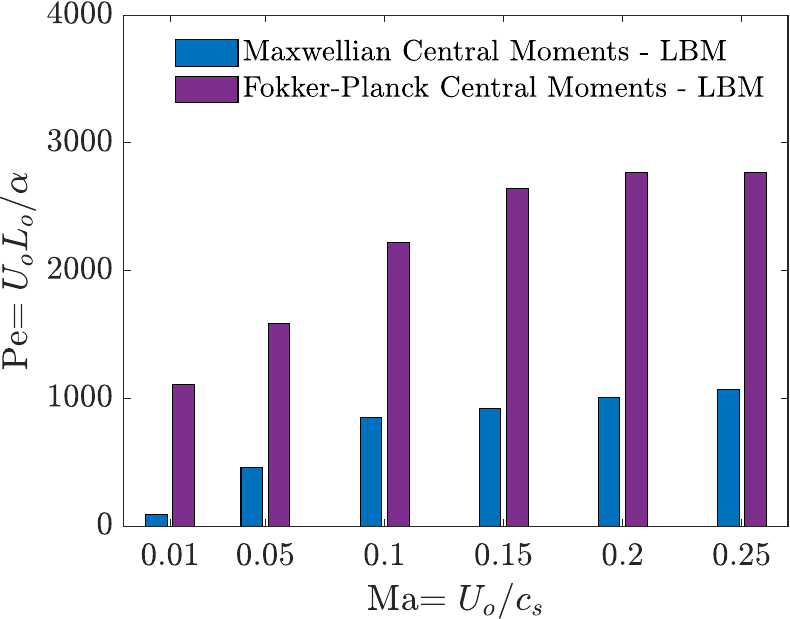}
\end{subfigure}
\centering
\begin{subfigure}{0.48\textwidth}
\includegraphics[trim = 0 0 0 0, clip, width =70mm]{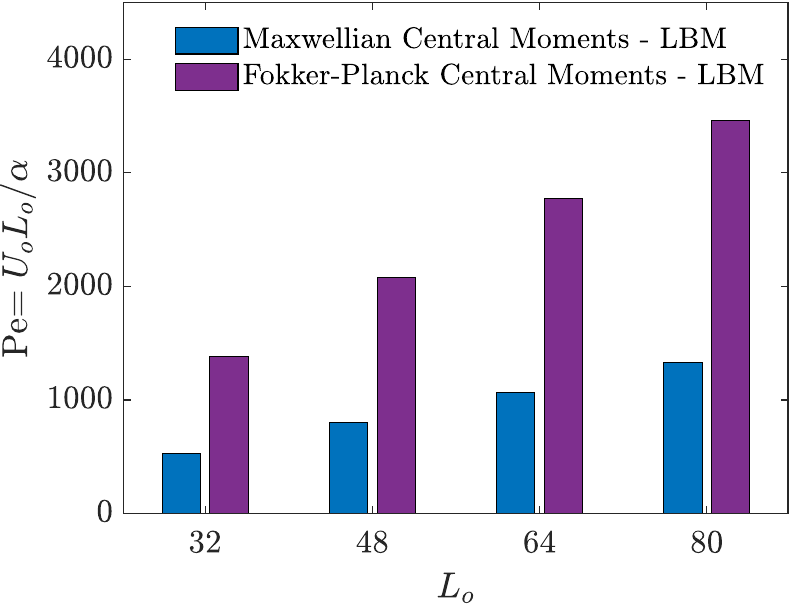}
\end{subfigure}

\vspace{5mm}
\centering
\begin{subfigure}{0.48\textwidth}
\includegraphics[trim = 0 0 0 0, clip, width =70mm]{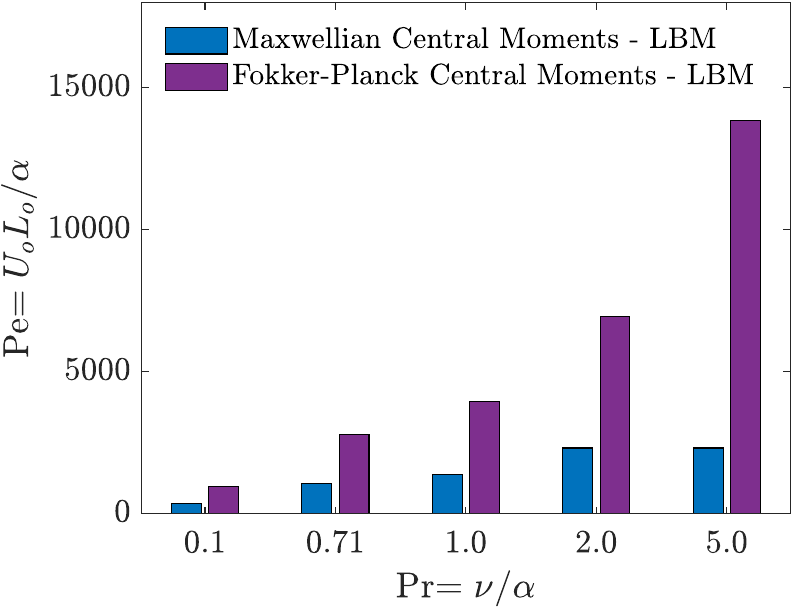}
\end{subfigure}
\caption{Results from three sets of numerical stability studies, each finding the maximum possible Peclet number for stable simulations using 3D FPC-LBM and the 3D central moment LBM using Maxwell distribution (referred to as the MCM-LBM) at various Mach numbers (top left), grid resolutions, where $L_o$ is the number of grid nodes resolving each side of the cavity (top right), and Prandtl numbers (bottom). Simulations were run $10^6$ time steps without encountering numerical instabilities in each case. The use of the renormalization principle in selecting the diffusion tensor parameter associated with the FPC-LBM (see e.g., Sec.~\ref{subsec:choice_of_diffusion_paramter}) facilitates its significant stability improvements over the MCM-LBM.}
\label{stabilitytestresults}
\end{figure}

Then, the three scenarios investigated are as follows: 1) For a fixed coarse resolution of $64^3$ and a Prandtl number of $\mbox{Pr} = 0.71$, we consider a range of Mach numbers given by $\mbox{Ma} = 0.01, 0.05, 0.10, 0.15, 0.2$, and $0.25$ and in each case the maximum achievable stable Peclet number $\mbox{Pe}$ by simulations using the FPC-LBM and MCM-LBM are noted. 2) For a fixed Prandtl number $\mbox{Pr}=0.71$ and Mach number $\mbox{Ma}=0.25$, we consider a range of grid resolutions including $32^3$, $48^3$, $64^3$, and $80^3$, and determine the maximum stable Peclet number $\mbox{Pe}$ achieved by FPC-LBM and MCM-LBM. 3) For a fixed resolution of $64^3$, Mach number $\mbox{Ma}=0.25$, we take a range of Prandtl numbers $\mbox{Pr}$ equal to $0.1$, $0.71$, $1.0$, $2.0$ and $5.0$, and deduce the maximum $\mbox{Pe}$ for both FPC-LBM and MCM-LBM. The numerical stability results obtained for these three scenarios are reported in Fig.~\ref{stabilitytestresults}. For all cases, the FPC-LBM is found to be significantly more stable in the simulation of mixed convective flows. The maximum stable Peclet number achieved by the FPC-LBM is seen to be several times larger when compared to that reached by the MCM-LBM and the improvements become greater at higher Prandtl numbers. Note that our previous study~\cite{schupbach2024fokker} showed that the FPC-LBM superior to MCM-LBM in terms of stability characteristics for simulating fluid motions without heat transfer. The present investigation further supports this conclusion for simulations of coupled thermal flows under a wide range of conditions.

To provide further perspective, we now discuss the relative computational costs of the LB methods associated with different collision models. As such, depending on the computing architecture used, the LB schemes are generally dominated by memory-bound operations rather than compute-bound calculations involved with their collide-and-stream based algorithms and hence using different collision models does not alter the overall computational cost significantly. The Maxwellian central moment (MCM)-LBM and Fokker-Planck central moment (FPC)-LBM have essentially the same computational efficiencies since modifying the attractors in the collision step results in negligible differences in terms of the overall operations involved. On the other hand, when comparing with the use of the simplest among the collision models, namely the SRT-LBM, the FPC-LBM has up to about 50\% computational overhead associated with the additional transformations involved with the pre- and post-collision operations; however, this is offset by the dramatic advantages of the FPC-LBM in delivering an order of magnitude or more improvements in numerical stability in simulating flows at high Reynolds numbers or Peclet numbers when compared to such collision models.

\begin{figure}[H]
\centering 
\includegraphics[trim = 0 0 0 0, clip, width =75mm]{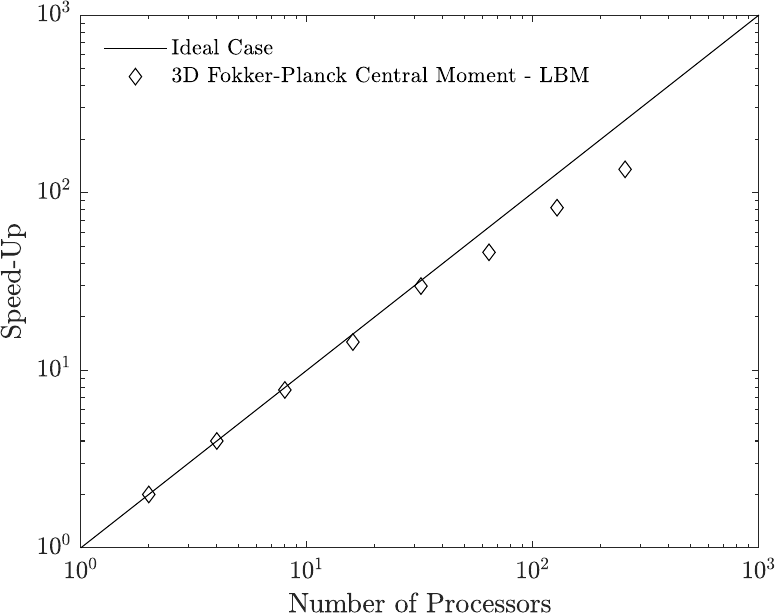}
\caption{Speed-up versus the number of processor cores for simulation of thermal convective flows in a cubic cavity with a domain resolved with $128^3$ nodes using the 3D FPC-LBM implemented on the parallel computer cluster INCLINE hosted at the University of Colorado at Colorado Springs (UCCS).}
\label{fig:speedup}
\end{figure}

Furthermore, the conventional numerical methods for incompressible flow simulations involve the solution of the Poisson-type equation for the pressure field which limits their scalability in using large parallel computing platforms effectively. On the other hand, the collide-and-stream steps of the LBM are inherently parallel in nature and well suited for implementations on such platforms. Our FPC-LBM for 3D thermal convective flow simulations have been parallelized using the Message Passing Interface (MPI) libraries in C++ and implemented using a domain decomposition strategy on the parallel computer cluster named INCLINE and hosted at the University of Colorado at Colorado Springs (UCCS). It was then tested with a domain resolved with $128^3$ grid nodes for measuring the speed-ups in using the $2$, $4$, $8$, $16$, $32$, $64$, $128$, and $256$ computing cores over its implementation on a single core on this cluster. Figure~\ref{fig:speedup} presents a plot of the speed-up versus the number of processor cores which shows that the 3D FPC-LBM implementation is scaling up quite well and almost linearly thereby demonstrating its parallel computing advantages.
Such natural scaling on parallel computer clusters can be exploited in simulating large-scale problems. In particular, the 3D FPC-LBMs can be adapted for performing large-eddy simulations of turbulent convective flows via utilizing appropriate subgrid scale (SGS) turbulence models for the unresolved turbulent stresses and heat fluxes via eddy viscosity and eddy thermal diffusivity, respectively. In this regard, the local eddy viscosity is used to augment the relaxation parameters of the second order moments involved in the collision step of the FPC-LBM for fluid motions and the eddy thermal diffusivity for adapting the relaxation parameters of the first order moments utilized in the collision step of the FPC-LBM for energy transport. These considerations pave the way towards simulating more complex thermal and flow transport problems of engineering interest.

\section{Summary and Conclusions\label{sec:summary}}
Originally developed to characterize Brownian dynamics, the Fokker-Planck (FP) equation represents an important element to represent non-equilibrium processes via the evolution of particle density distribution functions in statistical mechanics and various other applications. It has also been suggested as a drift and diffusion based stochastic model for the collision term of the Boltzmann equation. In our recent work, we developed a central moment formulation of the FP collision operator with its implementation in the lattice Boltzmann (LB) framework capable of robust numerical simulation of fluid dynamics especially at high Reynolds numbers.

In this work, we extend this approach for effective simulation of thermal convective flows. We utilize a double distribution functions (DDF)-based formulation, where one continuous kinetic equation is used to recover the equations of fluid motions and the other for the energy transport while endowing each of them with a FP central moment collision operator. By matching the changes in the discrete central moments under collision with those of the continuous FP-based central moment collision model for each of these kinetic equations, we develop novel LB schemes, referred to as the FPC-LBMs, for hydrodynamic fields and temperature fields in 2D using the D2Q9 lattice and in 3D using the D3Q27 lattice for the former and D3Q15 lattice for the latter. In each case, they involve the relaxations of different central moments to their ‘equilibria’, which are referred to as the Markovian central moment attractors reflecting the repeated randomness inherent in microscopic collision processes in fluids; such attractors depend on the products of lower central moments and diffusion parameter tensors. The latter are parameterized as fixed points of certain renormalization group equations involving the variations of the second order moments under collision to enable effective simulations of hydrodynamics for a wide range of transport coefficients without numerical artifacts due to the interference of higher kinetic moments that can arise when there are disparities in the relaxation rates. Since the number of collision invariants of the kinetic equations for recovering the fluid motions and energy transport in the DDF approach are different, the resulting expressions of such Markovian central moment attractors in the respective cases are quite distinct.

The 2D and 3D FPC-LBMs are validated against prior numerical/experimental data for the simulations of natural convection in square cavity and cubic cavity, respectively, over a wide range of Rayleigh numbers, Ra. The computed velocity and temperature profiles as well as the heat transfer rates based on the Nusselt number using the DDF-based FP-LBMs are shown be in good quantitative agreement with existing benchmark results, including those at higher Ra. Furthermore, we demonstrate that significant improvements in numerical stability can be achieved when using the FPC-LBMs in lieu of other existing central moment LBMs that use the Maxwellian distribution to prescribe equilibria in the collision models for simulation of a shear-driven mixed thermal convective flow in a cubic cavity. In particular, the FPC-LBMs are shown to be more stable in reaching markedly higher Peclet numbers than that using the latter.

The outlook on future work of this study has a number of possible directions for further investigations. In addition to being promising for simulation of flows with thermal transport, the FPC-LBMs as developed in this work are directly applicable for simulating the transport of any passive scalar (e.g., concentration of a tracer species or surfactants) under diffusion and advection due to fluid motions. Since the FPC-LBM is capable of avoiding hyperviscosity artifacts when simulating flows at relatively low physical viscosities, it can be utilized effectively in large-eddy simulations of anisotropic turbulent flows including heat transfer at high Reynolds numbers. Moreover, the method can be potentially extended for other complex flows such as those involving non-Newtonian fluids and multiphase flows by appropriate modifications of the FP-based central moment collision operator.

\section*{Acknowledgements}
Early stages of this research were presented at the 2023 American Physical Society March Meeting in Las Vegas, Nevada (https://meetings.aps.org/Meeting/MAR23/Session/T18.4). The authors would like to acknowledge the support of the US National Science Foundation (NSF) for research under Grant CBET-1705630. Simulations were performed using the high performance computing clusters Alderaan hosted at the Center for Computational Mathematics at the University of Colorado Denver and funded by NSF under Grant OAC-2019089, and INCLINE hosted at the University of Colorado Colorado Springs and funded by NSF under Grant Number OAC-2017917.

\appendix

\section{Implementation of 2D FPC-LBM with D2Q9 lattice for computing fluid motions\label{sec:appendixA}}
We present the implementation details of solving the 2D NSE using the FPC-LBM with the D2Q9 lattice in the following.

\noindent
Step 1: Transform the pre-collision distribution functions into pre-collision raw moments via $\m = \PP\mathbf{f}$ as
\begin{eqnarray}
\Kps{00} = \f{0} + \f{1} + \f{2} + \f{3} + \f{4} + \f{5} + \f{6} + \f{7} + \f{8},
\end{eqnarray}
\vspace{-9mm}
\begin{eqnarray}
\Kps{10} = \f{1} - \f{3} + \f{5}- \f{6} - \f{7} + \f{8}, \nonumber
\end{eqnarray}
\vspace{-9mm}
\begin{eqnarray}
\Kps{01} = \f{2} - \f{4} + \f{5} + \f{6} - \f{7} - \f{8},\nonumber
\end{eqnarray}
\vspace{-9mm}
\begin{eqnarray}
\Kps{20} = \f{1} + \f{3} + \f{5} + \f{6} + \f{7} + \f{8},\nonumber
\end{eqnarray}
\vspace{-9mm}
\begin{eqnarray}
\Kps{02} = \f{2} + \f{4} + \f{5} + \f{6} + \f{7} + \f{8},\nonumber
\end{eqnarray}
\vspace{-9mm}
\begin{eqnarray}
\Kps{11} = \f{5} - \f{6} + \f{7} - \f{8},\nonumber
\end{eqnarray}
\vspace{-9mm}
\begin{eqnarray}
\Kps{21} = \f{5} + \f{6} - \f{7} - \f{8},\nonumber
\end{eqnarray}
\vspace{-9mm}
\begin{eqnarray}
\Kps{12} = \f{5} - \f{6} - \f{7} + \f{8},\nonumber
\end{eqnarray}
\vspace{-9mm}
\begin{eqnarray}
\Kps{22} = \f{5} + \f{6} + \f{7} + \f{8}.\nonumber
\end{eqnarray}

\noindent
Step 2: Convert the pre-collision raw moments into pre-collision central moments via $\mc = \F \m$ as
\begin{eqnarray}
\Ks{00} = \Kps{00},
\end{eqnarray}
\vspace{-9mm}
\begin{eqnarray}
\Ks{10} = \Kps{10} - \ux \Kps{00}, \nonumber
\end{eqnarray}
\vspace{-9mm}
\begin{eqnarray}
\Ks{01} = \Kps{01} - \uy  \Kps{00} \nonumber
\end{eqnarray}
\vspace{-9mm}
\begin{eqnarray}
\Ks{20} =\Kps{20} - 2\ux  \Kps{10} +\uxx  \Kps{00}, \nonumber
\end{eqnarray}
\vspace{-9mm}
\begin{eqnarray}
\Ks{02} = \Kps{02} - 2\uy  \Kps{01} + \uyy \Kps{00}, \nonumber
\end{eqnarray}
\vspace{-9mm}
\begin{eqnarray}
\Ks{11} = \Kps{11} -\uy  \Kps{10} - \ux   \Kps{01} + \ux \uy \Kps{00}, \nonumber
\end{eqnarray}
\vspace{-9mm}
\begin{eqnarray}
\Ks{21} = \Kps{21} - 2\ux \Kps{11} + \uxx \Kps{01} - \uy  \Kps{20} +2 \ux\uy  \Kps{10}- \uxx  \uy  \Kps{00}, \nonumber
\end{eqnarray}
\vspace{-9mm}
\begin{eqnarray}
\Ks{12} = \Kps{12} -2u_y  \Kps{11} + \uyy \Kps{10} - \ux \Kps{02} + 2 \ux \uy  \Kps{01} - \ux  \uyy  \Kps{00}, \nonumber
\end{eqnarray}
\vspace{-9mm}
\begin{eqnarray}
\Ks{22} \! \! \! & =& \! \! \! \Kps{22} - 2\ux \Kps{12} +\uxx  \Kps{02} - 2 \uy \Kps{21} + 4 \ux \uy  \Kps{11} - \uxx  2 \uy  \Kps{01} + \uyy  \Kps{20}  \nonumber \\
& & -2 \ux  \uyy  \Kps{10} + \uxx \uyy  \Kps{00}. \nonumber
\end{eqnarray}

Step 3: Collision update - Perform relaxation of all the 9 independent central moments to their attractors for the D2Q9 lattice with additional contributions from source terms due to the body force. First, we combine the diagonal second order central moments in order to independently evolve the effects of shear and bulk viscosities as
\begin{eqnarray}
\Ks{2s} = \Ks{20} + \Ks{02},  \qquad \qquad
\Ks{2d} = \Ks{20} - \Ks{02}.
\end{eqnarray}
Then, we calculate the lower order Markovian central moment attractors based on the following:
\begin{gather}
\KMv{00} = \rho,   \qquad \qquad
\KMv{10} = 0,  \qquad \qquad
\KMv{01} = 0, \qquad \qquad
\KMv{20} = \css \rho, \\[2mm]
\KMv{02} = \css \rho, \qquad \qquad
\KMv{11} = 0, \qquad \qquad
\KMv{21} = 0, \qquad \qquad
\KMv{12} = 0, \nonumber
\end{gather}
where $c_s^2=1/3$, and also combine the corresponding attractors and the sources (see Eq.~(\ref{eq:sourceCM2D})) in the same manner as
\begin{eqnarray*}
\KMv{2s} &=& \KMv{20} + \KMv{02},  \qquad \qquad
\KMv{2d} = \KMv{20} - \KMv{02}. \nonumber\\
\Ss{2s} &=& \Ss{20} + \Ss{02},  \qquad \qquad\quad
\Ss{2d} = \Ss{20} - \Ss{02}. \nonumber
\end{eqnarray*}
Next, perform relaxations under collision with source updates to obtain the post-collision central moments  $\Kts{mn}$ up to the third order (i.e., $m+n\le 3$) as
\begin{eqnarray}
\Kts{00} = \Ks{00} + \omega\sbs{0} (\KMv{00} - \Ks{00})+ \left(1-\omega\sbs{0}/2 \right)\Ss{00}\delta\sbs{t},
\end{eqnarray}
\vspace{-9mm}
\begin{eqnarray}
\Kts{10} = \Ks{10} + \omega\sbs{1} (\KMv{10} - \Ks{10})+ \left(1-\omega\sbs{1}/2 \right)\Ss{10}\delta\sbs{t},\nonumber
\end{eqnarray}
\vspace{-9mm}
\begin{eqnarray}
\Kts{01} = \Ks{01} + \omega\sbs{2} (\KMv{01} - \Ks{01})+ \left(1-\omega\sbs{2}/2 \right)\Ss{01}\delta\sbs{t},\nonumber
\end{eqnarray}
\vspace{-9mm}
\begin{eqnarray}
\Kts{2s} = \Ks{2s} + \omega\sbs{3 } (\KMv{2s}- \Ks{2s})+ \left(1-\omega\sbs{3}/2 \right)\Ss{2s}\delta\sbs{t},\nonumber
\end{eqnarray}
\vspace{-9mm}
\begin{eqnarray}
\Kts{2d} =\Ks{2d} + \omega\sbs{4 } (\KMv{2d}- \Ks{2d})+ \left(1-\omega\sbs{4}/2 \right)\Ss{2d}\delta\sbs{t},\nonumber
\end{eqnarray}
\vspace{-9mm}
\begin{eqnarray}
\Kts{11} = \Ks{11} + \omega\sbs{5}  (\KMv{11} - \Ks{11})+ \left(1-\omega\sbs{5}/2 \right)\Ss{11}\delta\sbs{t},\nonumber
\end{eqnarray}
\vspace{-9mm}
\begin{eqnarray}
\Kts{21} = \Ks{21} + \omega\sbs{6}  (\KMv{21} - \Ks{21})+ \left(1-\omega\sbs{6}/2 \right)\Ss{21}\delta\sbs{t},\nonumber
\end{eqnarray}
\vspace{-9mm}
\begin{eqnarray}
\Kts{12} = \Ks{12} + \omega\sbs{7 } (\KMv{12} - \Ks{12})+ \left(1-\omega\sbs{7}/2 \right)\Ss{12}\delta\sbs{t},\nonumber
\end{eqnarray}
Here, $\Ss{mn}$ is provided in Eq.~(\ref{eq:sourceCM2D}). Then, split the post-collision second order combined central moments to obtain their individual contributions through
\begin{eqnarray}
\tilde{\K}\sbs{20} = 0.5(\tilde{\K}\sbs{2s} +\tilde{\K}\sbs{2d}), \qquad \qquad
\tilde{\K}\sbs{02} = 0.5(\tilde{\K}\sbs{2s} -\tilde{\K}\sbs{2d}).
\end{eqnarray}
Using the above, we then specify the Markovian attractor for the fourth order central moment based on post-collision second order central moments as
\begin{eqnarray}
\KMv{22} = \frac{1}{\rho} (\tilde{\K}\sbs{20} \tilde{\K}\sbs{02} + 2 \tilde{\K}\sbs{11} \tilde{\K}\sbs{11}),
\end{eqnarray}
Finally, the fourth order central moment is updated under relaxation as
\begin{eqnarray}
\tilde{\K}\sbs{22} = \Ks{22} + \omega\sbs{8} (\KMv{22} - \Ks{22})+ \left(1-\omega\sbs{8}/2 \right)\sigma\sbs{22}.
\end{eqnarray}
\noindent
Step 4: Map post-collision central moments into post-collision raw moments using $\tilde{\m} = \Fi \tilde{\mathbf{m}}\sps{c}$ as
\begin{eqnarray}
\Ktps{00} =  \Kts{00},
\end{eqnarray}
\vspace{-9mm}
\begin{eqnarray}
\Ktps{10} = \Kts{10} + \ux\Kts{00},\nonumber
\end{eqnarray}
\vspace{-9mm}
\begin{eqnarray}
\Ktps{01} = \Kts{01} + \uy\Kts{00},\nonumber
\end{eqnarray}
\vspace{-9mm}
\begin{eqnarray}
\Ktps{20}=  \Kts{20} + 2  \ux\Kts{10} +\uxx \Kts{00},\nonumber
\end{eqnarray}
\vspace{-9mm}
\begin{eqnarray}
\Ktps{02} =  \Kts{02} + 2 \uy\Kts{01} + \uyy\Kts{00},\nonumber
\end{eqnarray}
\vspace{-9mm}
\begin{eqnarray}
\Ktps{11} =  \Kts{11} + \ux   \Kts{01} + \uy\Kts{10} + \ux \uy \Kts{00},\nonumber
\end{eqnarray}
\vspace{-9mm}
\begin{eqnarray}
\Ktps{21} =  \Kts{21} + 2  \ux\Kts{11} + \uy\Kts{20} + \uxx  \Kts{01} + 2 \ux \uy  \Kts{10} + \uxx \uy  \Kts{00}, \nonumber
\end{eqnarray}
\vspace{-9mm}
\begin{eqnarray}
\Ktps{12} =  \Kts{12} + 2  \uy\Kts{11} + \ux\Kts{02} + 2 \ux \uy \Kts{01} + \uyy \Kts{10} + \ux \uyy \Kts{00},\nonumber
\end{eqnarray}
\vspace{-9mm}
\begin{eqnarray}
\Ktps{22}  \! \! \! & =& \! \! \!  \Kts{22} + 2  \ux\Kts{12} + 2 \uy\Kts{21} + 4 \ux \uy \Kts{11} + \uyy \Kts{20} + \uxx \Kts{02} + 2 u_x^2 \uy\Kts{01}\nonumber \\
& & + 2 \ux \uyy \Kts{10} + \uxx \uyy \Kts{00}.\nonumber
\end{eqnarray}

\noindent
Step 5: Convert post-collision raw moments into post-collision distribution functions via $\tilde{\mathbf{f}} = \PP\sps{\sm1}\tilde{\m}$ as
\begin{eqnarray}
\tilde{f}\sbs0 = \Ktps{00} - \Ktps{20} - \Ktps{02} + \Ktps{22},
\end{eqnarray}
\vspace{-9mm}
\begin{eqnarray}
\tilde{f}\sbs1 =(\Ktps{10} + \Ktps{20} - \Ktps{12} - \Ktps{22})/2,\nonumber
\end{eqnarray}
\vspace{-9mm}
\begin{eqnarray}
\tilde{f}\sbs2 = (\Ktps{01} + \Ktps{02} - \Ktps{21} - \Ktps{22})/2,\nonumber
\end{eqnarray}
\vspace{-9mm}
\begin{eqnarray}
\tilde{f}\sbs3 = (-\Ktps{10} + \Ktps{20} + \Ktps{12} - \Ktps{22})/2,\nonumber
\end{eqnarray}
\vspace{-9mm}
\begin{eqnarray}
\tilde{f}\sbs4 = (-\Ktps{01}+\Ktps{02} + \Ktps{21} - \Ktps{22})/2,\nonumber
\end{eqnarray}
\vspace{-9mm}
\begin{eqnarray}
\tilde{f}\sbs5 = (\Ktps{11} + \Ktps{21} + \Ktps{12} + \Ktps{22})/4,\nonumber
\end{eqnarray}
\vspace{-9mm}
\begin{eqnarray}
\tilde{f}\sbs6 = (-\Ktps{11} + \Ktps{21} - \Ktps{12} + \Ktps{22})/4,\nonumber
\end{eqnarray}
\vspace{-9mm}
\begin{eqnarray}
\tilde{f}\sbs7 = (\Ktps{11} - \Ktps{21} - \Ktps{12} + \Ktps{22})/4,\nonumber
\end{eqnarray}
\vspace{-9mm}
\begin{eqnarray}
\tilde{f}\sbs8 = (-\Ktps{11} - \Ktps{21} + \Ktps{12} + \Ktps{22})/4.\nonumber
\end{eqnarray}
\noindent
Step 6: Perform streaming step using the post-collision distribution functions for all the 9 directions of the D2Q9 lattice as
\begin{eqnarray}
  f\sbs{\alpha}(\bm{x},t+\delta\sbs{t}) &=& \tilde{f}\sbs{\alpha}(\bm{x}-\bm{e}\sbs{\alpha}\delta\sbs{t},t), \quad \alpha = 0,1,2,\ldots, 8,
\end{eqnarray}
\noindent
and implement the boundary conditions on the distribution functions as necessary. \newline
Step 7: Calculate the fluid dynamical variables (i.e., the density and velocity) using the zeroth and first moments of the distribution function as
\begin{eqnarray}
\rho = \sum\sbs{\alpha=0}\sps{8} f\sbs{\alpha}, \qquad \rho \bm{u} = \sum\sbs{\alpha=0}\sps{8} f\sbs{\alpha}\bm{e}\sbs{\alpha} + \frac{1}{2}\bm{F} \delta\sbs{t},
\end{eqnarray}
and pressure $P = \rho \css$. The choice of the relaxation parameters $\omega_j$, which satisfy $0 < \omega_j < 2$, in the above are as follows. We select the relaxation parameters for the second order moments $\omega_3$ and $\omega_4=\omega_5$ based on the transport coefficients of the fluid, viz., its bulk and shear viscosities, $\zeta$ and $\nu$, respectively, according to (see~\cite{schupbach2024fokker})
\begin{eqnarray}\label{eqn:transportcoeff2DFPCLBM}
\zeta = c_s^2\left(\frac{1}{\omega_{3}}-\frac{1}{2}\right)\delta_t, \quad \nu =c_s^2\left(\frac{1}{\omega_{4}}-\frac{1}{2}\right)\delta_t = c_s^2\left(\frac{1}{\omega_{5}}-\frac{1}{2}\right)\delta_t.
\end{eqnarray}
To recover the 2D NSE in the incompressible limit, the shear viscosity $\nu$ is the only relevant physical property, which is, in turn, based on the Reynolds number obtained from the flow problem statement; hence $\omega_4$ and $\omega_5$ is chosen using $\nu$ based on Eq.~(\ref{eqn:transportcoeff2DFPCLBM}). The remaining $\omega_j$, where $j = 0, 1, 2, 6, 7, 8$, are free parameters and are selected based on numerical stability considerations. We set them to be equal to 1.0 in this work. As such, the Steps 1 through 7 represent one complete step of the FPC-LBM with a time step $\delta_t$ to simulate hydrodynamics.

\section{Implementation of 2D FPC-LBM with D2Q9 lattice for computing temperature field\label{sec:appendixB}}
We present the implementation details of solving the 2D energy equation using the FPC-LBM with the D2Q9 lattice in the following.

\noindent
Step 1: Transform the pre-collision distribution functions into pre-collision raw moments via $\n = \PP\mathbf{g}$ as
\begin{eqnarray}
\Nps{00} = \g{0} + \g{1} + \g{2} + \g{3} + \g{4} + \g{5} + \g{6} + \g{7} + \g{8},
\end{eqnarray}
\vspace{-9mm}
\begin{eqnarray}
\Nps{10} = \g{1} - \g{3} + \g{5}- \g{6} - \g{7} + \g{8}, \nonumber
\end{eqnarray}
\vspace{-9mm}
\begin{eqnarray}
\Nps{01} = \g{2} - \g{4} + \g{5} + \g{6} - \g{7} - \g{8},\nonumber
\end{eqnarray}
\vspace{-9mm}
\begin{eqnarray}
\Nps{20} = \g{1} + \g{3} + \g{5} + \g{6} + \g{7} + \g{8},\nonumber
\end{eqnarray}
\vspace{-9mm}
\begin{eqnarray}
\Nps{02} = \g{2} + \g{4} + \g{5} + \g{6} + \g{7} + \g{8},\nonumber
\end{eqnarray}
\vspace{-9mm}
\begin{eqnarray}
\Nps{11} = \g{5} - \g{6} + \g{7} - \g{8},\nonumber
\end{eqnarray}
\vspace{-9mm}
\begin{eqnarray}
\Nps{21} = \g{5} + \g{6} - \g{7} - \g{8},\nonumber
\end{eqnarray}
\vspace{-9mm}
\begin{eqnarray}
\Nps{12} = \g{5} - \g{6} - \g{7} + \g{8},\nonumber
\end{eqnarray}
\vspace{-9mm}
\begin{eqnarray}
\Nps{22} = \g{5} + \g{6} + \g{7} + \g{8}.\nonumber
\end{eqnarray}

\noindent
Step 2: Convert the pre-collision raw moments into pre-collision central moments via $\nc = \F \n$ as
\begin{eqnarray}
\Ns{00} = \Nps{00},
\end{eqnarray}
\vspace{-9mm}
\begin{eqnarray}
\Ns{10} = \Nps{10} - \ux \Nps{00}, \nonumber
\end{eqnarray}
\vspace{-9mm}
\begin{eqnarray}
\Ns{01} = \Nps{01} - \uy  \Nps{00} \nonumber
\end{eqnarray}
\vspace{-9mm}
\begin{eqnarray}
\Ns{20} =\Nps{20} - 2\ux  \Nps{10} +\uxx  \Nps{00}, \nonumber
\end{eqnarray}
\vspace{-9mm}
\begin{eqnarray}
\Ns{02} = \Nps{02} - 2\uy  \Nps{01} + \uyy \Nps{00}, \nonumber
\end{eqnarray}
\vspace{-9mm}
\begin{eqnarray}
\Ns{11} = \Nps{11} -\uy  \Nps{10} - \ux   \Nps{01} + \ux \uy \Nps{00}, \nonumber
\end{eqnarray}
\vspace{-9mm}
\begin{eqnarray}
\Ns{21} = \Nps{21} - 2\ux \Nps{11} + \uxx \Nps{01} - \uy  \Nps{20} +2 \ux\uy  \Nps{10}- \uxx  \uy  \Nps{00}, \nonumber
\end{eqnarray}
\vspace{-9mm}
\begin{eqnarray}
\Ns{12} = \Nps{12} -2\uy  \Kps{11} + \uyy \Nps{10} - \ux \Nps{02} + 2 \ux \uy  \Nps{01} - \ux  \uyy  \Nps{00}, \nonumber
\end{eqnarray}
\vspace{-9mm}
\begin{eqnarray}
\Ns{22} \! \! \! & =& \! \! \! \Nps{22} - 2\ux \Nps{12} +\uxx  \Nps{02} - 2 \uy \Nps{21} + 4 \ux \uy  \Nps{11} - \uxx  2 \uy  \Nps{01} + \uyy  \Nps{20}  \nonumber \\
& & -2 \ux  \uyy  \Nps{10} + \uxx \uyy  \Nps{00}. \nonumber
\end{eqnarray}
\noindent
Step 3: Collision update - Perform relaxation of all the 9 independent central moments to their attractors for the D2Q9 lattice. First, we calculate the lower order Markovian central moment attractors based on the following:
\begin{eqnarray}
&&\NMv{00} = T,   \qquad \qquad
\NMv{10} = 0,  \qquad \qquad
\NMv{01} = 0, \qquad \nonumber\\[2mm]
&&\NMv{20} = \chss T, \qquad \quad
\NMv{02} = \chss T, \qquad \quad
\NMv{11} = 0,
\end{eqnarray}
where $\chss =1/3$. Next, perform relaxations under collision with source updates to obtain the post-collision central moments  $\Nts{mn}$ up to the second order (i.e., $m+n\le 2$) as
\begin{eqnarray}
\Nts{00} = \Ns{00} + \omega\sbs{0} (\NMv{00} - \Ns{00}),
\end{eqnarray}
\vspace{-9mm}
\begin{eqnarray}
\Nts{10} = \Ns{10} + \omega\sbs{1} (\NMv{10} - \Ns{10}),\nonumber
\end{eqnarray}
\vspace{-9mm}
\begin{eqnarray}
\Nts{01} = \Ns{01} + \omega\sbs{2} (\NMv{01} - \Ns{01}),\nonumber
\end{eqnarray}
\vspace{-9mm}
\begin{eqnarray}
\Nts{2s} = \Ns{2s} + \omega\sbs{3 } (\NMv{2s}- \Ns{2s}),\nonumber
\end{eqnarray}
\vspace{-9mm}
\begin{eqnarray}
\Nts{2d} =\Ns{2d} + \omega\sbs{4 } (\NMv{2d}- \Ns{2d}),\nonumber
\end{eqnarray}
\vspace{-9mm}
\begin{eqnarray}
\Nts{11} = \Ns{11} + \omega\sbs{5}  (\NMv{11} - \Ns{11}).\nonumber
\end{eqnarray}
\noindent
Using the above, we then specify the Markovian attractors for the third order central moments based on post-collision first order moments, and for the fourth order central moment based on post-collision second order central moments as follows:
\begin{eqnarray}
&&\Ns{21}\sps{Mv} = \frac{2}{3}\chss\tilde{\N}\sbs{01}, \; \; \;  \; \; \;
\Ns{12}\sps{Mv} = \frac{2}{3}\chss\tilde{\N}\sbs{10}, \; \; \;  \; \; \;
\Ns{22}\sps{ Mv} = \frac{1}{2}\chss\left(\tilde{\N}\sbs{20}+\tilde{\N}\sbs{02}\right).
\end{eqnarray}
Following these, finally, the third and fourth order central moments are updated under relaxation as
\begin{eqnarray}
\Nts{21} = \Ns{21} + \omega\sbs{6}  (\NMv{21} - \Ns{21}),\nonumber
\end{eqnarray}
\vspace{-9mm}
\begin{eqnarray}
\Nts{12} = \Ns{12} + \omega\sbs{7 } (\NMv{12} - \Ns{12}),\nonumber
\end{eqnarray}
\vspace{-9mm}
\begin{eqnarray}
\tilde{\N}\sbs{22} = \Ns{22} + \omega\sbs{8} (\NMv{22} - \Ns{22}).
\end{eqnarray}
\noindent
Step 4: Map post-collision central moments into post-collision raw moments using $\tilde{\n} = \Fi \tilde{\mathbf{n}}\sps{c}$ as
\begin{eqnarray}
\Ntps{00} =  \Nts{00},
\end{eqnarray}
\vspace{-9mm}
\begin{eqnarray}
\Ntps{10} = \Nts{10} + \ux\Nts{00},\nonumber
\end{eqnarray}
\vspace{-9mm}
\begin{eqnarray}
\Ntps{01} = \Nts{01} + \uy\Nts{00},\nonumber
\end{eqnarray}
\vspace{-9mm}
\begin{eqnarray}
\Ntps{20}=  \Nts{20} + 2  \ux\Nts{10} +\uxx \Nts{00},\nonumber
\end{eqnarray}
\vspace{-9mm}
\begin{eqnarray}
\Ntps{02} =  \Nts{02} + 2 \uy\Nts{01} + \uyy\Nts{00},\nonumber
\end{eqnarray}
\vspace{-9mm}
\begin{eqnarray}
\Ntps{11} =  \Nts{11} + \ux   \Nts{01} + \uy\Nts{10} + \ux \uy \Nts{00},\nonumber
\end{eqnarray}
\vspace{-9mm}
\begin{eqnarray}
\Ntps{21} =  \Nts{21} + 2  \ux\Nts{11} + \uy\Nts{20} + \uxx  \Nts{01} + 2 \ux \uy  \Nts{10} + \uxx \uy  \Nts{00}, \nonumber
\end{eqnarray}
\vspace{-9mm}
\begin{eqnarray}
\Ntps{12} =  \Nts{12} + 2  \uy\Nts{11} + \ux\Nts{02} + 2 \ux \uy \Nts{01} + \uyy \Nts{10} + \ux \uyy \Nts{00},\nonumber
\end{eqnarray}
\vspace{-9mm}
\begin{eqnarray}
\Ntps{22}  \! \! \! & =& \! \! \!  \Nts{22} + 2  \ux\Nts{12} + 2 \uy\Nts{21} + 4 \ux \uy \Nts{11} + \uyy \Nts{20} + \uxx \Nts{02} + 2 u_x^2 \uy\Nts{01}\nonumber \\
& & + 2 \ux \uyy \Nts{10} + \uxx \uyy \Nts{00}.\nonumber
\end{eqnarray}

\noindent
Step 5: Convert post-collision raw moments into post-collision distribution functions via $\tilde{\mathbf{g}} = \PP\sps{\sm1}\tilde{\n}$ as
\begin{eqnarray}
\tilde{g}\sbs0 = \Ntps{00} - \Ntps{20} - \Ntps{02} + \Ntps{22},
\end{eqnarray}
\vspace{-9mm}
\begin{eqnarray}
\tilde{g}\sbs1 = (\Ntps{10} + \Ntps{20} - \Ntps{12} - \Ntps{22})/2,\nonumber
\end{eqnarray}
\vspace{-9mm}
\begin{eqnarray}
\tilde{g}\sbs2 = (\Ntps{01} + \Ntps{02} - \Ntps{21} - \Ntps{22})/2,\nonumber
\end{eqnarray}
\vspace{-9mm}
\begin{eqnarray}
\tilde{g}\sbs3 = (-\Ntps{10} + \Ntps{20} + \Ntps{12} - \Ntps{22})/2,\nonumber
\end{eqnarray}
\vspace{-9mm}
\begin{eqnarray}
\tilde{g}\sbs4 = (-\Ntps{01}+\Ntps{02} + \Ntps{21} - \Ntps{22})/2,\nonumber
\end{eqnarray}
\vspace{-9mm}
\begin{eqnarray}
\tilde{g}\sbs5 = (\Ntps{11} + \Ntps{21} + \Ntps{12} + \Ntps{22})/4,\nonumber
\end{eqnarray}
\vspace{-9mm}
\begin{eqnarray}
\tilde{g}\sbs6 = (-\Ntps{11} + \Ntps{21} - \Ntps{12} + \Ntps{22})/4,\nonumber
\end{eqnarray}
\vspace{-9mm}
\begin{eqnarray}
\tilde{g}\sbs7 = (\Ntps{11} - \Ntps{21} - \Ntps{12} + \Ntps{22})/4,\nonumber
\end{eqnarray}
\vspace{-9mm}
\begin{eqnarray}
\tilde{g}\sbs8 = (-\Ntps{11} - \Ntps{21} + \Ntps{12} + \Ntps{22})/4.\nonumber
\end{eqnarray}
\noindent
Step 6: Perform streaming step using the post-collision distribution functions for all the 9 directions of the D2Q9 lattice as
\begin{eqnarray}
  g\sbs{\alpha}(\bm{x},t+\delta\sbs{t}) &=& \tilde{g}\sbs{\alpha}(\bm{x}-\bm{e}\sbs{\alpha}\delta\sbs{t},t), \quad \alpha = 0,1,2,\ldots, 8,
\end{eqnarray}
and implement the boundary conditions on the distribution functions as necessary. \newline
Step 7: Compute the temperature field via zeroth moment of the distribution function as
\begin{eqnarray}
T = \sum\sbs{\alpha=0}\sps{8} g\sbs{\alpha}.
\end{eqnarray}
\noindent
The choice of the relaxation parameters $\omega\sbs{j}$, which satisfy $0 < \omega_j < 2$, in the above are as follows. We select the relaxation parameters for the first order moments $\omega\sbs{1}=\omega\sbs{2}$ based on the thermal transport coefficient of the fluid, viz., thermal diffusivity, $\alpha$ (which is related to the Rayleigh number or Peclet number based on the thermal flow problem statement) according to
\begin{eqnarray}\label{eqn:thermaltransportcoeff2DFPCLBM}
\alpha =\chss\left(\frac{1}{\omega\sbs{1}}-\frac{1}{2}\right)\delta_t = \chss\left(\frac{1}{\omega\sbs{2}}-\frac{1}{2}\right)\delta_t.
\end{eqnarray}
so that the 2D equation of the convection and diffusion transport of energy is recovered. The remaining $\omega_j$, where $j = 0, 3, 4, 5, 6, 7, 8$, are free parameters and are selected based on numerical stability considerations. We set them to be equal to 1.0 in this work. As such, the Steps 1 through 7 represent one complete step of the FPC-LBM with a time step $\delta_t$ to simulate energy transport.

\section{Implementation of 3D FPC-LBM with D3Q15 lattice for computing temperature field\label{sec:appendixC}}
We present the implementation details of solving the 3D energy equation using the FPC-LBM with the D3Q15 lattice in the following.

\noindent
Step 1: Transform the pre-collision distribution functions into pre-collision raw moments via $\n = \PP\mathbf{g}$ as
\begin{eqnarray}
\Nps{000} = \g{0} + \g{1} + \g{2} + \g{3} + \g{4} + \g{5} + \g{6} + \g{7} + \g{8} + \g{9} + \g{10}+ \g{11} + \g{12} + \g{13} + \g{14}, \nonumber
\end{eqnarray}
\vspace{-9mm}
\begin{eqnarray}
\Nps{100} = \g{1} - \g{2} + \g{7} - \g{8} + \g{9} - \g{10} + \g{11} - \g{12} + \g{13} - \g{14},\nonumber
\end{eqnarray}
\vspace{-9mm}
\begin{eqnarray}
\Nps{010} = \g{3} - \g{4} + \g{7} + \g{8} - \g{9} - \g{10} + \g{11} + \g{12} - \g{13} - \g{14},\nonumber
\end{eqnarray}
\vspace{-9mm}
\begin{eqnarray}
\Nps{001} = \g{5} - \g{6} + \g{7} + \g{8} + \g{9} + \g{10} - \g{11} - \g{12} - \g{13} - \g{14},\nonumber
\end{eqnarray}
\vspace{-9mm}
\begin{eqnarray}
\Nps{110} = \g{7} - \g{8} - \g{9} + \g{10} + \g{11} - \g{12} - \g{13} + \g{14},\nonumber
\end{eqnarray}
\vspace{-9mm}
\begin{eqnarray}
\Nps{101} = \g{7} - \g{8} + \g{9} - \g{10} - \g{11} + \g{12} - \g{13} + \g{14},\nonumber
\end{eqnarray}
\vspace{-9mm}
\begin{eqnarray}
\Nps{011} = \g{7} + \g{8} - \g{9} - \g{10} - \g{11} - \g{12} + \g{13} + \g{14},\nonumber
\end{eqnarray}
\vspace{-9mm}
\begin{eqnarray}
\Nps{200} = \g{1} + \g{2} + \g{7} + \g{8} + \g{9} + \g{10} + \g{11} + \g{12} + \g{13} + \g{14},\nonumber
\end{eqnarray}
\vspace{-9mm}
\begin{eqnarray}
\Nps{020} = \g{3} + \g{4} + \g{7} + \g{8} + \g{9} + \g{10} + \g{11} + \g{12} + \g{13} + \g{14},\nonumber
\end{eqnarray}
\vspace{-9mm}
\begin{eqnarray}
\Nps{002} = \g{5} + \g{6} + \g{7} + \g{8} + \g{9} + \g{10} + \g{11} + \g{12} + \g{13} + \g{14},\nonumber
\end{eqnarray}
\vspace{-9mm}
\begin{eqnarray}
\Nps{120} = \g{7} - \g{8} + \g{9} - \g{10} + \g{11} - \g{12} + \g{13} - \g{14},\nonumber
\end{eqnarray}
\vspace{-9mm}
\begin{eqnarray}
\Nps{012} = \g{7} + \g{8} - \g{9} - \g{10} + \g{11} + \g{12} - \g{13} - \g{14},\nonumber
\end{eqnarray}
\vspace{-9mm}
\begin{eqnarray}
\Nps{201} = \g{7} + \g{8} + \g{9} + \g{10} - \g{11} - \g{12} - \g{13} - \g{14},\nonumber
\end{eqnarray}
\vspace{-9mm}
\begin{eqnarray}
\Nps{111} = \g{7} - \g{8} - \g{9} + \g{10} - \g{11} + \g{12} + \g{13} - \g{14},\nonumber
\end{eqnarray}
\vspace{-9mm}
\begin{eqnarray}
\Nps{220} = \g{7} + \g{8} + \g{9} + \g{10} + \g{11} + \g{12} + \g{13} + \g{14}.\nonumber
\end{eqnarray}

\noindent
Step 2: Convert the pre-collision raw moments into pre-collision central moments via $\nc = \F \n$ as
\begin{eqnarray}
\Ns{000} = \Nps{000},
\end{eqnarray}
\vspace{-9mm}
\begin{eqnarray}
\Ns{100} = \Nps{100} - \ux\Nps{000}, \nonumber
\end{eqnarray}
\vspace{-9mm}
\begin{eqnarray}
\Ns{010} = \Nps{010} - \uy\Nps{000},\nonumber
\end{eqnarray}
\vspace{-9mm}
\begin{eqnarray}
\Ns{001} = \Nps{001} - \uz\Nps{000},\nonumber
\end{eqnarray}
\vspace{-9mm}
\begin{eqnarray}
\Ns{110} = \Nps{110} - \ux \Nps{010} - \uy \Nps{100} + \ux\uy \Nps{000},\nonumber
\end{eqnarray}
\vspace{-9mm}
\begin{eqnarray}
\Ns{101} = \Nps{101} - \ux \Nps{001} - \uz \Nps{100} + \ux\uz \Nps{000},\nonumber
\end{eqnarray}
\vspace{-9mm}
\begin{eqnarray}
\Ns{011} = \Nps{011} - \uy \Nps{001} - \uz \Nps{010} + \uy\uz \Nps{000},\nonumber
\end{eqnarray}
\vspace{-9mm}
\begin{eqnarray}
\Ns{200} = \Nps{200} - 2\ux \Nps{100} + \uxx \Nps{000},\nonumber
\end{eqnarray}
\vspace{-9mm}
\begin{eqnarray}
\Ns{020} = \Nps{020} - 2 \uy \Nps{010} + \uyy \Nps{000},\nonumber
\end{eqnarray}
\vspace{-9mm}
\begin{eqnarray}
\Ns{002} = \Nps{002} - 2 \uz \Nps{001}  + \uzz \Nps{000},\nonumber
\end{eqnarray}
\vspace{-9mm}
\begin{eqnarray}
\Ns{120} = \Nps{120} - \ux\Nps{020} - 2 \uy\Nps{110} + \uyy \Nps{100}   + 2 \ux\uy \Nps{010}  -  \ux  \uyy\Nps{000} ,\nonumber
\end{eqnarray}
\vspace{-9mm}
\begin{eqnarray}
\Ns{012} = \Nps{012} - \uy\Nps{002} - 2 \uz \Nps{011}  + \uzz\Nps{010}  + 2 \uy\uz \Nps{001} -  \uy \uzz \Nps{000} ,\nonumber
\end{eqnarray}
\vspace{-9mm}
\begin{eqnarray}
\Ns{201} = \Nps{201} - 2\ux \Nps{101}  - \uz\Nps{200}  + \uxx\Nps{001} + 2  \ux\uz \Nps{100}  -\uxx \uz \Nps{000} ,\nonumber
\end{eqnarray}
\vspace{-9mm}
\begin{eqnarray}
\Ns{111} = \Nps{111} - \ux\Nps{011}  - \uy\Nps{101} - \uz\Nps{110}  + \ux\uy\Nps{001} + \ux\uz\Nps{010} +  \uy\uz\Nps{100} -  \ux\uy\uz \Nps{000},\nonumber
\end{eqnarray}
\vspace{-9mm}
\begin{eqnarray}
\Ns{220} =\Nps{220} - 2\ux \Nps{120} - 2 \uy \Nps{012}+ \uxx\Nps{020} +\uyy \Nps{200}+ 4\ux\uy\Nps{110} - 2\ux \uyy \Nps{100} - 2\uxx\uy \Nps{010} +\uxx \uyy \Nps{000}.\nonumber
\end{eqnarray}

\noindent
Step 3: Collision update - Perform relaxation of all the 9 independent central moments to their attractors for the D2Q9 lattice. First, we calculate the lower order Markovian central moment attractors based on the following:
\begin{eqnarray}
&&\NMv{000} = T,\qquad \qquad
\NMv{100} = 0,\qquad \qquad
\Ns{010}\sps{ Mv} = 0, \qquad \qquad
\NMv{001} = 0, \qquad \qquad
\NMv{110} = 0, \nonumber \\[2mm]
&&\NMv{101} = 0,\qquad \qquad
\NMv{011} = 0, \qquad \qquad
\NMv{200} =  \chss T,\qquad \quad
\NMv{020} =\chss T, \qquad \quad
\NMv{002} =\chss T,\nonumber
\end{eqnarray}
where $\chss = 1/3$. Next, perform relaxations under collision with source updates to obtain the post-collision central moments $\Nts{mnp}$ up to the second order (i.e., $m+n+p\le 2$) as
\begin{eqnarray}
\Nts{000} = \Ns{000} + \omega\sbs{0} (\NMv{000} - \Ns{000}),\nonumber
\end{eqnarray}
\vspace{-9mm}
\begin{eqnarray}
\Nts{100} = \Ns{100} + \omega\sbs{1} (\NMv{100} - \Ns{100}),\nonumber
\end{eqnarray}
\vspace{-9mm}
\begin{eqnarray}
\Nts{010} = \Ns{010} + \omega\sbs{2} (\NMv{010} - \Ns{010}),\nonumber
\end{eqnarray}
\vspace{-9mm}
\begin{eqnarray}
\Nts{001} = \Ns{001} + \omega\sbs{3} (\NMv{001} - \Ns{001}),\nonumber
\end{eqnarray}
\vspace{-9mm}
\begin{eqnarray}
\Nts{110} = \Ns{110} + \omega\sbs{4} (\NMv{110} - \Ns{110}),\nonumber
\end{eqnarray}
\vspace{-9mm}
\begin{eqnarray}
\Nts{101} = \Ns{101} + \omega\sbs{5} (\NMv{101} - \Ns{101}),\nonumber
\end{eqnarray}
\vspace{-9mm}
\begin{eqnarray}
\Nts{011} = \Ns{011} + \omega\sbs{6} (\NMv{011} - \Ns{011}),\nonumber
\end{eqnarray}
\vspace{-9mm}
\begin{eqnarray}
\Nts{200} = \Ns{200} + \omega\sbs{7} (\NMv{200} - \Ns{200}),\nonumber
\end{eqnarray}
\vspace{-9mm}
\begin{eqnarray}
\Nts{020} = \Ns{020} + \omega\sbs{8} (\NMv{020} - \Ns{020}),\nonumber
\end{eqnarray}
\vspace{-9mm}
\begin{eqnarray}
\Nts{002} = \Ns{002} + \omega\sbs{9} (\NMv{002} - \Ns{002}),\nonumber
\end{eqnarray}
Using the above, we then specify the Markovian attractor for the third order central moments based on post-collision first order moments, and for the fourth order central moment based on post-collision second order central moments as follows:
\begin{eqnarray}
&&\NMv{120} = \frac{2}{3}\chss\Nts{100},\qquad \qquad
\NMv{012} = \frac{2}{3}\chss\Nts{010},\qquad \qquad
\NMv{201} = \frac{2}{3}\chss\Nts{001},\qquad \qquad
\NMv{111} = 0,  \nonumber\\[2mm]
&&\NMv{220} = \frac{1}{2}\chss\left(\Nts{200}+\Nts{020}\right).
\end{eqnarray}
\noindent
Following these, finally, the third and fourth order central moments are updated under relaxation as
\begin{eqnarray}
\Nts{120} = \Ns{120} + \omega\sbs{10} (\NMv{120} - \Ns{120}),\nonumber
\end{eqnarray}
\vspace{-9mm}
\begin{eqnarray}
\Nts{012} = \Ns{012} + \omega\sbs{11} (\NMv{012} - \Ns{012}),\nonumber
\end{eqnarray}
\vspace{-9mm}
\begin{eqnarray}
\Nts{201} = \Ns{201} + \omega\sbs{12} (\NMv{201} - \Ns{201}),\nonumber
\end{eqnarray}
\vspace{-9mm}
\begin{eqnarray}
\Nts{111} = \Ns{111} + \omega\sbs{13} (\NMv{111} - \Ns{111}),\nonumber
\end{eqnarray}
\vspace{-9mm}
\begin{eqnarray}
\Nts{220} = \Ns{220} + \omega\sbs{14} (\NMv{220} - \Ns{220}).\nonumber
\end{eqnarray}

\noindent
Step 4: Map post-collision central moments into post-collision raw moments using $\tilde{\n} = \Fi \tilde{\n}\sps{c}$ as
\begin{eqnarray}
\Ntps{000} = \Nts{000},\nonumber
\end{eqnarray}
\vspace{-9mm}
\begin{eqnarray}
\Ntps{100} = \Nts{100} + \ux\Nts{000},\nonumber
\end{eqnarray}
\vspace{-9mm}
\begin{eqnarray}
\Ntps{010} = \Nts{010} + \uy\Nts{000},\nonumber
\end{eqnarray}
\vspace{-9mm}
\begin{eqnarray}
\Ntps{001} = \Nts{001} + \uz\Nts{000},\nonumber
\end{eqnarray}
\vspace{-9mm}
\begin{eqnarray}
\Ntps{110} = \Nts{110} + \ux\Nts{010} + \uy\Nts{100} + \ux\uy\Nts{000},\nonumber
\end{eqnarray}
\vspace{-9mm}
\begin{eqnarray}
\Ntps{101} = \Nts{101} + \ux\Nts{001} + \uz\Nts{100} + \ux\uz\Nts{000},\nonumber
\end{eqnarray}
\vspace{-9mm}
\begin{eqnarray}
\Ntps{011} = \Nts{011} + \uy\Nts{001} + \uz\Nts{010} + \uy\uz\Nts{000},\nonumber
\end{eqnarray}
\vspace{-9mm}
\begin{eqnarray}
\Ntps{200} = \Nts{200} + 2  \ux\Nts{100}+\uxx\Nts{000},\nonumber
\end{eqnarray}
\vspace{-9mm}
\begin{eqnarray}
\Ntps{020} = \Nts{020} + 2 \uy \Nts{010}+\uyy\Nts{000},\nonumber
\end{eqnarray}
\vspace{-9mm}
\begin{eqnarray}
\Ntps{002} = \Nts{002} + 2 \uz \Nts{001}+\uzz\Nts{000},\nonumber
\end{eqnarray}
\vspace{-9mm}
\begin{eqnarray}
\Ntps{120} = \Nts{120} + \ux\Nts{020} + 2\uy\Nts{110} + \uyy\Nts{100} + 2\ux\uy\Nts{010} + \ux\uyy \Nts{000},\nonumber
\end{eqnarray}
\vspace{-9mm}
\begin{eqnarray}
\Ntps{012} = \Nts{012} + \uy\Nts{002} + 2\uz\Nts{011} + \uzz\Nts{010} + 2 \uy\uz\Nts{001}  + \uy\uzz\Nts{000},\nonumber
\end{eqnarray}
\vspace{-9mm}
\begin{eqnarray}
\Ntps{201} = \Nts{201} + 2\ux\Nts{101} + \uz\Nts{200} + \uxx\Nts{001} + 2\ux\uz\Nts{100} + \uxx\uz\Nts{000},\nonumber
\end{eqnarray}
\vspace{-9mm}
\begin{eqnarray}
\Ntps{111} = \Nts{111} + \ux\Nts{011} + \uy\Nts{101} + \uz\Nts{110} + \ux\uy \Nts{001} + \ux\uz \Nts{010} + \uy\uz\Nts{100} + \ux\uy\uz\Nts{000},\nonumber
\end{eqnarray}
\vspace{-9mm}
\begin{eqnarray}
\Ntps{220} \! \! \! & =& \! \! \! \Nts{220}   + 2\ux\Nts{120} + 2\uy\Nts{012}  - \uyy\Nts{200}+ \uxx\Nts{020} + 2\uyy\Nts{002} + 4\uy\uz\Nts{011} - 2\ux\uyy\Nts{100} \nonumber\\
 & & + 2\uy\uzz \Nts{010}+ 4\uyy\uz\Nts{001} - (\uxx\uyy - 2 \uyy\uzz)\Nts{000} . \nonumber
\end{eqnarray}

\noindent
Step 5: Convert post-collision raw moments into post-collision distribution functions via $\tilde{\mathbf{g}} = \PP\sps{\sm1}\tilde{\n}$ as
\begin{eqnarray}
\gt{0} = \Ntps{000} - \Ntps{002} - \Ntps{020} - \Ntps{200} + 2\Ntps{220}, \nonumber
\end{eqnarray}
\vspace{-9mm}
\begin{eqnarray}
\gt{1} = (\Ntps{100} - \Ntps{120} + \Ntps{200} - \Ntps{220})/2,\nonumber
\end{eqnarray}
\vspace{-9mm}
\begin{eqnarray}
\gt{2}= (\Ntps{120} - \Ntps{100} + \Ntps{200} - \Ntps{220})/2,\nonumber
\end{eqnarray}
\vspace{-9mm}
\begin{eqnarray}
\gt{3} = (\Ntps{010} - \Ntps{012} + \Ntps{020} - \Ntps{220})/2,\nonumber
\end{eqnarray}
\vspace{-9mm}
\begin{eqnarray}
\gt{4}= (\Ntps{012} - \Ntps{010} + \Ntps{020} - \Ntps{220})/2,\nonumber
\end{eqnarray}
\vspace{-9mm}
\begin{eqnarray}
\gt{5} = (\Ntps{001} + \Ntps{002} - \Ntps{201} - \Ntps{220})/2,\nonumber
\end{eqnarray}
\vspace{-9mm}
\begin{eqnarray}
\gt{6} = (\Ntps{002} - \Ntps{001} + \Ntps{201} - \Ntps{220})/2,\nonumber
\end{eqnarray}
\vspace{-9mm}
\begin{eqnarray}
\gt{7} = (\Ntps{011} + \Ntps{012} + \Ntps{101} + \Ntps{110} + \Ntps{111} + \Ntps{120} + \Ntps{201} + \Ntps{220})/8,\nonumber
\end{eqnarray}
\vspace{-9mm}
\begin{eqnarray}
\gt{8} = (\Ntps{011} + \Ntps{012} - \Ntps{101} - \Ntps{110} - \Ntps{111} - \Ntps{120} + \Ntps{201} + \Ntps{220})/8,\nonumber
\end{eqnarray}
\vspace{-9mm}
\begin{eqnarray}
\gt{9}= (\Ntps{101} - \Ntps{012} - \Ntps{011} - \Ntps{110} - \Ntps{111} + \Ntps{120} + \Ntps{201} + \Ntps{220})/8,\nonumber
\end{eqnarray}
\vspace{-9mm}
\begin{eqnarray}
\gt{10}= (\Ntps{110} - \Ntps{012} - \Ntps{101} - \Ntps{011} + \Ntps{111} - \Ntps{120} + \Ntps{201} + \Ntps{220})/8,\nonumber
\end{eqnarray}
\vspace{-9mm}
\begin{eqnarray}
\gt{11} = (\Ntps{012} - \Ntps{011} - \Ntps{101}+ \Ntps{110} - \Ntps{111} + \Ntps{120} - \Ntps{201} + \Ntps{220})/8,\nonumber
\end{eqnarray}
\vspace{-9mm}
\begin{eqnarray}
\gt{12} = (\Ntps{012} - \Ntps{011} + \Ntps{101} - \Ntps{110} + \Ntps{111} - \Ntps{120} - \Ntps{201} + \Ntps{220})/8,\nonumber
\end{eqnarray}
\vspace{-9mm}
\begin{eqnarray}
\gt{13} = (\Ntps{011} - \Ntps{012} - \Ntps{101} - \Ntps{110} + \Ntps{111} + \Ntps{120} - \Ntps{201} + \Ntps{220})/8,\nonumber
\end{eqnarray}
\vspace{-9mm}
\begin{eqnarray}
\gt{14}= (\Ntps{011} - \Ntps{012} + \Ntps{101} + \Ntps{110} - \Ntps{111} - \Ntps{120} - \Ntps{201} + \Ntps{220})/8.\nonumber
\end{eqnarray}
					
\noindent
Step 6: Perform streaming step using the post-collision distribution functions for all the 15 directions of the D3Q15 lattice as
\begin{eqnarray}
  g\sbs{\alpha}(\bm{x},t+\delta\sbs{t}) &=& \tilde{g}\sbs{\alpha}(\bm{x}-\bm{e}\sbs{\alpha}\delta\sbs{t},t), \quad \alpha = 0,1,2,\ldots, 14,
\end{eqnarray}
and implement the boundary conditions on the distribution functions as necessary. \newline
Step 7: Compute the temperature field via zeroth moment of the distribution function as
\begin{eqnarray}
T = \sum\sbs{\alpha=0}\sps{14} g\sbs{\alpha}.
\end{eqnarray}
\noindent
The choice of the relaxation parameters $\omega\sbs{j}$, which satisfy $0 < \omega_j < 2$, in the above are as follows. We select the relaxation parameters for the first order moments $\omega\sbs{1}=\omega\sbs{2}=\omega\sbs{3}$ based on the thermal transport coefficient of the fluid, viz., thermal diffusivity, $\alpha$ (which is related to the Rayleigh number or Peclet number based on the thermal flow problem statement) according to
\begin{eqnarray}\label{eqn:thermaltransportcoeff3DFPCLBM}
\alpha =\chss\left(\frac{1}{\omega\sbs{1}}-\frac{1}{2}\right)\delta_t = \chss\left(\frac{1}{\omega\sbs{2}}-\frac{1}{2}\right)\delta_t= \chss\left(\frac{1}{\omega\sbs{3}}-\frac{1}{2}\right)\delta_t.
\end{eqnarray}
so that the 3D equation of the convection and diffusion transport of energy is recovered. The remaining $\omega_j$, where $j = 0, 3, 4, 5, 6, 7, 8, 9, 10, 11, 12, 13, 14$, are free parameters and are selected based on numerical stability considerations. We set them to be equal to 1.0 in this work. As such, the Steps 1 through 7 represent one complete step of the FPC-LBM with a time step $\delta_t$ to simulate energy transport.

\newpage

\end{document}